\documentclass[AEJ]{AEA}
  \usepackage[spanish,english]{babel}
  \usepackage{relsize}\usepackage{xcolor}
 \usepackage{amsfonts,multirow,amsmath,adjustbox,natbib,bbm, bm}
 \usepackage{accents}\usepackage{booktabs}\usepackage{lscape}
 \usepackage{animate, lscape}
 \usepackage{graphicx}
 \usepackage{media9}
\usepackage{tikz}
  \usepackage[utf8]{inputenc}
 \usepackage{subcaption}
 \usepackage{longtable}
 \usepackage{bm,bigints}
 \usepackage[textsize=tiny]{todonotes}
\makeatletter
\AtBeginDocument{
\g@addto@macro{\appendix}{\renewcommand{\p@subsection}{}}
}
\usepackage[noend]{algpseudocode}
\usepackage{pseudocode}

\makeatother
 \usepackage{color,soul}
 
\usepackage{soul,color}
\usepackage[para,online,flushleft]{threeparttable}\usepackage{multirow}\usepackage{multicol}
 \usepackage{graphicx, comment}
\definecolor{ahjcolor}{rgb}{0.0, 0.13, 0.40}			
\usepackage[colorlinks	=true,										
            linkcolor	=ahjcolor,									%
            urlcolor	=ahjcolor,									%
            citecolor	=ahjcolor,
            bookmarksnumbered=true]{hyperref}
\draftSpacing{1.5}
\begin{document}
\title{Semi-parametric estimation of the EASI model: Welfare implications of taxes identifying clusters due to unobserved preference heterogeneity\thanks{We acknowledge supercomputing resources made available by the Centro de Computación Científica Apolo at Universidad EAFIT (http://www.eafit.edu.co/apolo) to conduct the research reported in this scientific product. All errors are our own.}}
\author{Andr\'es Ramírez--Hassan\thanks{Department of Economics, School of Economics and Finance, Universidad EAFIT, Medellín, Colombia. \href{aramir21@eafit.edu.co}{E-mail: aramir21@eafit.edu.co}} and Alejandro L\'opez-Vera\thanks{Department of Economics, School of Economics and Finance, Universidad EAFIT, Medell\'{i}n, Colombia, \href{E-mail: calopezv@eafit.edu.co}{E-mail: calopezv@eafit.edu.co}}}
\date{\today}\pubMonth{}\pubYear{}\pubVolume{}\pubIssue{}

\begin{abstract}
We provide a novel inferential framework to estimate the exact affine Stone index (EASI) model, and analyze welfare implications due to price changes caused by taxes. Our inferential framework is based on a non-parametric specification of the stochastic errors in the EASI incomplete demand system using Dirichlet processes. Our proposal enables to identify consumer clusters due to unobserved preference heterogeneity taking into account, censoring, simultaneous endogeneity and non-linearities. We perform an application based on a tax on electricity consumption in the Colombian economy. Our results suggest that there are four clusters due to unobserved preference heterogeneity; although 95\% of our sample belongs to one cluster. This suggests that observable variables describe preferences in a good way under the EASI model in our application. We find that utilities seem to be inelastic normal goods with non-linear Engel curves. Joint predictive distributions indicate that electricity tax generates substitution effects between electricity and other non-utility goods. These distributions as well as Slutsky matrices suggest good model assessment. We find that there is a 95\% probability that the equivalent variation as percentage of income of the representative household is between 0.60\% to 1.49\% given an approximately 1\% electricity tariff increase. However, there are heterogeneous effects with higher socioeconomic strata facing more welfare losses on average. This highlights the potential remarkable welfare implications due taxation on inelastic services.
\end{abstract}
\maketitle
\section{Introduction}\label{sec:introduction}

\textit{Ex ante} welfare analysis of new taxes should be a basic economic principle. Thus, policy makers have a baseline framework to make decisions that potentially minimize welfare losses due to tax distortions as taxation can only be a second-best option \citep{baumol1970optimal}.

This paper provides a novel inferential framework to analyze potential welfare implications due to a consumption tax based on the equivalent variation. Our inferential approach is done using an econometric framework based on a non-parametric specification of the stochastic perturbations in the exact affine Stone index incomplete demand system using Dirichlet processes to identify clusters of consumers due to unobserved preference heterogeneity taking censoring, simultaneous endogeneity and non-linearity into account. 


Consumer responses to changes in prices is instrumental for any assessment of welfare consequences of taxation, and demand models play an essential role in assessing such consequences. Our specification is based on an extension of the exact affine Stone index (EASI) demand system \citep{Lewbel2009} as this satisfies the axioms of choice such that additivity, homogeneity, and symmetry restrictions can easily be imposed in a simple setting to perform estimation. In addition, the rank in the function space spanned by the Engel curves is flexible (it can be more than three), allows the Engel curves to take arbitrary shapes, and the stochastic errors can be interpreted as unobserved consumer heterogeneity. These are nice features that other well known demand systems do not take into account, for instance, the almost ideal demand system (AIDS, \cite{Deaton1980}), and quadratic AIDS (QAIDS, \cite{Banks1997}). These properties are particularly relevant when working with micro level data due to variation in expenditure being not smoothed out by aggregation, and much of the demand variation may not be explained by observable variables \citep{Blundeletal2007, zhen2014predicting}. 

To the best of our knowledge, we are the first to propose a non-parametric specification of the stochastic errors in the EASI demand system. We extend \cite{Ramirez2021}'s inferential approach to an incomplete demand system in a semi-parametric specification, and \cite{Conley2008} and \cite{Jensen2014}'s proposals to a high dimensional system of simultaneous equations with cross-equation restrictions. Our approach easily allows to check and impose symmetry, strict cost monotonicity and concavity, and perform inference of the equivalent variation as a by-product of the posterior chains.

Our application is based on utilities demand, particularly electricity, gas, water and sewerage. These goods are approximately 5\% of households budget share in developed countries like Australia (4.6\%), Germany (5.8\%), Japan (5.7\%), South Korea (3.7\%), Spain (5.7\%) and United States (4.2\%). The average is equal to 5.9\% in developing countries like Colombia (5.8\%), Chile (5.5\%), Georgia (4.8\%), Jordan (4.7\%), Maldives (7.7\%),  Mexico (4.8\%), Peru (5.9\%) and Serbia (6.5\%). Electricity represents the highest share (3.1\%), followed by gas (1.3\%) and water (1.1\%) in these countries. In particular, we analyze the welfare implications of an approximately 1\% electricity consumption tax rate on middle to high income households using Colombian data. This electricity demand tax was implemented by the national Colombian government in July 2019 charging these households to subsidy recovery of the electricity system attending the population in the north-western Colombian region after bankruptcy of its former provider.\footnote{This tax was declined in December 2020 by order of the Colombian National Constitutional Court because \href{https://www.eltiempo.com/justicia/cortes/las-razones-de-la-corte-para-tumbar-sobretasa-de-energia-en-estratos-4-5-y-6-para-electricaribe-553358}{``this extra charge is against equity, equality, efficiency and progresivity''} (Authors translation).} This provider attended about one third of the total residential subscribers in Colombia, that is, 2.4 million users, most of them belonging to the poorest segment, about 84\% of its total subscribers. 

     There are remarkable welfare analysis due to price changes using demand systems, for instance, \cite{Banks1997} and \cite{Lewbel2009}. In particular, previous studies analyzing implications associated with electricity price changes are \cite{Schulte2017,Tovar2018,Pereira2019,Ramirez-Hassan2019}. The latter two use univariate demand approaches, whereas the former two use demand systems. \cite{Schulte2017} use a quadratic expenditure system, whereas \cite{Tovar2018} use the EASI demand system. 

We matched household level data from different sources: \href{http://microdatos.dane.gov.co/index.php/catalog/566/get_microdata}{the Colombian national household budget survey}, \href{http://www.sui.gov.co/web/}{the utilities information system}, and the regulation councils of \href{https://www.creg.gov.co/}{energy and gas}, and \href{https://www.cra.gov.co/seccion/inicio.html}{water and sewerage} to perform our application. Our results seem to be consistent with previous literature in utilities, that is, electricity, water, gas and sewerage are inelastic normal services. It seems that Engel curves of utilities in our application are non-linear. This feature seems to be particularly relevant in the tails of the income distribution, and in sewerage. In addition, predictive distributions, which take simultaneous endogeneity into account, seem to be consistent with observable data, and predict substitutions effects between electricity and other non-utility goods given electricity price changes. Predictive p-values indicate good model assessment for inferring price effects based on the Slutsky matrix. Moreover, we identify four clusters due to unobserved preference heterogeneity, but evidence suggests that observable variables describe in good way preferences under the EASI model in our application. There is heterogeneity regarding welfare implications of this new tax on electricity demand; high socioeconomic households face the highest losses on average. In particular, there is a 95\% probability that the equivalent variation as percentage of income of the representative household is between 0.60\% to 1.49\% given an approximately 1\% electricity tariff increase. This result highlights the remarkable welfare implications due to tariff increases of inelastic goods.

After this introduction, we briefly sum up the incomplete demand EASI model for facility in exposition in Section \ref{sec:sumEASI}. Particular emphasis is on equivalent variation in Section \ref{sec:EqVar}. Section \ref{sec:econometrics} shows our econometric proposal. In particular, we set our specification in Section \ref{sec:spec}, sections \ref{sec:post} and \ref{sec:Predictive} show our inferential framework and construction of the predictive distribution, respectively. The hypothesis testing setting of microeconomic restrictions and model assessment are set in sections \ref{sec:micrest} and \ref{sec:pval}. Section \ref{sec:data} shows data construction and descriptive statistics, and Section \ref{sec:results} displays inferential and predictive results regarding price effects, Engel curves and welfare estimates. Section \ref{sec:concl} concludes and give guidelines for future research. We show algorithmic details and descriptive statistics by relevant clusters (observable and unobservable) in the Appendix (Section \ref{sec: appendix}).         

\section{Model}\label{sec:model}

The EASI incomplete demand system is an extension of the EASI demand system \citep{Lewbel2009} proposed by \cite{zhen2014predicting}. Incomplete demand systems give correct welfare change measures without invoking the weak separability assumption and the exogeneity of group expenditures \citep{zhen2014predicting}. This is done by introducing in the demand system a composite \textit{numéraire} good representing all other goods and services, and using household income rather than group expenditure.

\subsection{Summary: The EASI model}\label{sec:sumEASI}

Details of the EASI demand system, and its incomplete version, can be found in \cite{Lewbel2009} and \cite{zhen2014predicting}. However, we provide here the basic framework for exposition convenience. In our application, we model simultaneously five implicit Marshallian budget shares including the composite \textit{numéraire} good ($J=5$), $\bm\omega(\bm p_i, y_i, \bm z_i, \bm \epsilon_i) \equiv \bm w_i = \begin{bmatrix} \text{electricity}_i & \text{water}_i & \text{gas}_i & \text{sewerage}_i & \text{numéraire}_i\end{bmatrix}^{\top}$ for $i=1,2,\dots, N$ households,\footnote{We do not include interaction effects between socioeconomic characteristics and prices in our main specification following \cite{zhen2014predicting} and \cite{Tovar2018}. Main results are robust regarding these effects in our application.}
\begin{equation}\label{refM:eq1}
\bm{w}_i=\sum_{r=0}^R\bm{b}_ry^r_i+\bm{Cz}_i+\bm{Dz}_iy_i+\bm{A}\bm{p}_i+\bm{Bp}_iy_i+\bm{\epsilon}_i,
\end{equation}

\begin{equation}\label{refM:eq2}
y_i=\frac{x_i-\bm{p}^{\top}_i\bm{w}_i+\bm{p}^{\top}_i\bm{A}\bm{p}_i/2}{1-\bm{p}^{\top}_i\bm{B}\bm{p}_i/2},
\end{equation}
where the price index for the \textit{numéraire} good is $p_{iJ}=\frac{\log(\text{CPI}_i)-\sum_{j=1}^{J-1} w_{ij}p_{ij}}{w_{iJ}}$, $\text{CPI}$ is the consumer price index. The implicit utility, $y$, is an exact affine transformation of the (log) Stone index, $x-\bm{p}^{\top}\bm{w}$, $x$ is the (log) nominal income and $\bm p$ is the (log) Fisher ideal price index vector based on the sample average \citep{zhen2014predicting}. The space spanned by the price vectors can have any rank up to $4$ in our case. The system of equations \ref{refM:eq1} can involve any polynomial degree in $y$, which gives flexibility for the Engel curves, and take into account socioeconomic controls $l=1,2,\dots,L$ ($\bm z$, $z_0=1$ excluded from $\bm z$). In addition, the stochastic error ($\bm{\epsilon}$) can be interpreted as unobserved preference heterogeneity. Notice that this system (equations \ref{refM:eq1} and \ref{refM:eq2}) is endogenous and non-linear in parameters ($\bm b_r, \bm C, \bm D, \bm A, \bm B$). In our application, the main endogeneity concern is due to simultaneity between budget shares and the measure of real total income. We do not consider price endogeneity due to using household level data and regulated natural monopoly markets taking prices directly from provider records. The former prevents demand-supply simultaneity, and the latter prevents searching provider strategies and price measurement errors.\footnote{\cite{zhen2014predicting} use a weighted average price based on adjacent locations as the instrument of price. Observe that this approach is equivalent to ours as price in a region is the same for all households conditional on strata in our application.} 

To achieve cost function regularity in the EASI model, the sets of coefficients in the system of equations \ref{refM:eq1} have to satisfy $\bm{1}_J^{\top}\bm{b}_0=1$, $\bm{1}_J^{\top}\bm{b}_r=0$ for $r\neq 0$, $\bm{1}_J^{\top}\bm{A}=\bm{1}_J^{\top}\bm{B}=\bm{0}_J^{\top}$, $\bm{1}_J^{\top}\bm{C}=\bm{1}_J^{\top}\bm{D}=\bm{0}_L^{\top}$ and $\bm 1_J^{\top}\bm{\epsilon}=0$ for adding up constraints, $\bm{A}\bm{1}_J=\bm{B}\bm{1}_J=\bm{0}_J$ for cost function homogeneity, strict cost monotonicity requires $\bm p^{\top}\left[\sum_{r=0}^R \bm b_r r y^{r-1}+\bm D \bm z + \bm B \bm p / 2\right]>-1$, Slutsky symmetry is ensured by symmetry of $\bm A$ and $\bm B$, and a sufficient and necessary condition for concavity of the cost function is negative semidefiniteness of the normalized Slutsky matrix $\bm A+\bm B y+\bm w \bm w^{\top}- \bm W$, where $\bm W$ is the diagonal matrix composed by $\bm w$, and $\bm{1}_J$ is a $J$-dimensional vector of ones.

The compensated share price semi-elasticities are given by
\begin{equation}\label{refM:eq3}
\nabla_{\bm p}\bm\omega(\bm p, y, \bm z, \bm \epsilon)\equiv \bm\Gamma =\bm A+\bm By,
\end{equation}
whose value for the representative individual is $\bm A$ given that $y=0$ because baseline prices are equal to 1, and demeaned nominal income ($x=0$).     

The real income share semi-elasticities are 
\begin{equation}\label{refM:eq4}
\nabla_y\bm\omega(\bm p, y, \bm z, \bm \epsilon)=\sum_{r=1}^R \bm b_r r y^{r-1}+\bm D\bm z+\bm B \bm p.    
\end{equation} 

Household characteristics share semi-elasticities are given by

\begin{equation}\label{refM:eq5}
\nabla_{z_l}\bm \omega\bm{(p,z},y,\bm\epsilon)= \bm{c}_{l}+\bm{d}_{l}y, 
\end{equation}

where $\bm{c}_{l}$ and $\bm{d}_{l}$ are columns from $\bm C$ and $\bm D$, respectively.

The normalized Slutsky matrix is

\begin{equation}\label{refM:eq5A}
\bm S = \bm\Gamma+\bm w \bm w^{\top}-\bm W.
\end{equation}



The Marshallian demand functions in the EASI model are implicitly given by:
\begin{equation*}
\bm w (\bm{p},x, \bm{z}, \bm\epsilon)= \bm\omega \left(\bm{p}, \frac{x -\bm{p}^{\top} \bm{w}(\bm{p},x, \bm{z}, \bm\epsilon) + \bm{p}^{\top}  \bm{A} \bm p/2}{1-\bm{p}^{\top}  \bm{B} \bm p/2}, \bm{z}, \bm\epsilon\right). 
\end{equation*}

Solving for the Marshallian share semi-elasticities with respect to nominal income is

\begin{equation}\label{refM:eq6}
\nabla_{x}\bm{w}(\bm p, x, \bm{z}, \bm\epsilon)=\left(\bm{I}_{J} - \frac{\nabla_{y}\bm \omega (\bm{p}, \bm z,y,\bm\epsilon)\bm{p}^{\top}}{1-\bm{p}^{\top} \bm{B} \bm p/2}\right)^{-1}\left(\frac{(1-x) \nabla_{y}\bm \omega( \bm{p}, \bm{z},y,\bm\epsilon)}{1-\bm{p}^{\top} \bm{B} \bm p/2}\right).
\end{equation}

Marshallian share price semi-elasticities are recovered from Hicksian share price semi-elasticities (equation \ref{refM:eq3}) and the above Marshallian share income semi-elasticities using the Slutsky matrix.

\begin{equation}\label{refM:eq7}
\nabla_{\bm p}\bm{w}(\bm p, x, \bm{z}, \bm\epsilon)=
\nabla_{\bm p}\bm \omega(\bm{p},y,\bm {z},\bm\epsilon) -
\nabla_{x}\bm{w}(\bm{p}, x, \bm{z}, \bm\epsilon) \bm{\omega}(\bm{p},\bm{z},y,\bm\epsilon)^{\top}.
\end{equation}

Observe that using the causal path diagram (Figure \ref{fig:CausalDia}) we can decompose the causal effect of price on Marshallian shares into the direct effect ($\nabla_{\bm p}\bm \omega(\bm{p},y,\bm {z},\bm\epsilon)$), and the indirect (mediator) effect through the nominal income ($\bm B$).

In addition, the Hicksian demand price elasticities are

\begin{equation}\label{refM:eq8}
    \epsilon_{lj}^H=-\mathbbm{1}(l=j)+\frac{1}{w_l}\nabla_{\bm p}\bm\omega(\bm p, y, \bm z, \bm \epsilon)_{lj}+w_j,
\end{equation}
where $\mathbbm{1}(\cdot)$ is the indicator function and $\nabla_{\bm p}\bm\omega(\bm p, y, \bm z, \bm \epsilon)_{lj}$ is the $lj$-th element of the matrix $\nabla_{\bm p}\bm\omega(\bm p, y, \bm z, \bm \epsilon)$. The Marshallian demand income elasticities are

\begin{equation}\label{refM:eq9}
    \eta_j^M=\frac{1}{w_j}\nabla_{x}\bm{w}(\bm p, x, \bm{z}, \bm\epsilon)_j+1,
\end{equation}

where $\nabla_{x}\bm{w}(\bm p, x, \bm{z}, \bm\epsilon)_j$ is the $j$-th element of the vector $\nabla_{x}\bm{w}(\bm p, x, \bm{z}, \bm\epsilon)$. Using again the Slutsky decomposition with equations \ref{refM:eq7} and \ref{refM:eq8}, we obtain the Marshallian price elasticities,

\begin{equation}\label{refM:eq10}
    \epsilon_{lj}^M=-\mathbbm{1}(l=j) + \frac{1}{w_l}\nabla_{\bm p}\bm\omega(\bm p, y, \bm z, \bm \epsilon)_{lj}-\frac{w_j}{w_l}\nabla_{x}\bm{w}(\bm p, x, \bm{z}, \bm\epsilon)_l.
\end{equation}

The Marshallian Engel curves for household $i$ at the base prices $\bm p=\bm 0$ where $x_i=y_i$ are
\begin{equation}\label{EngelCurve}
    \bm{w}_i=\sum_{r=0}^R\bm{b}_r x^r_i+\bm{Cz}_i+\bm{Dz}_i x_i+\bm{\epsilon}_i.
\end{equation}

These functions summarize the parameter estimates associated with income.

\subsection{Welfare analysis}\label{sec:EqVar}

We use the equivalent variation (EV) which is the maximum amount a consumer would be willing to pay at income level $x$ to avoid the change from price vector $\bm{P}^{0}$ to $\bm{P}^{1}$ \citep[~ p. 82]{Mascollel1995}, where the superscripts 0 and 1 denote prices before and after change. \citet*{Chipman1980} showed that the equivalent variation (EV) is generally the relevant measure for performing welfare analysis in a context in which different tariff policies are ordered due to using the actual price vector as reference. In the case of the EASI model, this is \citep{Tovar2018}:

\begin{align}\label{refM:eq11}
EV & =  x - \exp \left\{ \log (x)-\sum_{j=1}^J\left( \log(P^{1}_j w_j^1) - \log(P^{0}_j w_j^0)\right) \right.\nonumber \\
& \left. + \frac{1}{2} \sum_{l=1}^J \sum_{j=1}^J a_{lj}\left(\log(P_l^1) \log(P_j^1)-\log(P_l^0)\log(P_j^0)\right)
\right\}\nonumber\\
& = x\left(1-\prod_{j=1}^J\left(\frac{P_j^1w_j^1}{P_j^0w_j^0}\right)^{-1}\exp\left(\frac{1}{2}\sum_{l=1}^J\sum_{j=1}^J a_{lj}(\log(P_l^1)\log(P_j^1)-\log(P_l^0)\log(P_j^0))\right)\right),
\end{align}

where $a_{lj}$ is the $lj$-th element of matrix $\bm A$, $P_j^1=(1+\Delta_{j})P_j^0$ and $w_j^1=w_j^0+\nabla_{\bm p}\bm w_{jl}\times \Delta_{l}$, $\nabla_{\bm p}\bm w_{jl}$ is the $jl$-th element of the Marshallian share price semi-elasticities (equation \ref{refM:eq7}), and $\Delta_{j}$ is the new tax rate on good $j$.

In the case where there is a new tax in just one good, say good $l$, equation \ref{refM:eq11} becomes
\begin{align}\label{refM:eq11a}
EV_l & = x\left(1-\frac{1}{1+\Delta_l}\prod_{j=1}^J\left(\frac{w_j^0}{w_j^0+\nabla_{\bm p}\bm w_{jl}\times \Delta_{l}}\right)\left(P_j^0\right)^{a_{lj}\log(1+\Delta_l)/2}\left(P_j^0\right)^{a_{jl}\log(1+\Delta_l)/2}(1+\Delta_l)^{a_{ll}\log(1+\Delta_l)/2}\right)\nonumber\\
& = x\left(1-(1+\Delta_l)^{a_{ll}\log(1+\Delta_l)/2-1}\prod_{j=1}^J\left(\frac{w_j^0}{w_j^0+\nabla_{\bm p}\bm w_{jl}\times \Delta_{l}}\right)\left(P_j^0\right)^{a_{lj}\log(1+\Delta_l)}\right),
\end{align}


where the second equality follows assuming symmetry.\footnote{We predict income shares in the denominator of equation \ref{refM:eq11a} using Marshallian elasticities to avoid singularity issues arising from using the predictive distribution, and improve precision of estimates (see Section \ref{sec:Pred} for more discussion). This approach seems plausible in our application given the small price change (0.8\%).}

Observe that this expression incorporates unobserved heterogeneous, substitution and income effects through $w_j^0$, $\bm{A}$ and $\nabla_{\bm p}\bm w_{jl}$, respectively.

\section{Econometric Framework}\label{sec:econometrics}

We propose a semi-parametric estimation of the EASI incomplete demand system extending \cite{Ramirez2021}'s parametric proposal. We base our approach on \cite{Conley2008}'s ideas, but in a high dimensional system of equations with cross-equation restrictions in latent variables. In particular, we model the stochastic errors using Dirichlet processes, this is particularly appealing in the EASI model as stochastic errors are interpreted as unobserved consumer heterogeneity, then we can potentially identify conditional clusters, and analyze welfare implications associated with particular consumer groups. Our inferential framework also allows to test and impose symmetry, strict cost monotonicity and concavity of the cost function, handle censored data, take parameter non-linearities, simultaneous endogeneity and obtain coherent predictive distributions including estimation errors in systems of simultaneous equations. \cite{Conley2008} show that a Bayesian semi-parametric specification is more efficient than parametric Bayesian or classical methods when the errors are non-normal.

\subsection{Specification}\label{sec:spec}

We assume specific equations in the ``structural latent forms'' (system \ref{ref:eq6} and equation \ref{refM:eq2}) and reduce form (``first stage system'' \ref{ref:eq7}), but stochastic errors distribute non-parametrically. In particular, we have the following system of simultaneous equations for $i=1,2,\dots, N$:
\begin{equation}\label{ref:eq6}
    \tilde{\bm w}_i^*=\bm F_i \bm \phi +\bm\epsilon_i,
\end{equation}
\begin{equation}\label{ref:eq7}
    \bm y_i^*=\bm G_i \bm \psi +\bm v_i.
\end{equation}

The structural form system \ref{ref:eq6} has the set of unobserved latent budget shares $\tilde{\bm w}_i^*=\left[w_{i1}^* \ w_{i2}^* \dots w_{iJ-1}^*\right]^{\top}$, $\bm F_i=\left[\bm Y_i \ \bm W_i^{w}\right]$, $\bm Y_i=\left[\left[\bm I_{J-1}\otimes (y_i \ y_i^2 \dots y_i^R \ \bm z_{i}^{\top} y_i)\right] \ \left[(\bm I_{J-1}\otimes \tilde{\bm p}_i^{\top})y_i\bm D_{J-1}\right]\right]$ where there is the log of price ratios $\tilde{\bm p}_i=\left[\log(P_{i1}/P_{iJ}) \ \log(P_{i2}/P_{iJ}) \dots \log(P_{iJ-1}/P_{iJ})\right]^{\top}$, $\bm D_{J-1}$ is the $(J-1)^2\times J(J-1)/2$ dimensional duplication matrix ($\bm D_{J-1}vech(\bm H^{\top})=vec(\bm H^{\top})$, $\bm H$ is a symmetric matrix), and $\otimes$ is the Kronecker product. Observe that $\bm Y_i$ has the endogenous regressors from the system of equations \ref{refM:eq1}. Meanwhile, the set of exogenous regressors is 
$\bm W_i^{w}=\left[\left[\bm I_{J-1}\otimes \bm z_{i}^{\top}\right] \ \left[(\bm I_{J-1}\otimes \tilde{\bm p}_i^{\top})\bm D_{J-1}\right]\right]$. Observe that we omit one latent share due to the adding up restriction implies that the $J$ dimensional joint density function is degenerate. 

In addition, $\bm{\phi}=\left[\bm{\beta}^{\top} \ \bm{\delta}^{\top}\right]^{\top}$, $\bm{\beta}=\left[b_{11}\dots b_{1R} \ D_{11} \dots D_{1L} \ b_{21} \dots D_{J-1L} \ vech(\bm B^{\top})\right]^{\top}$ and $\bm{\delta}=\left[C_{11} \dots C_{1L} \ C_{21} \dots C_{2L} \dots C_{J-1L} \ vech(\bm{A}^{\top})\right]^{\top}$ are the vectors of endogenous coefficients and exogenous coefficients, respectively. 

The system of reduced form equations is given by \ref{ref:eq7} where $\bm y_i^*=\left[y_i \ y_i^2 \dots y_i^R \ \bm z_{i}^{\top} y_i \ \tilde{\bm p}_i^{\top}y_i \right]^{\top}$ is a $q=R+L+J-1$ dimensional vector, and $\bm G_i=\left[\bm Z_i \ \bm W_i^y\right]$ where the matrix of exogenous regressors is $\bm W_i^y=\left[\left[\bm I_{q}\otimes \bm z_{i}^{\top}\right] \ \left[\bm I_{q}\otimes \tilde{\bm p}_i^{\top}\right]\right]$, and the matrix of instruments is $\bm Z_i=\left[\left[\bm I_{q}\otimes (\tilde{y}_i \ \tilde{y}_i^2 \dots \tilde{y}_i^R \ \bm z_{i}^{\top} \tilde{y}_i)\right] \ \left[(\bm I_{q}\otimes \tilde{\bm p}_i^{\top})\tilde{y}_i\right]\right]$ where $\tilde{y}_i=x_i-\bm{p}_i^{\top}\bar{\bm w}$ is proposed by \cite{Lewbel2009} as baseline instrument, $\bar{\bm w}$ is the sample average.

\begin{figure}[!h]
\caption{Wright-Burks-Pearl type causal path diagram for the EASI model.}
\includegraphics[width=5cm]{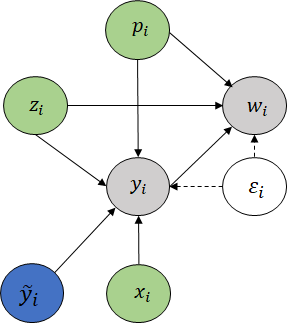}
\centering
\label{fig:CausalDia}
{\scriptsize \caption*{\textit{Notes}: Unmeasurable factors affect implicit utility and income shares. The instrumental variables built using $\tilde{y}_i$ help to identify the causal effects.}}
\end{figure}

Observe that we can give a causal interpretation of our econometric specification of the EASI model that can be represented using the Wright-Burks-Pearl causal diagram shown in Figure \ref{fig:CausalDia} (see \cite{Wright1921,burks1926inadequacy,pearl1995causal} for origins and details).

By construction, there is simultaneous causality between the income budget shares and the implicit utility. We can see this simultaneous causality as unmeasurable factors which influence both household's income shares and implicit utility. Therefore, the instruments built using $\tilde{y}_i$ help to identify the causal effect of the implicit utility on income shares, and as a consequence, identify direct and indirect effects associated with prices, nominal income and socioeconomic controls.

Regarding these unmeasurable factors, which are interpreted as unobserved preference heterogeneity in the EASI model, we assume the following: 
\begin{equation}\label{eq:errors}
\left.\begin{bmatrix} \bm\epsilon_i \\ \bm v_i \end{bmatrix}\right\vert \bm\mu_i, \bm\Sigma_i\stackrel{iid}{\sim} N(\bm\mu_i, \bm \Sigma_i),
\end{equation}
where $\bm\Sigma_i=\begin{bmatrix} \bm\Sigma_{\epsilon\epsilon,i} & \bm\Sigma_{\epsilon v,i}\\ \bm\Sigma_{v\epsilon,i} & \bm\Sigma_{vv,i} \end{bmatrix}$, $\bm\mu_i, \bm\Sigma_i| G\stackrel{i.i.d}\sim G$, and $G|G_0,\alpha\sim DP(\alpha,G_0(\cdot|\bm \kappa))$, that is, the unknown distribution $G$ has a Dirichlet process mixture prior (DPM) with base distribution $G_0(\cdot|\bm \kappa)$ with unknown parameters $\bm \kappa$, and precision parameter $\alpha$ \citep{Ferguson1973}. In particular, $\mathbb{E}\left[G(A)\right]=G_0(A)$ and $\mathbb{V}ar\left[G(A)\right]=G_0(A)\left[1-G_0(A)\right]/(1+\alpha)$ for a given event $A$ \cite[p. ~8]{Muller2015}; observe that $\alpha  \rightarrow \infty$ then $G$ concentrates at $G_0$. We assume conjugate prior specifications for computational convenience \citep{Conley2008, Jensen2014},

\begin{equation}\label{eq:priors}
\bm\mu_i|\bm\Sigma_i\sim N\left(\bm\mu_0,\frac{1}{\tau_0} \bm\Sigma_i\right), \  \bm\Sigma_i\sim IW(r_0, \bm R_0), \ \alpha\sim G(\alpha_0, \beta_0), \ \bm\phi\sim N(\bm\phi_0, \bm\Phi_0) \ \text{and} \ \bm\psi\sim N(\bm\psi_0, \bm\Psi_0),     
\end{equation}

that is the base distribution $G_0$ is a normal-inverse Wishart distribution.

This is a very flexible way to model the stochastic errors in our specification with a nice economic interpretation. The DPM establishes that the unknown distribution of $\left[\bm\epsilon_i \ \bm v_i\right]$ is an infinite mixture \citep{Antoniak1974} $\int F_N(\bm \mu, \bm \Sigma) dG(\bm \theta)$ where $\bm\theta = \left\{\bm \mu, \bm \Sigma\right\}$, $dG(\bm \theta)=\sum_{m=1}^{\infty}\pi_m\delta_{\bm\theta_m}(\cdot)$ \citep{Sethuraman1994}, $F_N$ is a Normal distribution with mean $\bm \mu$ and variance matrix $\bm \Sigma$ (equation \ref{eq:errors}), $\delta_{\bm\theta_m}(\cdot)$ denotes the measure given mass one to the atom $\bm\theta_m$, $\pi_1=v_1$, $\pi_m=v_m\prod_{h<m}(1-v_h)$, $v_m\sim Be(1, \alpha)$. Observe that non-zero mean stochastic errors allows greater flexibility in this distribution than fixing it at zero.\footnote{This implies that we do not require intercepts in the system (\ref{ref:eq6} and \ref{ref:eq7}).} In addition, the hierarchical representation of the Dirichlet process \citep{Antoniak1974} induces a probability model on clusters. This implies two advantages in our setting: first, it facilities posterior computation, and second, the clusters can be interpreted as groups of households with same conditional unobserved preferences (random utility parameters). However, we should acknowledge that DPM is not consistent for the number of clusters, but there is posterior asymptotic concentration in the other model's components such as structural parameters and predictive densities \citep{Miller2014}. 

The hierarchical representation implies that there are latent assignment variables $s_i=m$ such that when $\bm\theta_i$ is equal to $m$-th unique $\bm\theta_m^*$, that is, $\bm\theta_i=\bm\theta_m^*$ then $s_i=m$. Given a DP prior, 
\begin{equation}
  s_i\sim \sum_{m=0}^{\infty} \pi_m\delta_m,  
\end{equation}
and
\begin{equation}
    \left.\begin{bmatrix} \bm\epsilon_i \\ \bm v_i \end{bmatrix}\right\vert s_i, \bm\theta_{s_i}\stackrel{iid}{\sim} N(\bm\mu_{s_i}, \bm \Sigma_{s_i}),
\end{equation}
where $\pi_m=P(\bm\theta_i=\bm\theta_m^*)$.

Following \cite{Kasteridis2011}, the link between the latent shares and the observed shares is
\begin{equation}\label{ref:eq8}
    w_{ij}=\begin{Bmatrix}
    w_{ij}^*/\sum_{j\in J_i}w_{ij}^*, & w_{ij}^*>0\\
    0, & w_{ij}^*\leq 0
    \end{Bmatrix}, \ J_i=\left\{j: w_{ij}^*>0\right\}, \sum_{j=1}^J w_{ij}^*=1.
\end{equation}

\cite{Wales1983} note that the transformation \ref{ref:eq8} has the property that the resulting density function is independent of which set of $J-1$ latent shares are used in its derivation. 

\subsection{Posterior inference}\label{sec:post}

Setting $\bm{Data}=\bm w, \bm p, \bm z, \bm x$, the posterior distribution is proportional to
\small
\begin{align}\label{ref:eq9}
    \pi(\bm\phi, \bm\psi,\left\{\bm\Sigma_i\right\},\left\{\bm\mu_i\right\},\bm s,\alpha,\tilde{\bm w}^*, \bm y^*|\bm{Data})\propto&\prod_{s_m=1}^M\bigintss\left\{\prod_{i\in s_m}\prod_{j=1}^{J-1}\left\{\mathbbm{1}(w_{ij}=0)\mathbbm{1}(w_{ij}^*\leq 0)+\mathbbm{1}\left(w_{ij}=w_{ij}^*/\sum_{j\in J_i}w_{ij}^*\right)\mathbbm{1}(w_{ij}^*> 0)\right\}\right.\nonumber\\
    &\times \left.\pi(\tilde{\bm w}_i^*,\bm y^*_i|\bm\phi,\bm\psi,\left\{\bm\Sigma_{i}\right\},\left\{\bm\mu_{i}\right\},s_i)\right\}dG_0(\left\{\bm\Sigma_{i}\right\},\left\{\bm\mu_{i}\right\})\pi(s_i|\alpha)\pi(\alpha)\pi(\bm\phi)\pi(\bm\psi),
\end{align}

\normalsize
where $M$ is the number of clusters.

Equation \ref{ref:eq9} does not have a closed analytical expression. Then, we group parameters in order to obtain standard conditional posterior distributions that facilitate posterior computations due to the Pólya urn characterization of the DPM \citep{Blackwell1973}.
\begin{align*}
    &\pi(s_i|\left\{\bm\Sigma_{i'},\bm\mu_{i'},\bm s_{i'}:i'\neq i\right\}, \bm u_i, \alpha),\\
    &\pi(\bm\Sigma_l|\left\{\bm u_{i'}:i'\in s_i=m\right\}, \bm s),\\  
    &\pi(\bm\mu_l|\left\{\bm u_{i'}:i'\in s_i=m\right\}, \bm\Sigma_l, \bm s),\\
    &\pi(\alpha|\bm s, N),\\
    &\pi(\bm\phi|\bm\psi,\left\{\bm\Sigma_{i}\right\},\left\{\bm\mu_{i}\right\}, \bm s,\tilde{\bm w}^*, \bm y^*),\\
    &\pi(\bm\psi|\bm\phi,\left\{\bm\Sigma_{i}\right\},\left\{\bm\mu_{i}\right\}, \bm s,\tilde{\bm w}^*, \bm y^*),\\
    &\pi(\tilde{\bm w}^*| \bm\psi,\bm\phi,\left\{\bm\Sigma_{i}\right\},\left\{\bm\mu_{i}\right\}, \bm s, \bm y^*),\\
    &\pi( \bm y^*|\bm\phi,\bm{Data}),
\end{align*}

where $\bm u_i=\begin{bmatrix} \tilde{\bm w}^*\\ \bm y^* \end{bmatrix}-\begin{bmatrix} \bm F_i\bm\phi\\ \bm G_i\bm\psi \end{bmatrix}$.

These blocks allow a Gibbs sampling algorithm. See Appendix \ref{sec:MCMC} for expressions of the posterior conditional distributions and algorithmic details.

We used non-informative hyperparameters, that is, hyperparameters that are consistent with sample information in the set of prior distributions (expression \ref{eq:priors}). In particular, $\bm\phi_0=\bm 0_{\dim\left\{\bm\phi\right\}}$, $\bm\psi_0=\bm 0_{\dim\left\{\bm\psi\right\}}$, $\bm\Phi_0=100\bm I_{\dim\left\{\bm\phi\right\}}$, $\bm\Psi_0=100\bm I_{\dim\left\{\bm\psi\right\}}$, $r_0=q+J+1$, $\bm R_0=\bm I_{q+J-1}$, $\bm\mu_0=\bm 0_{q+J-1}$, $\tau_0 = 0.01$, and $\alpha_0=\beta_0=0.1$. In general, structural parameters and predictive densities are not sensitive to hyperparameters \citep{Escobar1995,Conley2008,Jensen2014}. However, higher values of the smoothing hyperparameter ($\tau_0$) affects the number of modes in the stochastic errors, although, from a practical point of view, the differences are small \citep{Escobar1995}. We decided to choose a small value to avoid this issue. The number of clusters is affected by $\alpha_0$ and $\beta_0$ \citep{Escobar1995,Conley2008,Jensen2014}, but in general there is not posterior consistency regarding this not matter the values of the hyperparameters are \citep{Miller2014}.

\subsection{Predictive density}\label{sec:Predictive}

Observe that we based our inferential framework on conditional statements of each household's income shares to sample the latent shares. This is not possible in the predictive analysis due to not observing budget shares for the potentially unobserved household $0$. We follow ideas of \cite{Jensen2014} extending their approach to a high dimensional system of contemporaneous simultaneous equations with cross-equation restrictions.

Taking into account that $\begin{bmatrix} 
    \tilde{\bm w}_{0}^*\\
    \bm y_{0}^*
    \end{bmatrix}\perp \begin{bmatrix} 
    \tilde{\bm w}_i^*\\
    \bm y_i^*
    \end{bmatrix}  \Bigg|\Lambda$, where $\Lambda=\left\{\bm\phi, \bm\psi,\bm s,\left\{\bm\Sigma_i\right\},\left\{\bm\mu_i\right\},\alpha\right\}$, the marginal predictive density of $\tilde{\bm w}_{0}^*$ at the potentially unobserved household $0$ is given by\footnote{Observe that we can get the posterior predictive distribution based on an observed household. We perform this in our application based on the representative household to perform model assessment based on predictive p-values (see section \ref{sec:pval}).}
    
\begin{align}\label{ref:margpred1}
    f(\tilde{\bm w}_{0}^*|\bm{Data})&=\int f(\tilde{\bm w}_{0}^*,y_0|\bm{Data})dy_0\nonumber\\
    &=\int \int f(\tilde{\bm w}_{0}^*,y_0|\bm\Lambda)\pi(\bm\Lambda|\bm{Data})d\bm\Lambda dy_0\nonumber\\
    &=\int \int f(\tilde{\bm w}_{0}^*|y_0,\bm\Lambda)f(y_0|\bm\Lambda)dy_0\pi(\bm\Lambda|\bm{Data})d\bm\Lambda \nonumber\\
    &=\mathbb{E}_{\bm\Lambda}\left[\mathbb{E}_{y_0|\bm\Lambda}\left[f(\tilde{\bm w}_{0}^*|y_0,\bm\Lambda)\right]\right]\nonumber\\
    &\approx \frac{1}{S^1}\sum_{s^1=1}^{S^1} \left[\frac{1}{S^2}\sum_{s^2=1}^{S^2} f(\tilde{\bm w}_{0}^*|y_0^{(s^2)},\bm\Lambda^{(s^1)})\right],
\end{align}

where for each random draw from the posterior distribution of $\bm\Lambda^{(s^1)}, s^1=1,2,\dots,S^1$, there are $S^2$ random draws from the posterior predictive distribution of $y_0^{(s^2)}, s^2=1,2,\dots,S^2$.

The conditional predictive density of $\tilde{\bm w}_{0}^*$ is 

\begin{align}\label{ref:margpred2}
    f(\tilde{\bm w}_{0}^*|y_0^{(s)},\bm\Lambda^{(s)})&=\sum_{m=1}^{M^{(s)}}\frac{N_m^{(s)}}{\alpha^{(s)}+N} f_N(\tilde{\bm w}_{0}^*|\bm\mu_{m,\tilde{\bm w}^*}^{(s)}+\bm F_0^{(s)}\bm\phi^{(s)},\bm\Sigma_{\bm\epsilon\bm\epsilon,m}^{(s)})\\
    &+\frac{\alpha^{(s)}}{\alpha^{(s)}+N}f_{MSt}\left(\tilde{\bm w}_{0}^*\Big|\bm\mu_{0,\tilde{\bm w}^*}+\bm F_0^{(s)}\bm\phi^{(s)},\frac{1+\tau_0}{\tau_0}\frac{\bm R_{11,0}}{v},v\right),\nonumber
\end{align}

where $\bm R_{11,0}$ is the upper-right $J-1 \times J-1$ block matrix of $\bm R_{0}$, $\bm\mu_{k,\tilde{\bm w}_{0}^*}$ is the first $J-1$ elements of $\bm\mu_k$, $k=\left\{m,0\right\}$, $N_m$ is the number of observations such that $s_i=m$, $v=r_0+1-(q+J-1)$, and $F_0^{(s)}$ has the same structure than $\bm F_i$, but using values for the household 0 regarding prices and socioeconomic characteristics, and the predictive distribution of $y_0^{(s)}$ is

\begin{align}\label{eq:y0}
    f(y_0|\bm\Lambda^{(s)})&=\sum_{m=1}^M \frac{N_m^{(s)}}{\alpha^{(s)}+N}f_N(y_0|\mu_{m,y}^{(s)}+\bm G_{1,0}\bm \psi^{(s)},\sigma_{1,vv,m}^{(s)})\nonumber\\
    &+\frac{\alpha^{(s)}}{\alpha^{(s)}+N}f_{St}\left(y_0|\mu_{0,y}+\bm G_{1,0}\bm \psi^{(s)},\frac{1+\tau_0}{\tau_0}\frac{r_{JJ,0}}{v},v\right).
\end{align}
 
where $\mu_{m,y}$ is the $J$-th element of $\bm\mu_k$, $\bm G_{1,0}$ is the first row $\bm G_0$, which takes same structure as $\bm G_i$, but using values associated with household $0$, $\sigma_{1,vv,m}$ is the $11$-th element of the matrix $\bm\Sigma_{vv,m}$, and $r_{JJ,0}$ is the $JJ$-th element of the matrix $\bm R_0$. 

In addition, $w_{0J}^*$ has a degenerate density with probability one at $1-\sum_{j=1}^{J-1}w_{0j}^*$. Observe that the latent budget shares will not necessarily satisfy the requirement of being between zero and one, although they sum to one.

We obtain the predictive distribution of $\bm w_0$ taking into account that 

\begin{align}
    f(w_{0J}|\tilde{\bm w}_0)&=\begin{Bmatrix}
    P\left(w_{0J}=1-\sum_{j=1}^{J-1}w_{0j}\right)=1, & w_{0J}^*>0\\
    P\left(w_{0J}=0\right)=1, & w_{0J}^*\leq 0\\
    \end{Bmatrix},\nonumber
\end{align}

and using the transformation \citep{Wales1983}

\begin{align}
    w_{0j}&=\begin{Bmatrix}
    w_{0j}^*/\left(1-\sum_{l\notin J_0}w_{0l}^*\right), & w_{0j}^*>0\\
    0, & w_{0j}^* \leq 0
    \end{Bmatrix},\\
    &J_0=\left\{l:w_{0l}^*>0\right\}, j=1,2,\dots,J-1 \ \text{and} \ l=1,2,\dots,J.\nonumber
\end{align}

Letting $\bm w_0 = \left[\tilde{\bm w}^{\top}_0 \ w_{0J}\right]^{\top} = \left[\tilde{\bm w}^{+\top}_0 \ \tilde{\bm w}^{-\top}_0 \ w_{0J}\right]^{\top}$, where $\tilde{\bm w}^{+}_0$ is a $K$ dimensional vector of positive shares ($0\leq K \leq J-1$), $\tilde{\bm w}^{-}_0$ is a $J-1-K$ dimensional vector of 0's, $\tilde{\bm w}^{+}_0\perp \tilde{\bm w}^{-}_0|\bm w_0^*$, $w_{0J}\perp \tilde{\bm w}^*_0|\tilde{\bm w}_0$, using the change of variable theorem, and taken into account that given $K$ positive shares there are $(J-1)!/(K!(J-1-K)!)$ cases to consider, 

\small
\begin{align}\label{ref:predw}
    f(\bm w_{0}|\bm w_0^*) & = f(\tilde{\bm w}^{+}_0|\bm w_0^*)\times f({\bm 0}_{J-1-K}|\bm w_0^*)\times f(w_{0J}|\tilde{\bm w}_0)\nonumber\\
    & = \frac{\frac{(J-1)!}{K!(J-1-K)!}}{\sum_{j=1}^{J-1}\frac{(J-1)!}{j!(J-1-j)!}}\nonumber\\
    & \times \begin{Bmatrix}
    P_{\tilde{\bm w}_0^*}(\tilde{\bm w}_0^{*}\in (-\bm\infty, \bm 0]_{J-1})\times \mathbbm{1}\left(w_{0J}=1\right), & K=0\\
    \begin{Bmatrix}
    f_{\tilde{\bm w}_0^*}\left(\tilde{\bm w}_0^{+}\left(1-\sum_{l\notin J_0}\tilde{w}_{0l}^{*}\right)\right)\left(1-\sum_{l\notin J_0}\tilde{w}_{0l}^*\right)^{K}\times \mathbbm{1}(w_{0J}=0), & w_{0J}^*\leq 0\\
    f_{\tilde{\bm w}_0^*}\left(\tilde{\bm w}_0^{+}\left(1-\sum_{l\notin J_0}\tilde{w}_{0l}^{*}\right)\right)\left(1-\sum_{l\notin J_0}\tilde{w}_{0l}^*\right)^{K}\times \mathbbm{1}(w_{0J}=1-\sum_{j}w_{0j}), & w_{0J}^* > 0\\
    \end{Bmatrix}, & 0<K<J-1\\
    \begin{Bmatrix}
    f_{\tilde{\bm w}_0^*}(\tilde{\bm w}_0(1-w_{0J}^*))\left(1-w_{0J}^*\right)^{K}\times \mathbbm{1}(w_{0J}=0), & w_{0J}^*\leq 0\\
    f_{\tilde{\bm w}_0^*}(\tilde{\bm w}_0)\times \mathbbm{1}(w_{0J}=1-\sum_{j}w_{0j}), & w_{0J}^* > 0
    \end{Bmatrix}, & K=J-1
    \end{Bmatrix},
\end{align}
\normalsize

where $(-\bm\infty, \bm 0]_{J-1}$ is the negative orthant in $\mathbb{R}^{J-1}$, $P_{{\bm w}_0^{*}}(\cdot$) and $f_{{\bm w}_0^{*}}(\cdot)$ are the probability and density function induced by the distribution of $\tilde{\bm w}_{0}^*|\bm\Lambda^{(s)}$. 

Equations \ref{ref:margpred1} to \ref{ref:predw} help us to obtain the predictive density function for $\bm w_0$. Observe that this predictive density is coherent by construction, that is, the marginal predictive distributions ($f(w_{0j}|\bm{Data}), j=1,2,\dots,J$) are in the probability simplex set,
\begin{equation*}\label{ref:Simplex}
    S_{\bm w}=\left\{\bm w\in \mathbb{R}^J:\sum_{j=1}^J w_{j}=1,w_{j}\geq 0, j=1,2,\dots,J\right\}.
\end{equation*}


\subsection{Testing microeconomic restrictions}\label{sec:micrest}

We test the microeconomic restrictions following the formal Bayesian framework, model posterior odds, which reduces to Bayes factors given a priori equal model probabilities, that is,  $PO_{01}=BF_{01}=\frac{m(\bm w|\mathcal{M}_0)}{m(\bm w|\mathcal{M}_1)}$ where $m({\bm w}|\mathcal{M}_r)$ is the marginal likelihood, $r=0,1$, $\mathcal{M}_0$ is the restricted model and $\mathcal{M}_1$ is the unrestricted (encompassing) model. 

Evaluation of the marginal likelihood in DPMs is challenging due to requiring integration over the infinite dimensional space of $G$ to obtain the likelihood function \citep{Basu2003}. Fortunately, we can take advantage of the nested structure of the microeconomic restrictions, and use the Savage-Dickey density ratio (or variations) to calculate the Bayes factors of $H_0$ against $H_1$ given that we use the same configuration of priors in both models \citep{dickey1971weighted}. Then, we can estimate posterior model probabilities, $P(\mathcal{M}_0|\bm{Data})=\frac{BF_{01}}{1+BF_{01}}$. 

\subsubsection{Slutsky symmetry}
Our point of departure is the unrestricted model. This specification follows exactly same stages as in the restricted model, it is just that in the specification of the system \ref{ref:eq6} we set $\bm D_{J-1}=\bm I_{(J-1)^2}$. Slutsky symmetry is ensured by symmetry of $\bm A$ and $\bm B$. The null hypothesis of symmetry restrictions can be written as $H_0. \ \bm R\bm\phi=\bm 0_{(J-1)(J-2)}$ versus the alternative $H_1. \ \bm R\bm\phi\neq\bm 0_{(J-1)(J-2)}$, where $\bm\phi$ in this case is the vector of unrestricted coefficients, $\bm R$ is the $(J-1)(J-2)\times (J-1)[R+L+L+(J-1)+(J-1)]$ dimensional restriction matrix, and $(J-1)(J-2)$ is the number of symmetry restrictions in our specification.  


Following the Savage-Dickey density ratio \citep{dickey1971weighted,verdinelli1995computing},   

\begin{align}\label{eq:SD}
    BF_{01}(\bm {Data})&=\frac{\pi(\bm R\bm\phi=\bm 0|\bm {Data},\mathcal{M}_1)}{\pi(\bm R\bm\phi=\bm 0|\mathcal{M}_1)}.
\end{align}

We show in Appendix \ref{sec:MCMC} that $\bm\phi|\bm\psi,\left\{\bm \Sigma_i\right\},\left\{\bm \mu_i\right\},\bm s,\tilde{\bm w}^*,\bm y^* \sim MN(\bar{\bm\phi},\bar{\bm\Phi})$, where 

\begin{align*} 
\bar{\bm{\Phi}}=\left(\sum_{m=1}^M\sum_{i:s_i=m} \left\{\bm F_i^{\top}(\bm\Sigma_{\epsilon\epsilon,i}-\bm\Sigma_{\epsilon v,i}\bm\Sigma_{vv,i}^{-1}\bm\Sigma_{v\epsilon,i})^{-1}\bm F_i\right\}+\bm\Phi^{-1}_0\right)^{-1},
\end{align*} and 
    \begin{align*}
        \bar{\bm\phi}=&\bar{\bm\Phi}\left[\sum_{m=1}^M\sum_{i:s_i=m}\left\{\bm F_i^{\top}(\bm\Sigma_{\epsilon\epsilon,i}-\bm\Sigma_{\epsilon v,i}\bm\Sigma_{vv,i}^{-1}\bm\Sigma_{v\epsilon,i})^{-1}\right.\right.\\
        &\left.\left.(\tilde{\bm w}_i^*-\bm \mu_{\tilde{\bm w}_i^*}-\bm\Sigma_{\epsilon v,i}\bm\Sigma_{vv,i}^{-1}(\bm y_i^*-\bm\mu_{y_i^*}-\bm G_i\bm\psi))\right\}+\bm\Phi_0^{-1}\bm\phi_0\right].
    \end{align*}   

Then, $\bm R\bm\phi|\bm\psi,\left\{\bm \Sigma_i\right\},\left\{\bm \mu_i\right\},\bm s,\tilde{\bm w}^*,\bm y^* \sim MN(\bm R\bar{\bm\phi},\bm R\bar{\bm\Phi}\bm R^{\top})$.

Observe that the denominator in Equation \ref{eq:SD} is easy to calculate given our prior assumptions (see expression \ref{eq:priors}). We take into account the following facts to calculate the numerator: 

\begin{align*}
    \pi(\bm R\bm\phi=\bm 0|\bm {Data},\mathcal{M}_1)=&\int \pi(\bm R\bm\phi=\bm 0|\bm{\Lambda}_{(-\bm{\phi})},\bm {Data},\mathcal{M}_1)\times \pi(\bm{\Lambda}_{(-\bm{\phi})}|\bm {Data},\mathcal{M}_1)d\bm{\Lambda}_{(-\bm{\phi})}\\
    =&\mathbb{E}[\pi(\bm R\bm\phi=\bm 0|\bm{\Lambda}_{(-\bm{\phi})},\bm {Data},\mathcal{M}_1)]\\
    \approx & \frac{1}{S}\sum_{s=1}^S \pi(\bm R\bm\phi=\bm 0|\bm{\Lambda}_{(-\bm{\phi})}^{(s)},\bm {Data},\mathcal{M}_1),
\end{align*}

where $\bm{\Lambda}_{(-\bm{\phi})}=\bm\psi,\left\{\bm \Sigma_i\right\},\left\{\bm \mu_i\right\},\bm s,\tilde{\bm w}^*,\bm y^*$ are draws from the posterior distribution.

\subsubsection{Slutsky negative semidefiniteness}

We follow common practice in empirical demand systems to estimate the model without imposing inequality restrictions, and then check the inequalities associated with utility function regularity using the Slutsky matrix estimates. In particular, we use the encompassing approach to check inequality restrictions proposed by \cite{klugkist2007bayes} to test negative semidefiniteness of the Slutsky matrix, $\bm A + \bm By + \bm w\bm w^{\top}-\bm W$. Observe that the this condition is equivalent to $\lambda_{J-1}\leq \lambda_{J-2}\leq \dots\leq \lambda_{1}\leq 0$, where $\lambda_{\cdot}$ are the eigenvalues of the normalized symmetric Slutsky matrix.\footnote{See \cite{wetzels2010encompassing} for a nice proof of the relationship between the encompassing approach for inequality and about restrictions, and the Savage-Dickey density ratio.}

Given the encompassing model $\mathcal{M}_1$, where there is no requirement of inequality restrictions, and the restricted model $\mathcal{M}_0$ subject to the inequality restrictions, the Bayes factor is

\begin{align}\label{eq:BFIneq}
    BF_{01}&=\frac{\frac{1}{d_0}}{\frac{1}{c_0}}\nonumber\\
    &\approx\frac{\frac{1}{S}\sum_{s=1}^S \mathbbm{1}(\mathcal{M}_0)\times\pi(\bm\Lambda^{(s)}|\bm{Data},\mathcal{M}_1)}{\frac{1}{S}\sum_{s=1}^S \mathbbm{1}(\mathcal{M}_0)\times\pi(\bm\Lambda^{(s)}|\mathcal{M}_1)}
\end{align}

where $1/d_0$ and  $1/c_0$ are the proportions of the encompassing posterior and encompassing prior that are in agreement with the constraints of model $\mathcal{M}_0$, respectively, and $\mathbbm{1}(\mathcal{M}_0)$ is the indicator function regarding model $\mathcal{M}_0$.

\subsection{Model assessment: Predictive p-value}\label{sec:pval}
Observe that we can use the predictive density to generate pseudo-shares from our model ($\bm w^{(s)}, s=1,2,\dots,S$), and calculate a discrepancy function $D(\bm w^{(s)}, \bm\Lambda^{(s)})$ to estimate $p_D(\bm w)=P[D(\bm w^{(s)}, \bm\Lambda)\geq D(\bm w, \bm\Lambda)]$ using the proportion of the $S$ pairs for which $D(\bm w^{(s)}, \bm\Lambda^{(s)})\geq D(\bm w, \bm\Lambda^{(s)})$, where $\bm w$ are the observed shares. We also use the average discrepancy statistic $D(\bm w)=\mathbb{E}[D(\bm w, \bm\Lambda)]=\int D(\bm w, \bm\Lambda) \pi(\bm\Lambda|\bm{Data})d\bm\Lambda\approx \frac{1}{S}\sum_{s=1}^S D(\bm w, \bm\Lambda^{(s)})$ to estimate $p_{\mathbb{E}[D]}(\bm w)=P[D(\bm w^{(s)}, \bm\Lambda)\geq D(\bm w)]$ based on the posterior distribution of $D(\bm w^{(s)},\bm\lambda^{(s)})$  (see Appendix \ref{subsec: p-value} for algorithm details). Those are posterior predictive p-values \citep{gelman1996posterior} which are measures of realized discrepancy assessments between our model and the data. Extreme tail probabilities ($p_D(\bm w)  \leq 0.05 \ \text{or} \ p_D(\bm w) \geq 0.95$) suggest potential discrepancies. In particular, our assessments are based on the observed income shares and the normalized Slutsky matrix (Equation \ref{refM:eq5A}).

\subsection{Summary}

Observed that we used data augmentation strategy \citep{Tanner1987} to handle censoring at zero. The endogeneity is tackled using different ways to write the likelihood function 
\begin{align*}
p(\tilde{\bm w}^*,\tilde{\bm y}^*|\bm\phi, \bm\psi,\left\{s_m\right\},\left\{\bm\theta_m\right\})&=p(\tilde{\bm w}^*|\tilde{\bm y}^*,\bm\phi, \bm\psi,\left\{s_m\right\},\left\{\bm\theta_m\right\})p(\tilde{\bm y}^*|\bm\phi, \bm\psi,\left\{s_m\right\},\left\{\bm\theta_m\right\})\\
&=p(\tilde{\bm y}^*|\tilde{\bm w}^*,\bm\phi, \bm\psi,\left\{s_m\right\},\left\{\bm\theta_l\right\})p(\tilde{\bm w}^*|\bm\phi, \bm\psi,\left\{s_m\right\},\left\{\bm\theta_m\right\}),
\end{align*}
and accordingly deduce particular conditional posterior distributions. In addition, cost homogeneity is imposed by using $\tilde{\bm p}$ rather than ${\bm p}$, and adding up constraints are used to get coefficients associated with the excluded good ($w^*_{J}$). Slutsky symmetry is imposed using the duplication matrix. The non-linearity issue is tackled iteratively updating $\bm y$ using equation \ref{refM:eq2} and draws of $\bm A$ and $\bm B$ from $\bm\phi$. Unobserved preference heterogeneity is introduced using the Dirichlet process mixture, and the coherent predictive distribution of the shares is done through the predictive distribution of the latent shares using equations \ref{ref:margpred1}, \ref{ref:margpred2} and \ref{eq:y0}. Slutsky symmetry and negative semidefiniteness are tested based on the Savage-Dickey density ratio \citep{dickey1971weighted}, and the encompassing approach with inequality restrictions \citep{klugkist2007bayes}. Finally, model assessment is based on predictive p-values \citep{gelman1996posterior}.


\section{Data}\label{sec:data}

We merge different data sets involving utility providers of electricity, water, gas and sewerage, and households in strata 4, 5, and 6, which are the households subject to this new electricity tax.\footnote{Colombian households are classified by the government into socioeconomic strata; the aim is to implement cross subsidies and social programs. There are 6 categories, from low-low to high-high, stratum 1 signals the most vulnerable socioeconomic households, whereas stratum 6 households are supposed to be the wealthiest.}  

We used the cross sectional \textit{Encuesta nacional de presupuesto de los hogares} (ENPH) - 2016, which is a representative national survey based on a probabilistic multistage, stratified and cluster sample, inquiring about household budget carried out by the Colombian national institute of statistics (\textit{DANE}) between July 2016 - June 2017.\footnote{This survey defines three stages. The fist stage has inclusion probability equal to one for 32 provinces with their metropolitan areas, and 6 important municipalities. The inclusion probability of other municipalities is stratified according to population size, urbanization, urban/rural population proportion, unsatisfied basic needs index, and strata. The second stage defines groups of contiguous blocks (urban area) and sections (rural area) as sampling units, each group has on average 10 blocks or sections that are proportionally drawn according to the population defined in the first stage. In the third stage an average group of 10 contiguous households (clusters) is randomly drawn.} We processed 5,780 households in strata 4, 5 and 6 getting information about monthly expenditures on each utility (electricity, water, gas and sewerage), total household monthly income (\$COP), municipality (location), gender and age of household head, highest level of education attained by any household member, number of people living in the household and strata.

Equation \ref{eqPr} shows expenditure ($E_{ismt}^u$) on utility $u=\left\{electricity, water, gas, sewerage\right\}$ at household $i$ in stratum $s$ with provider $m$ at time $t$.

\begin{equation}\label{eqPr}
E_{ismt}^u=V_{ismt}^u+F_{smt}^u=\sum_{c=1}^C P_{csmt}^u Q_{cismt}^u+F_{smt}^u.
\end{equation}

This expenditure is equal to payment for variable consumption ($V$) plus fixed charge ($F$). The former is equal to price ($P$) at consumption range $c$ times consumption in this range ($Q$) due to tariff depending on consumption level in some utilities by regulation.

Electricity does not have fixed charge for any strata, whereas gas does have for strata 4, 5 and 6. Tariffs depend on stratum and providers, households in strata 5 and 6 have to pay a contribution to subsidize households in strata 1, 2 and 3, whereas households in stratum 4 pay the reference cost.\footnote{Tariffs for households in strata 1, 2 and 3 depend on consumption levels of electricity and gas.} We used household municipality location to identify its providers, and therefore to get the electricity tariff (\$COP/kWh) and gas tariff (\$COP/m3) including contributions if apply. This information is available at \textit{Sistema Único de Información} (\href{www.sui.gov.co}{SUI}) and \textit{Comisión de Regulación de Energía y Gas} (\href{www.creg.gov.co}{CREG}). The former is a public available information repository where utility providers have to upload business information, and the latter is the electricity and gas regulatory counsil.   

Water and sewerage also have fixed charges which depend on providers and strata. This information is available at CREG for gas, and SUI for water and sewerage. Therefore, we obtain variable expenditure on these utilities subtracting fixed charges from total expenditures ($V_{ismt}^u=E_{ismt}^u-F_{smt}^u, u=\left\{water, gas, sewerage, \right\}$).

Water and sewerage tariffs (\$COP/m3) depend on municipality location (meters above sea level, m.a.s.l), providers, household consumption and strata. There are three ranges defined by the regulatory entity (CRA): basic, complementary and luxury, which depend on m.a.s.l.  \href{https://www.cra.gov.co/documents/Resolucion_CRA_750_de_2016-Edicion_y_copia.pdf}{(CRA 750 de 2016)}. The first range is for municipalities with an average altitude below 1,000 m.a.s.l., the second for municipalities with an average altitude between 1,000 and 2,000 m.a.s.l., and finally the last range for municipalities with average altitude above 2,000 meters. A municipality in a lower range can consume more cubic meters (m3) at lower consumption prices due to weather conditions; so, prices associated with basic water consumption is up to 16 m3, 13 m3 and 11 m3 for municipalities in first, second and third levels, respectively. Prices at complementary level is between upper basic range and 32 m3, 26 m3, and 22 m3, respectively, and luxury prices apply to additional consumption.

We can deduce average water and sewerage tariff per m3 using variable expenditure on these utilities, municipality location and strata from ENPH survey, consumption ranges from CRA, and provider marginal prices from SUI in conjunction with equation \ref{eqPr}. Observe that average prices are weighted averages of marginal prices, the former being the price signal that households perceive.

We use the utility shares based on variable expenditures because these are directly under household's control, whereas fixed charges are not.

Due to households were surveyed in different months between 2016 and 2017, prices will be expressed at June 2017 prices. Then, all prices, expenditures, and fixed charges are in dollars at the exchange rate of June 30, 2017, the month in which the survey ended.

\subsection{Descriptive Statistics}
Table \ref{sumstat} shows descriptive statistics of household income shares in utilities. It can be seen that in this group, on average, electricity share is the greatest (3\%), followed by water (2\%). On average, the variable expenditure in utilities represents 6\% of the income.\footnote{We omit households with shares equal to 0 to calculate sample means and standard deviations to avoid distortions.} We also report proportion of zeros in the sample, where electricity and water have the lowest figures (5\%), while sewerage has the highest (51\%). This highlights the importance of taking into account the censoring issue in the econometric framework.    

\begin{table}[htbp]\centering \caption{Summary statistics: Utility shares and prices \label{sumstat}}
\begin{threeparttable}
\resizebox{1\textwidth}{!}{\begin{minipage}{\textwidth}
\begin{tabular}{l c c c c c}\hline
\multicolumn{1}{c}{\textbf{Variable}} & \textbf{Mean}
 & \textbf{Std. Dev.} & \textbf{Zero Shares} & & \\ 
\hline
\textbf{Shares}& & & & & \\
Electricity &  0.03 & 0.05  & 0.05 & & \\
Water & 0.02 & 0.03  & 0.05 & & \\
Sewerage &  7e-03 & 0.01  & 0.51 & & \\
Natural Gas &  8e-03 & 0.01 & 0.20 & &\\
Numeraire  &  0.94 & 0.07 & 0.00 & &\\
\addlinespace[.75ex]
\hline
\textbf{Prices}& & & & Min & Max \\
\hline
\addlinespace[.75ex]
Electricity (USD/kWh) & 0.16 & 0.02 & & 0.13 & 0.27  \\
Water (USD/m3 ) & 0.67 & 0.25 & &  0.16 & 1.46 \\ 
Sewerage (USD/m3) & 0.57 & 0.24  & & 0.16 & 1.31  \\
Natural Gas (USD/m3) & 0.50 & 0.12 & & 0.09 & 0.91  \\
\addlinespace[.75ex]
\hline
\end{tabular}
\begin{tablenotes}[para,flushleft]
\footnotesize \textit{Notes}: Electricity is the most relevant utility expenditure followed by water. There is a high level of censoring, particularly in sewerage, and also a high variability regarding utility prices. Mean and standard deviation figures do not take zero shares into account.

Prices are converted to dollars using the exchange rate of 30/06/2017, equivalent to COP/USD 3,038.26. The sample size is 5,780. 

\textbf{Source:} Authors' calculations based on information from ENPH, CREG, CRA, SUI and Superintendencia Financiera de Colombia.
  \end{tablenotes}
  \end{minipage}}
  \end{threeparttable}
\end{table}

Electricity average tariff is 0.16 USD/kWh with a range from 0.13 USD/kWh to 0.27 USD/kWh. This high variability in electricity prices is also present in the other utilities: water prices range from 0.16 USD/m3 to 1.46 USD/m3, sewerage from 0.16 USD/m3 to 1.31 USD/m3, and gas from 0.09 USD/m3 to 0.91 USD/m3. This heterogeneity is due to different socioeconomic conditions, municipalities location, regulatory legislation and utility providers efficiency. We performed mean differences statistical tests comparing households at different strata, we reject the null hypothesis of equal share and price means. As expected by regulatory means, utility prices are increasing with strata, and there are also statistical differences regarding budget shares (see tables \ref{sumstat2} and \ref{ttestsp} in the Appendix \ref{subsec: descriptive}).

\begin{table}[h!]\centering \caption{Descriptive statistics: Household characteristics\label{demstat2}}
\begin{threeparttable}
\resizebox{1\textwidth}{!}{\begin{minipage}{\textwidth}
\begin{tabular}{l c c c c }\hline
\multicolumn{1}{c}{\textbf{Variable}} & \textbf{Mean}
 & \textbf{Std. Dev.}& \textbf{Min.} &  \textbf{Max.} \\ \hline

Household head age & 55.14 & 15.88   & 18 & 99 \\
Household head gender (female) & 0.41 &  & 0 & 1 \\
Household members & 2.84 & 1.39 & 1 & 10 \\

\textbf{Strata indicator}& & & & \\ 
Stratum 4 & 0.67 &  & 0 & 1 \\
Stratum 5 & 0.22 &  & 0 & 1  \\
Stratum 6 & 0.11 &  & 0 & 1  \\

\textbf{Education level}& & & & \\
Elementary school & 0.05 &  & 0 & 1 \\
High School & 0.15 &  & 0 & 1  \\
Vocational & 0.12 &  & 0 & 1 \\
Undergraduate & 0.40 & & 0 & 1 \\
Postgraduate & 0.28 & & 0 & 1 \\
 
\textbf{Municipality altitude}& & & & \\ 
Below 1,000 m.a.s.l & 0.44 &  & 0 & 1  \\
More than 1,000 m.a.s.l & 0.56 &  & 0 & 1  \\


\textbf{Income (USD)}& & & & \\ 
Total Income &  2,073.52 & 2,298.03 & 22.65 & 50,483.6 \\
 \hline
\end{tabular}
\begin{tablenotes}[para,flushleft]
 \footnotesize \textit{Notes}: The representative (mean) household has 3 members, classified in stratum 4, a 55 years-old man as household head, the highest education level is undergraduate, and it is located at an altitude more than 1,000 m.a.s.l. Total expenditure in utilities is around 6\% household total income, and there is a high level of variability regarding income.\\
 \textbf{Source:} Authors' calculations based on information from ENPH and Superintendencia Financiera de Colombia.
  \end{tablenotes}
    \end{minipage}}
  \end{threeparttable}
\end{table}

Table \ref{demstat2} presents descriptive statistics of household characteristics, which were selected according to data availability and literature review \citep{Deaton1980,Banks1997,Lewbel2009,Tovar2018}. We see from this table that the representative (modal) household has 3 members, classified in stratum 4, a 55 years-old man as household head, the highest education level is undergraduate, and it is located at an altitude more than 1,000 m.a.s.l. There is a high level of variability regarding income that shows high socioeconomic heterogeneity (Table \ref{demstat} in the Appendix \ref{subsec: descriptive} shows this information by strata, and tables \ref{ttestdem} and \ref{ttestdem2} show mean and factor difference tests).

\begin{figure}[h!]
\caption{Unconditional Engel curves: Level and derivatives}\label{Fig1}
\begin{subfigure}{.5\textwidth}
  \centering
  \includegraphics[width=1\linewidth]{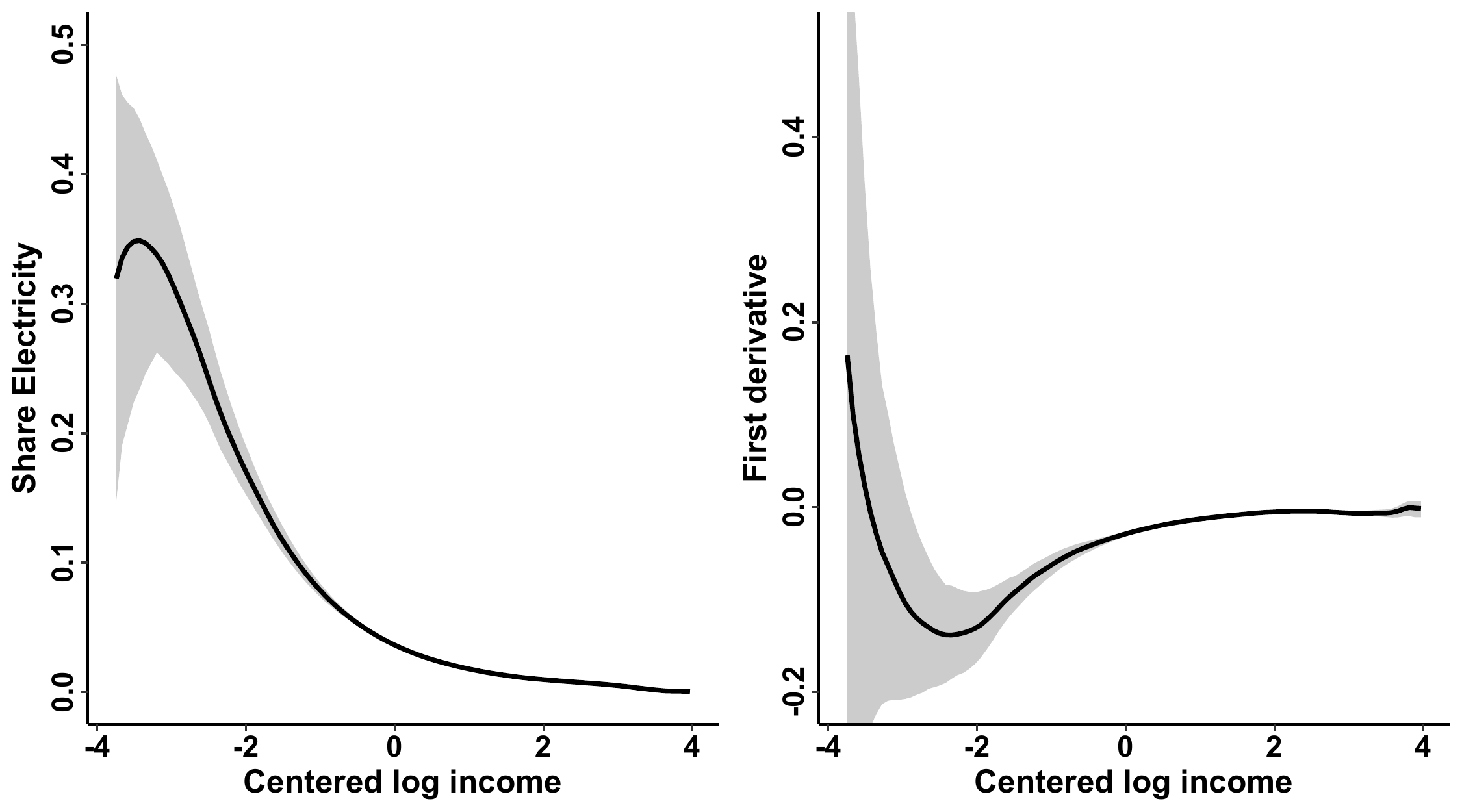}
  \caption{Electricity}
  \label{fig:2a}
\end{subfigure}%
\begin{subfigure}{.5\textwidth}
  \centering
  \includegraphics[width=1\linewidth]{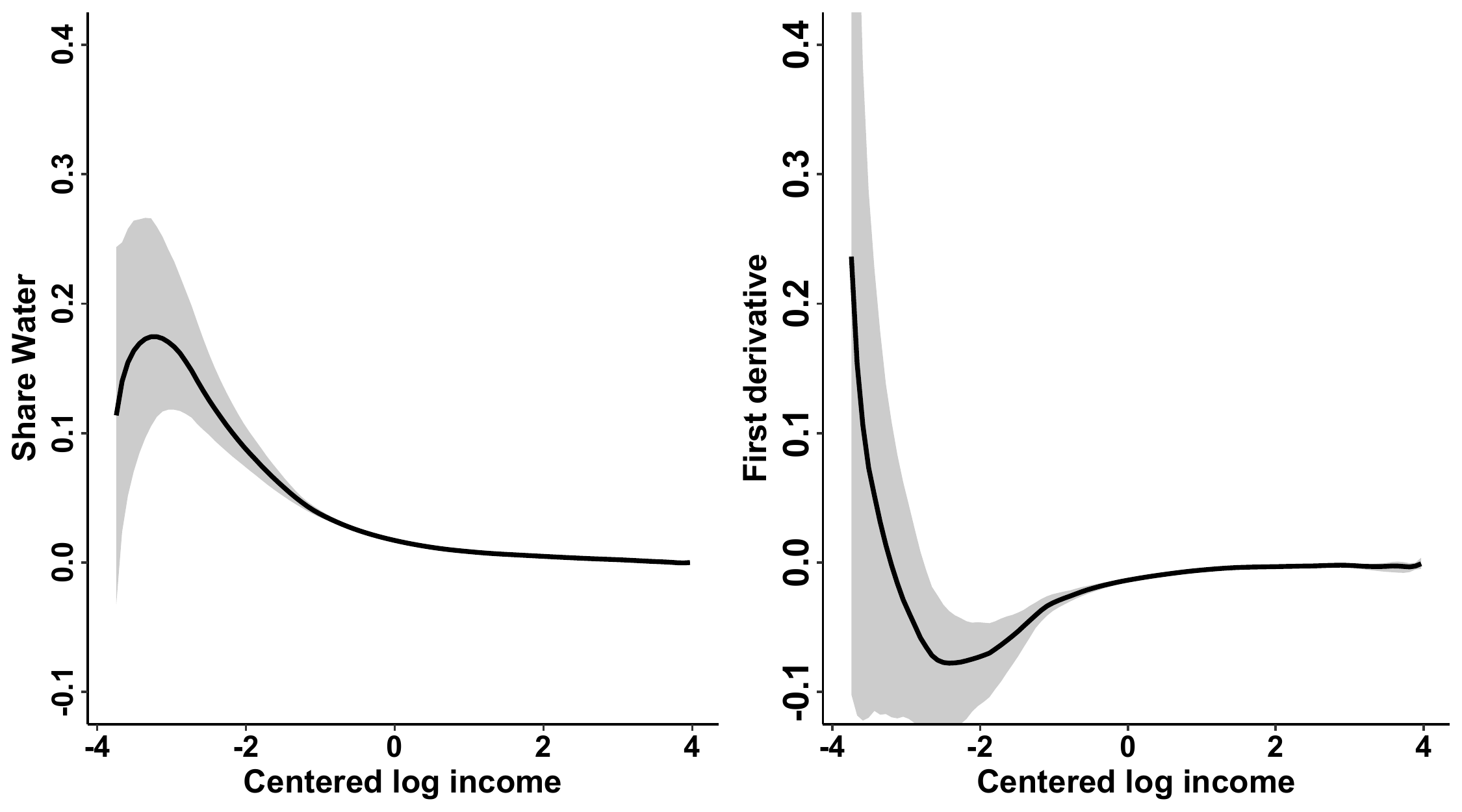}
  \caption{Water}
  \label{fig:2b}
\end{subfigure}
\begin{subfigure}{.5\textwidth}
  \centering
  \includegraphics[width=1\linewidth]{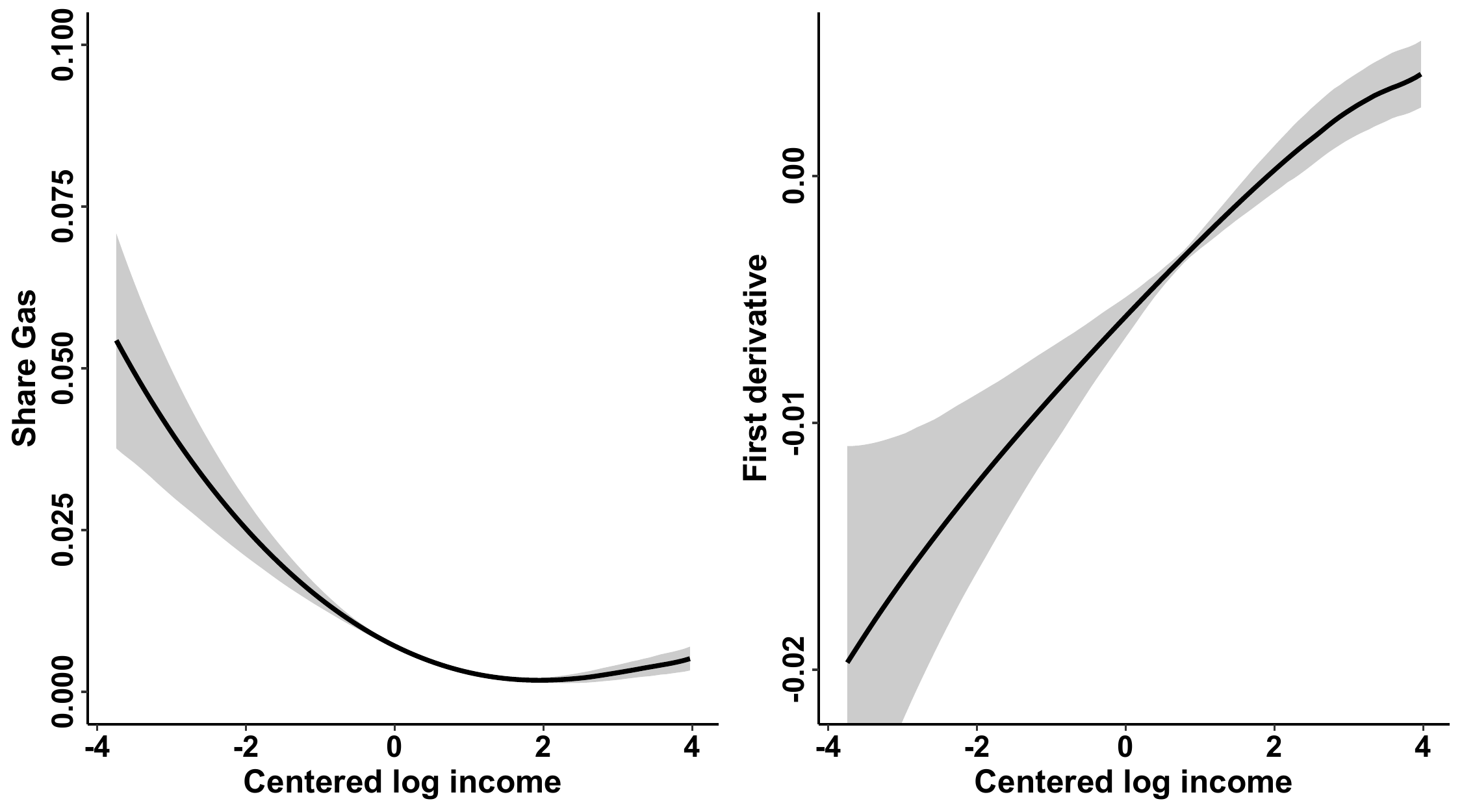}
  \caption{Gas}
  \label{fig:2c}
\end{subfigure}%
\begin{subfigure}{.5\textwidth}
  \centering
  \includegraphics[width=1\linewidth]{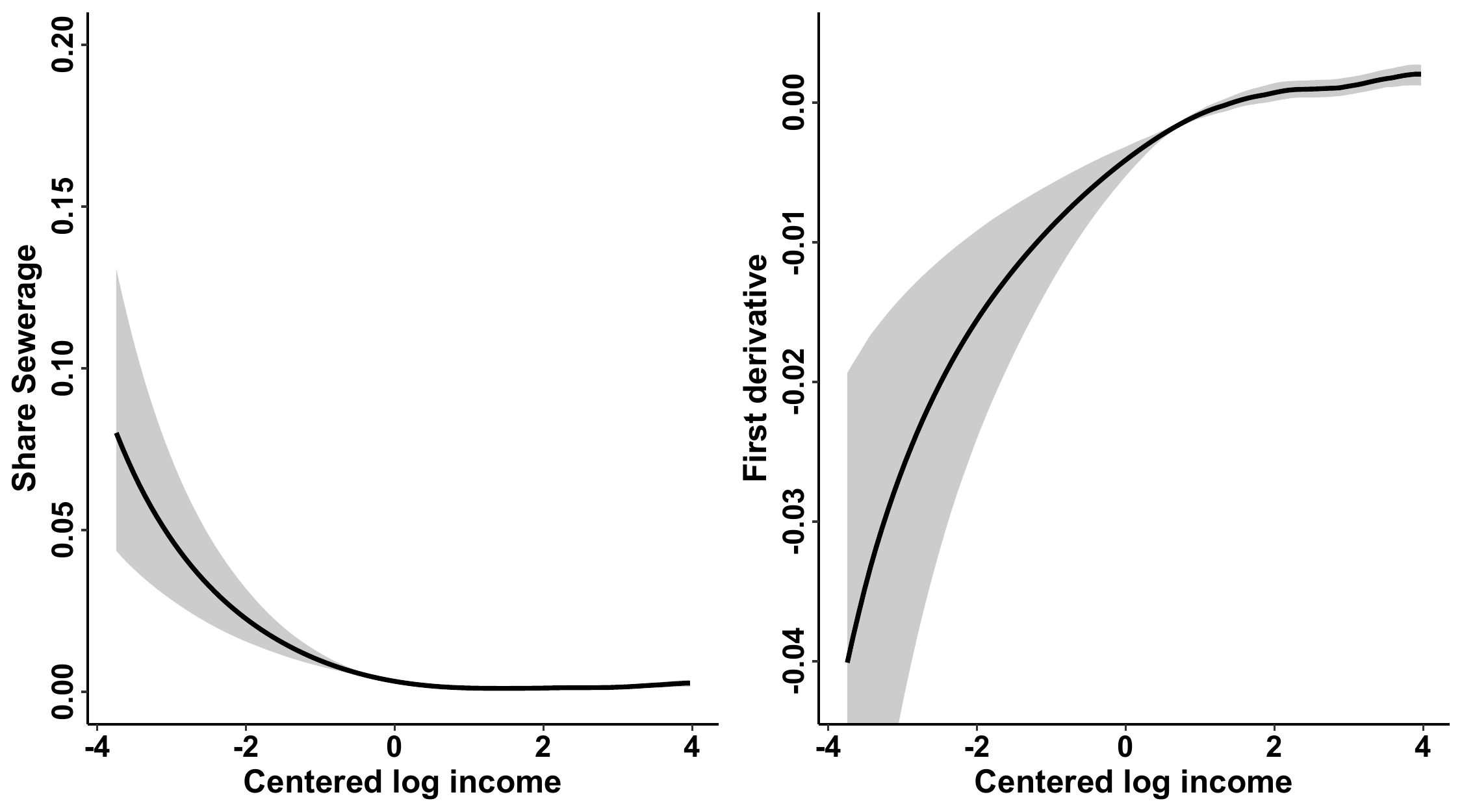}
  \caption{Sewerage}
  \label{fig:2d}
\end{subfigure}
\begin{subfigure}{.5\textwidth}
  \centering
  \includegraphics[width=1\linewidth]{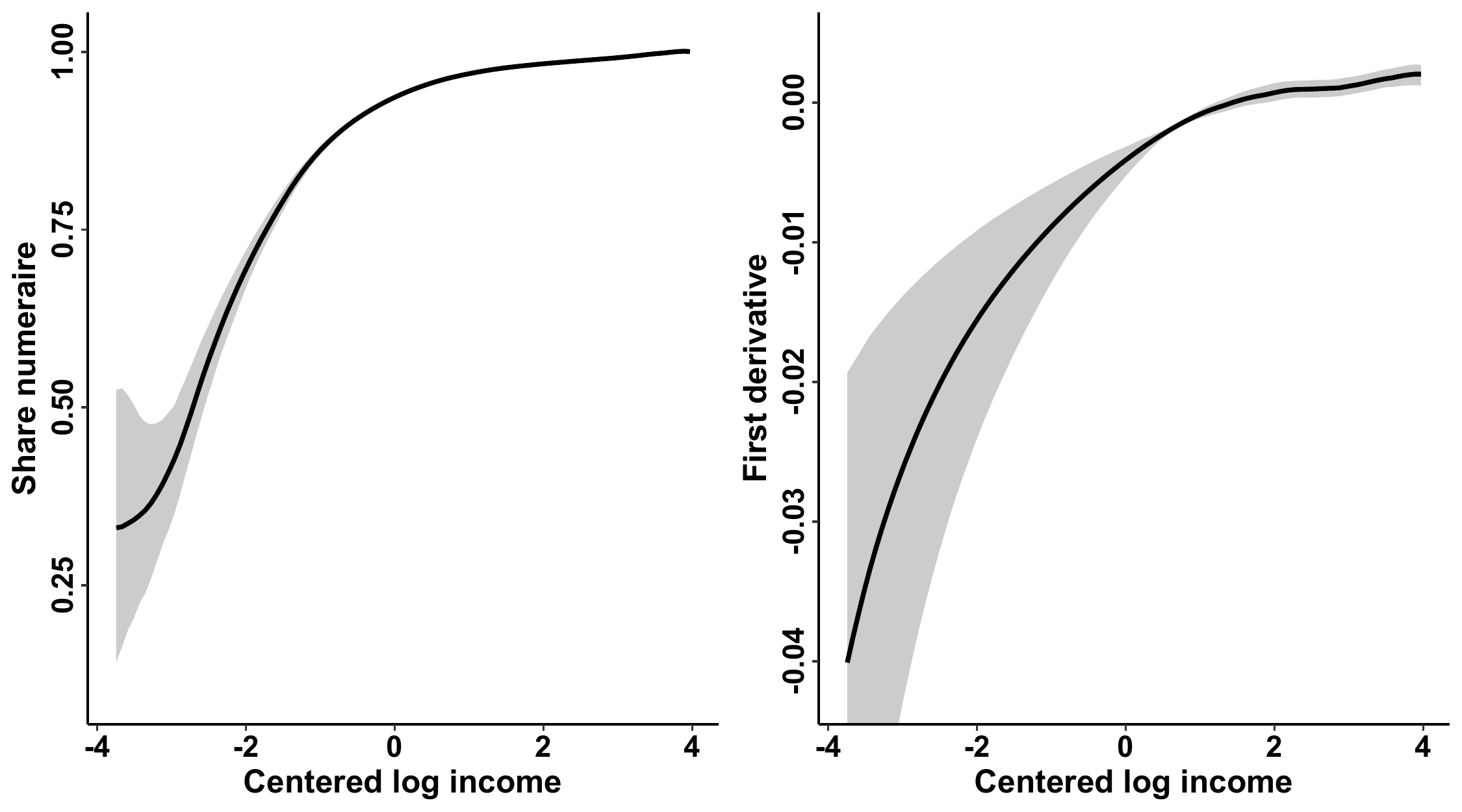}
  \caption{Numeraire}
  \label{fig:2e}
\end{subfigure}
\label{fig:2}
{\scriptsize \caption*{\textit{Notes}: Engel curves (left panels, solid black line), their derivatives (right panels, solid black line), and their predictive 95\% intervals (gray shaded area). There are polynomial complexities in Engel curves that cannot be recovered by linear \citep{Deaton1980} or quadratic \citep{Banks1997} Engel demand systems.}}
\end{figure}

Figure \ref{Fig1} shows unconditional Engel curves kernel estimations (left panels, solid black line), their derivatives (right panels, solid black line), and their predictive 95\% intervals (gray shaded area). It seems that there are polynomial complexities in these Engel curves that cannot be recovered using linear \citep{Deaton1980} or quadratic \citep{Banks1997} Engel demand systems. 

\section{Results}\label{sec:results}

We normalize prices using the representative household ($i= 4,334$) such that the baseline log price vector is equal to zero, and centered all  observable household characteristics using this household. Then, all socioeconomic controls are zero for the representative household. We also normalize income using this household to mitigate computational problems due to using high order polynomial in $y_i$. This transformation does not change model fit, and helps to perform predictive exercises for the representative household at baseline prices due to $y=x=0$, then $f(\tilde{\bm w}_{0}^*|y_0,\bm\Lambda)=f(\tilde{\bm w}_{0}^*|\bm\Lambda)$.

We run 1,600 iterations with a burn-in equal to 600 using parallel processing in a server composed by two processors each with 12 cores in an Intel(R) Xeon(R) CPU E5-2670 v3 @ 2.30GHz RAM 397 GB architecture x86\_64 CPU 64-bit. Total computing time 7 days 2.4 hours approx.

The parametric part of our structural equation \ref{ref:eq6} would have $(J-1)\left[2(L+J-1)+R\right]$ parameters without imposing symmetry on $\bm A$ and $\bm B$. This would be an exactly identified model. We set $R=5$ as \cite{Lewbel2009,zhen2014predicting}, this seems to give enough flexibility to handle Engel curves like those displayed in Figure \ref{Fig1}.\footnote{We perform robustness analysis regarding polynomial degree up to 5. We have computational issues using higher polynomial degrees.} In addition, we have $L=10$ and $J=5$; therefore, we would have 132 unrestricted parameters (33 for each equation). Imposing symmetry implies $(J-1)(J-2) = 12$ overidentifying restrictions. On the other hand, the reduced form equation has $q[2(L+J-1)+R]$ parameters, where $q=R+L+J-1$, this means 627 parameters. Therefore, our system of equations involves $(J-1+q)$ equations, that is, 23 equations, with 747 location parameters imposing symmetry restrictions, without taking unobserved heterogeneity parameters into account.

We compute several diagnostics to assess the convergence and stationarity of the posterior chains. In general, it seems that the posterior chains of the structural parameters in equation \ref{ref:eq6} are stable. In particular, 100\% of the parameters have dependence factors less than 5 using the \cite{Raftery1992}'s diagnostic with a 95\% probability of obtaining an estimate in the interval $5\% \pm 2.5\%$, just 2 out of 120 have dependence factors higher than 2. Regarding the \cite{Heidelberger1983}'s and \cite{Geweke1992}'s tests at 5\% significance level, 116 and 99 structural coefficients pass these tests, respectively. The former uses the Cramer-von-Mises statistic to test the null hypothesis that the sampled values come from a stationary distribution, and the latter tests for equality of the posterior means using the first 10\% and the last 50\% of the Markov chains. Results available upon authors request.

We check symmetry and negative semidefiniteness of the Slutsky matrix for the representative household using equations \ref{eq:SD} and \ref{eq:BFIneq}. We find that $2\log(BF_{01})$ are equal to 157.1 for the former, and 11.4 for the latter. These values suggest very strong evidence in favor of these hypothesis \citep{kass1995bayes}. This implies that the posterior model probabilities imposing each of these restrictions for the representative household are approximately equal to 1.\footnote{We use non-informative priors for testing Slutsky symmetry. Observe that the Savage-Dickey density ratio compares equal dimensional densities; thus the Bayes factor does not necessarily give overwhelming support to parsimonious models in this setting. In addition, the way that we standardized observations means that the Bayes factor does not depend upon units of measurement. We perform some simulation exercises to test these assessments.}   




\subsection{Posterior inference}

\begin{figure}[h!]
    \caption{Compensated and uncompensated elasticities: Point estimates and 95\% credible intervals for the representative household }\label{pi_compa}
\begin{subfigure}{.5\textwidth}
  \centering
  \includegraphics[width=1\linewidth]{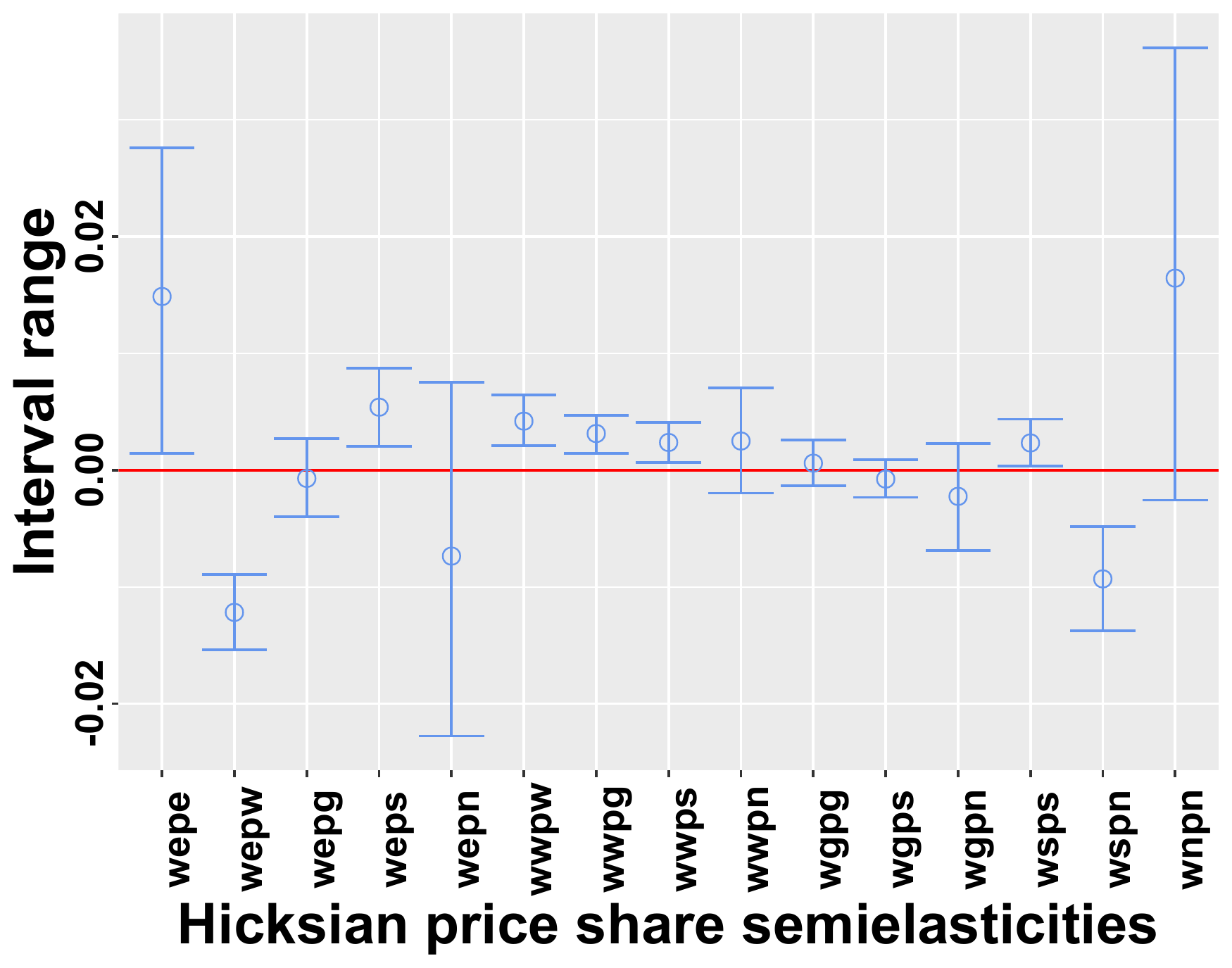}
   \label{fig:2a1}
   \caption{Hicksian share semi-elasticities}
\end{subfigure}%
\begin{subfigure}{.5\textwidth}
  \centering
  \includegraphics[width=1\linewidth]{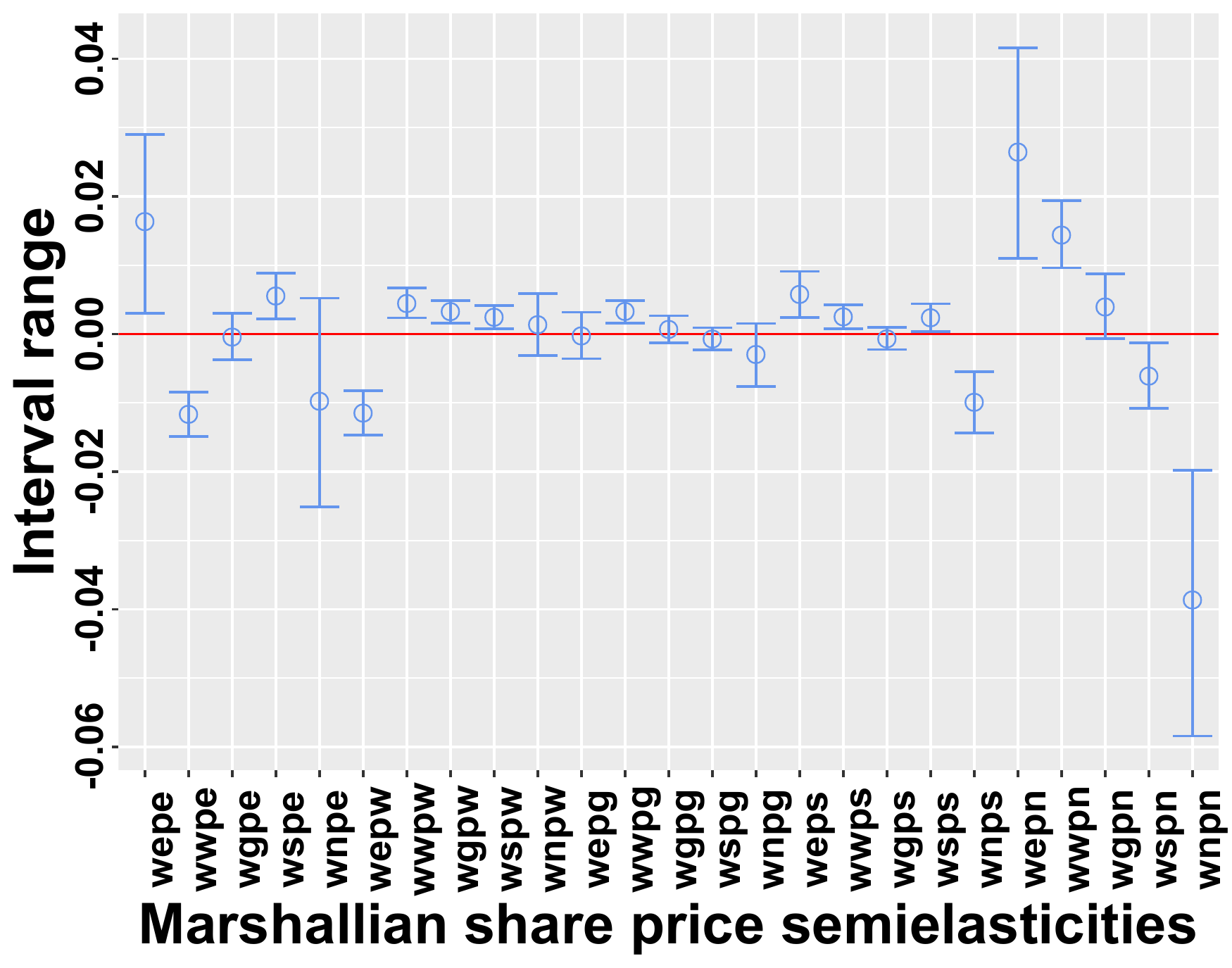}
   \label{fig:2b1}
   \caption{Marshallian share semi-elasticities}
\end{subfigure}%

\begin{subfigure}{.5\textwidth}
  \centering
  \includegraphics[width=1\linewidth]{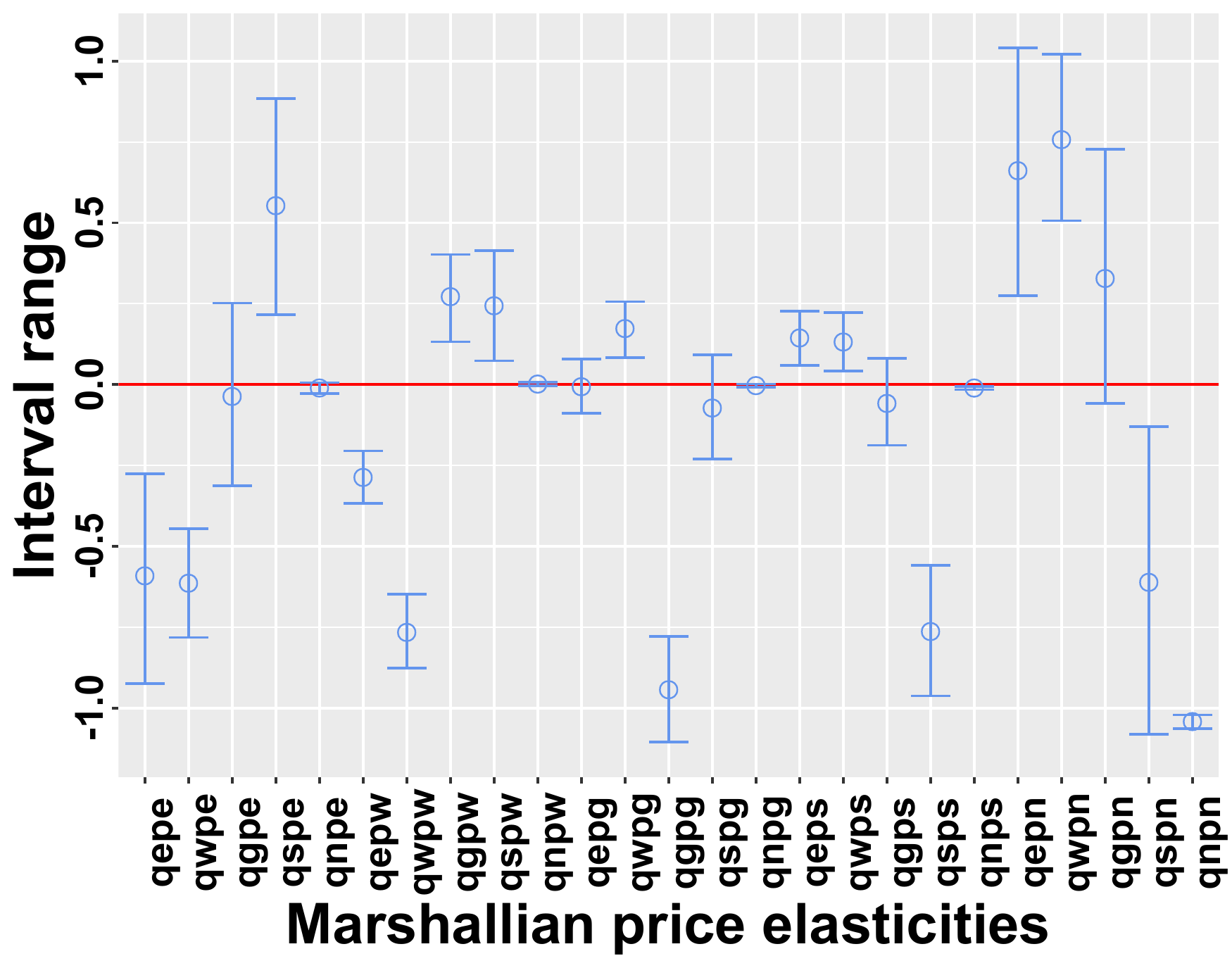}
   \label{fig:2c1}
   \caption{Marshallian price quantity elasticities}
\end{subfigure}%
\begin{subfigure}{.5\textwidth}
  \centering
  \includegraphics[width=1\linewidth]{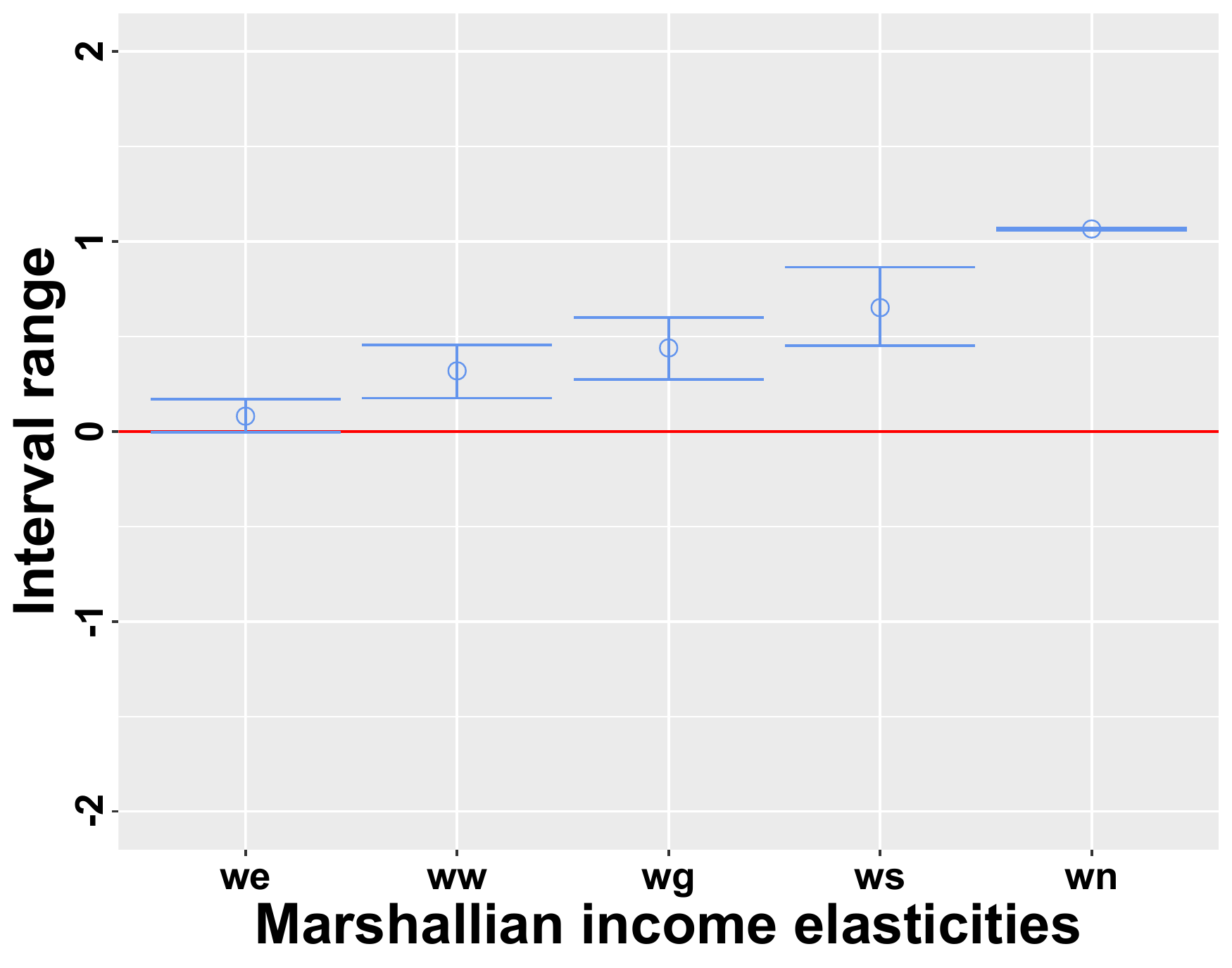}
   \label{fig:2d1}
   \caption{Marshallian income elasticities}
\end{subfigure}

{\scriptsize \caption*{\textit{Notes}: Circles are posterior mean values, and bars are 95\% symmetric credible intervals. Notation \textit{wupj} (\textit{qupj}) indicates the effect of percent change in price of good $j$ on share (quantity)for good $u$. For instance, \textit{wepw} in panel a indicates that 1 percent price increase in water would imply 0.016 percent points less income share for electricity.}}
\end{figure}

Figure \ref{pi_compa} shows posterior estimates for the representative household. In particular, Hicksian and Marshallian price share semi-elasticities (top-left and top-right panels, equations \ref{refM:eq3} and \ref{refM:eq7}), and Marshallian price quantity and income elasticities (bottom-left and bottom-right panels, equations  \ref{refM:eq10} and \ref{refM:eq9}).

We can see in Figure \ref{pi_compa} panel a that there are 8 out of 15 Hicksian price share semi-elasticities whose 95\% credible interval does not cross zero, 3 of them are own-price share semi-elasticities (electricity, water and sewerage). Electricity is the most sensitive income share utility, an electricity tariff increase of 10 percent would imply a share 0.17 percentage points higher when income is raised to equate utility with that in the initial situation. We also see that 5 of the cross-price semi-elasticities are ``statistically significant''. For instance, the water share compensated cross-price electricity semi-elasticity is -0.016, implying that a 10 percent water tariff increase would generate a statistically significant 0.16 percent points decrease in the income share for electricity, even when income is raised to hold utility constant.

Figure \ref{pi_compa} shows in panel b that there are 16 out of 25 Marshallian price share semi-elasticities whose 95\% credible interval does not cross zero. Observe that the numeraire ``good" is the most sensitive, a 10 percent price increase would imply approximately 0.4 percentage points real income share decrease when there is not compensation to keep utility level at the initial state. Uncompensated utility own-price share semi-elasticities are very similar to compensated ones. Observe that several uncompensated cross-price semi-elasticities are statistically significant, numeraire price increase causes the most relevant income share sensitivity. For instance, 10 percent price index numeraire increase implies 0.28, 0.15, 0.05 and -0.05 percent points change in the real income share for electricity, water, gas and sewerage, respectively.

Figure \ref{pi_compa} shows in panel c the Marshallian quantity demand elasticities. All 95\% credible intervals for own-price elasticities are negative. In particular, 95\% symmetric credible intervals for electricity, water and sewerage indicate inelastic goods, whereas gas own-price 95\% posterior credible interval is $(-1.2, -0.7)$. All posterior mean utility own-price estimates suggest inelastic goods. Observe that posterior mean own-price elasticity estimates of water and sewerage are very similar (-0.75). However, the uncertainty regarding the latter is higher due to its higher level of censoring.  On the other hand, the numeraire own-price elasticity is close to -1, although its 95\% credible interval does not embrace this value. We see in this figure that there are relevant substitution and complementary price effects.

We can see in Figure \ref{pi_compa} panel d the Marshallian income elasticities. All utilities seem to be normal goods with 95\% credible intervals less than 1, but greater than 0, except marginally for electricity. According to the posterior income elasticity mean estimates, sewerage is the most sensitive utility, followed by gas and water. The numeraire seems to be a luxury good, its very precise 95\% credible interval is higher than 1, although very close to this value.

Inelastic and normal results for utilities agree with previous literature of demand systems, most of them QAIDS. In particular, \cite{renzetti1999municipal,Blundell1999,Gundimeda2008, Schulte2017,moshiri2018welfare,Tovar2018,diaz2021price} found similar results for electricity, \cite{di2011consumers,Galves2016,suarez2020modeling} for water, and \cite{renzetti1999municipal,Blundell1999,Gundimeda2008,diaz2021price} for gas.\footnote{We did not find any references for sewerage. It should be because data limitations, but we suspect that sewerage elasticities should be similar to water elasticities as we got in our application.}

We also perform price and income effects analysis for representative households at strata level. We identify these households based on modal values of socioeconomic characteristics conditional on strata (see Table \ref{demstat} in the Appendix \ref{subsec: descriptive}). These representative households by stratum are the observations 4135, 4924 and 3611 for strata 4, 5 and 6. This is relevant in the Colombian economy as cross subsidies on utilities depend on this classification. We can see in Figure \ref{pi_compaN} in the Appendix \ref{subsec: descriptive} these results. In general, we observe same patter in Hicksian and Marshallian share semielasticities compared with the sample representative household; although uncertainty level increases with strata. This is expected as socioeconomic characteristics differences between strata increase. Regarding Marshallian quantity demand elasticities, it seems that there are some differences between strata 6 and the other two strata; particularly regarding the effect of the numeraire price on utilities. It seems that the representative household of strata 6 is more elastic. This higher sensitivity is also in evidence in the income elasticity, where we also observe more differences between representative households. To sum up, it seems that quantity and income Marshallian elasticities are more sensitive to socioeconomic differences than semi-elasticities.   

\begin{figure}[h!]
    \caption{Engel cuves: Point estimates and 90\%, 95\% and 99\% credible intervals for the representative household}\label{EngelCurveRep}
\begin{subfigure}{.45\textwidth}
  \centering
  \includegraphics[width=0.8\linewidth]{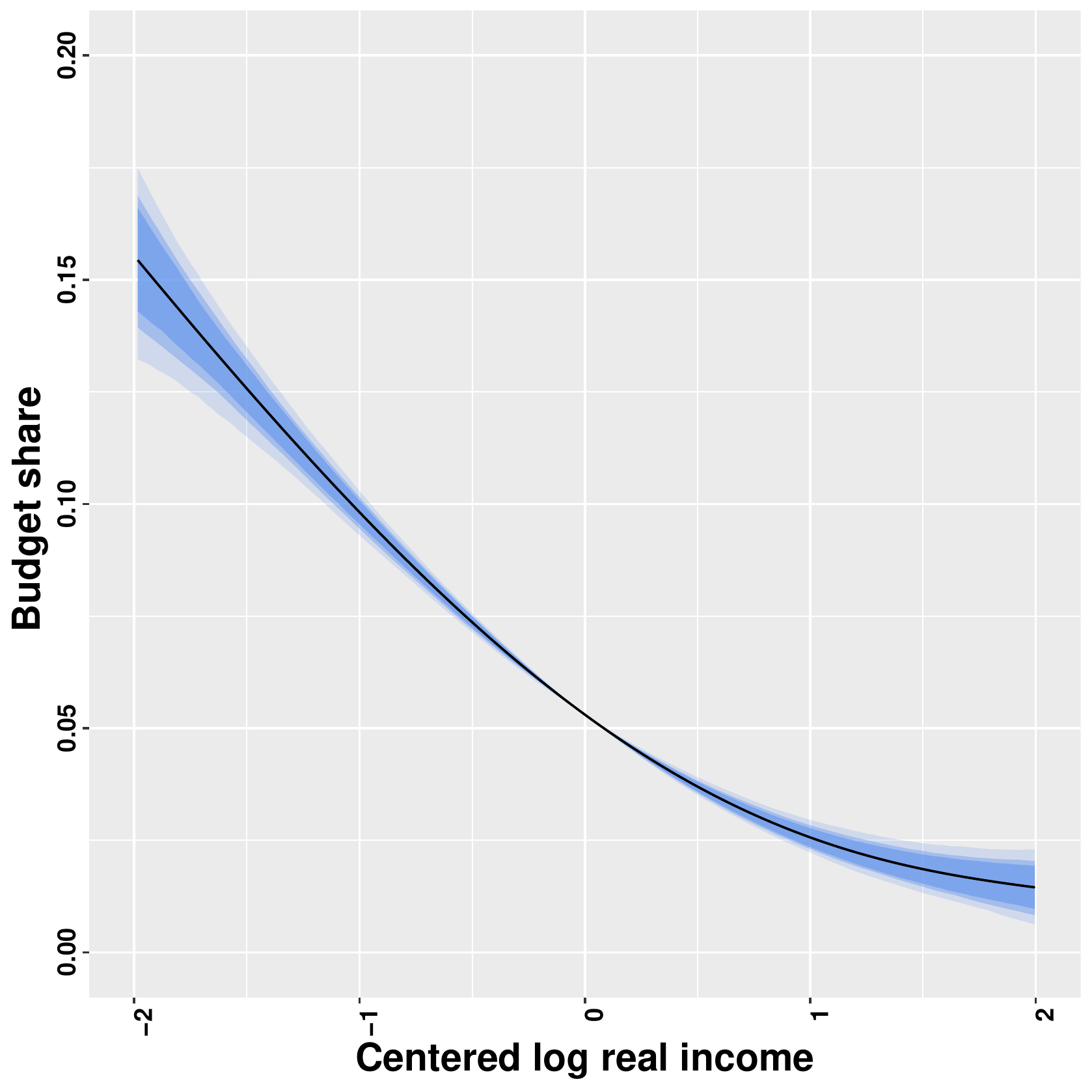}
    \caption{Electricity}
   \label{fig:2a2}
\end{subfigure}%
 \begin{subfigure}{.45\textwidth}
  \centering
  \includegraphics[width=0.8\linewidth]{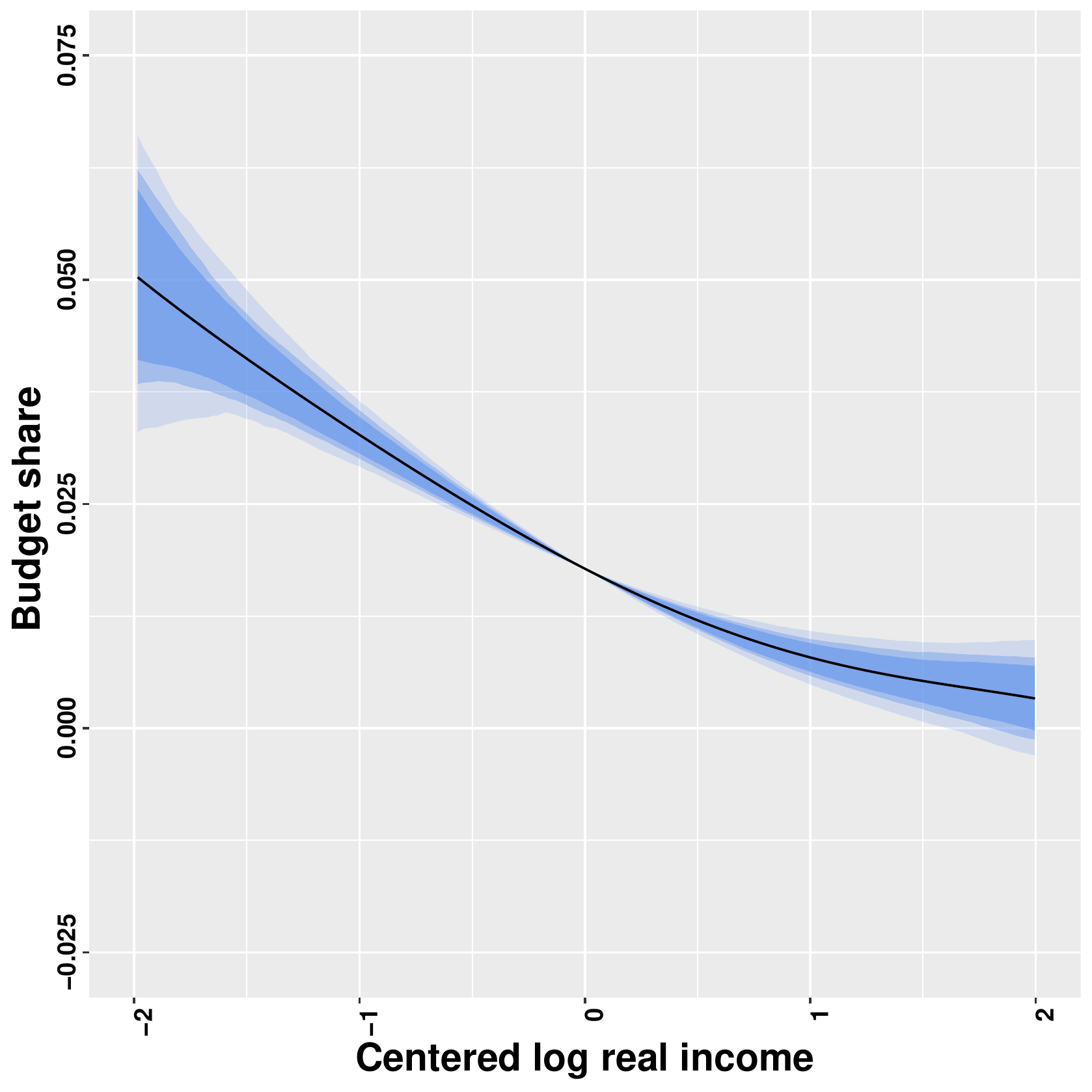}
      \caption{Water}
   \label{fig:2b2}
\end{subfigure}%

\begin{subfigure}{.45\textwidth}
  \centering
  \includegraphics[width=0.8\linewidth]{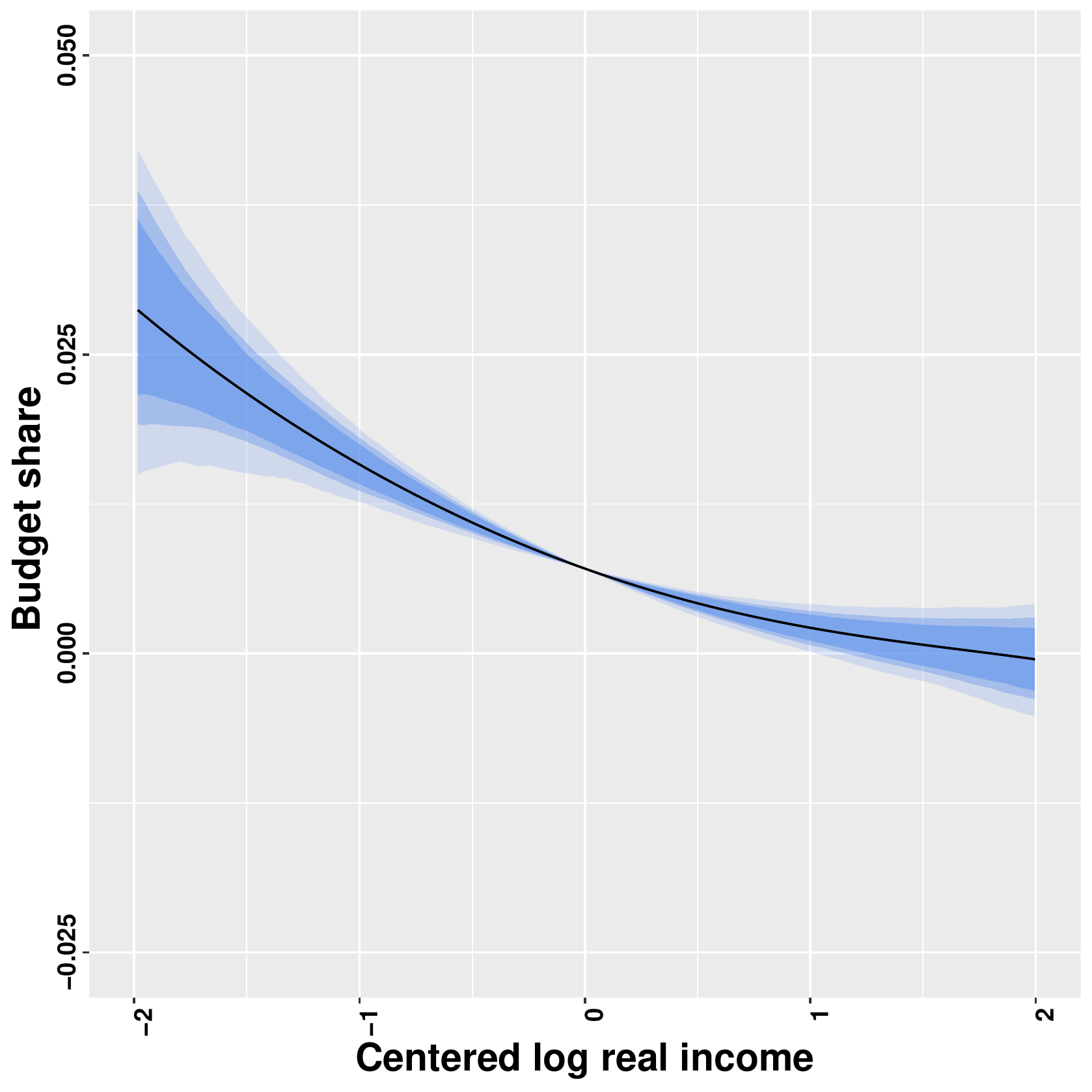}
      \caption{Gas}
   \label{fig:2c2}
 \end{subfigure}%
\begin{subfigure}{.45\textwidth}
  \centering
  \includegraphics[width=0.8\linewidth]{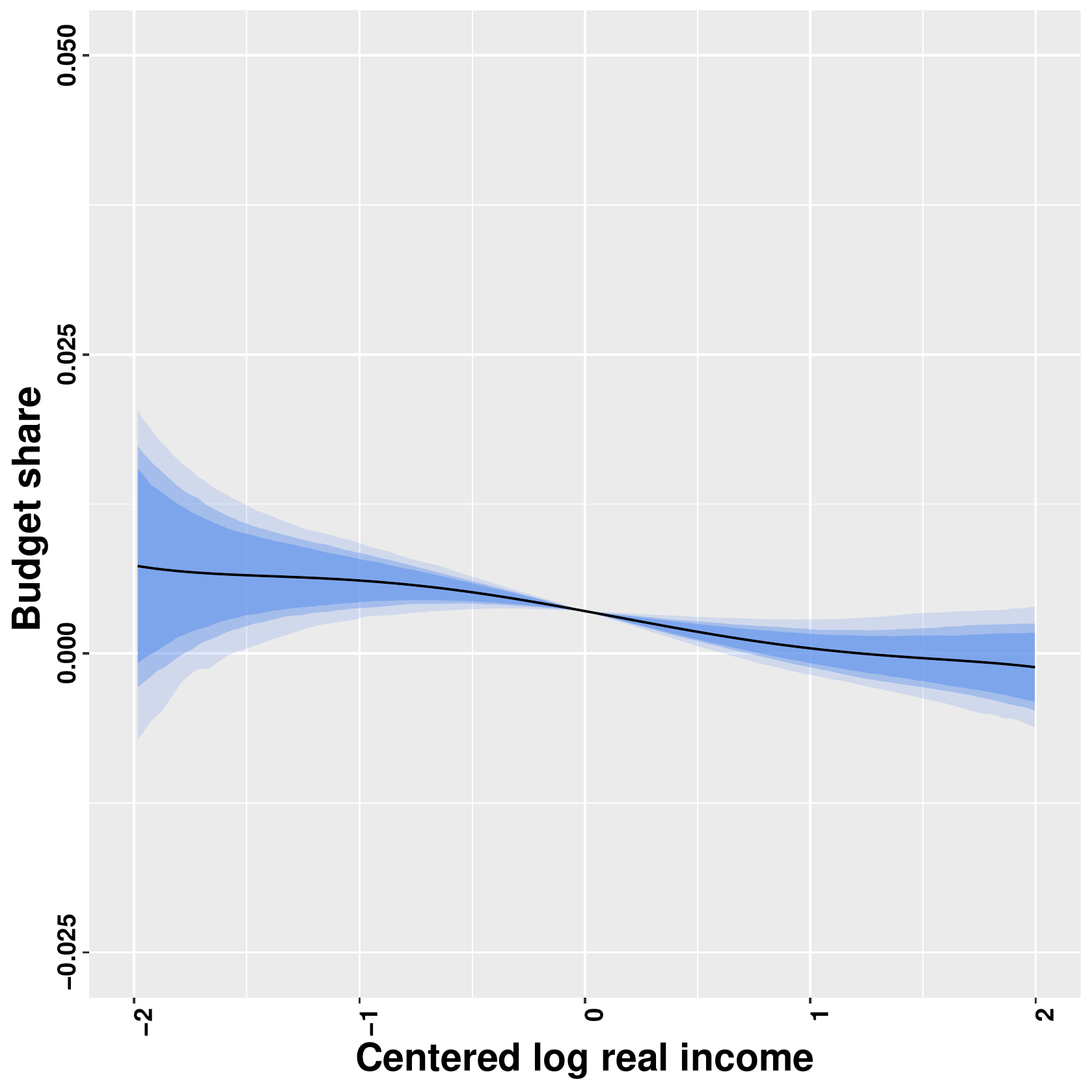}
      \caption{Sewerage}
   \label{fig:2d2}
 \end{subfigure}%

\begin{subfigure}{.45\textwidth}
  \centering
  \includegraphics[width=0.8\linewidth]{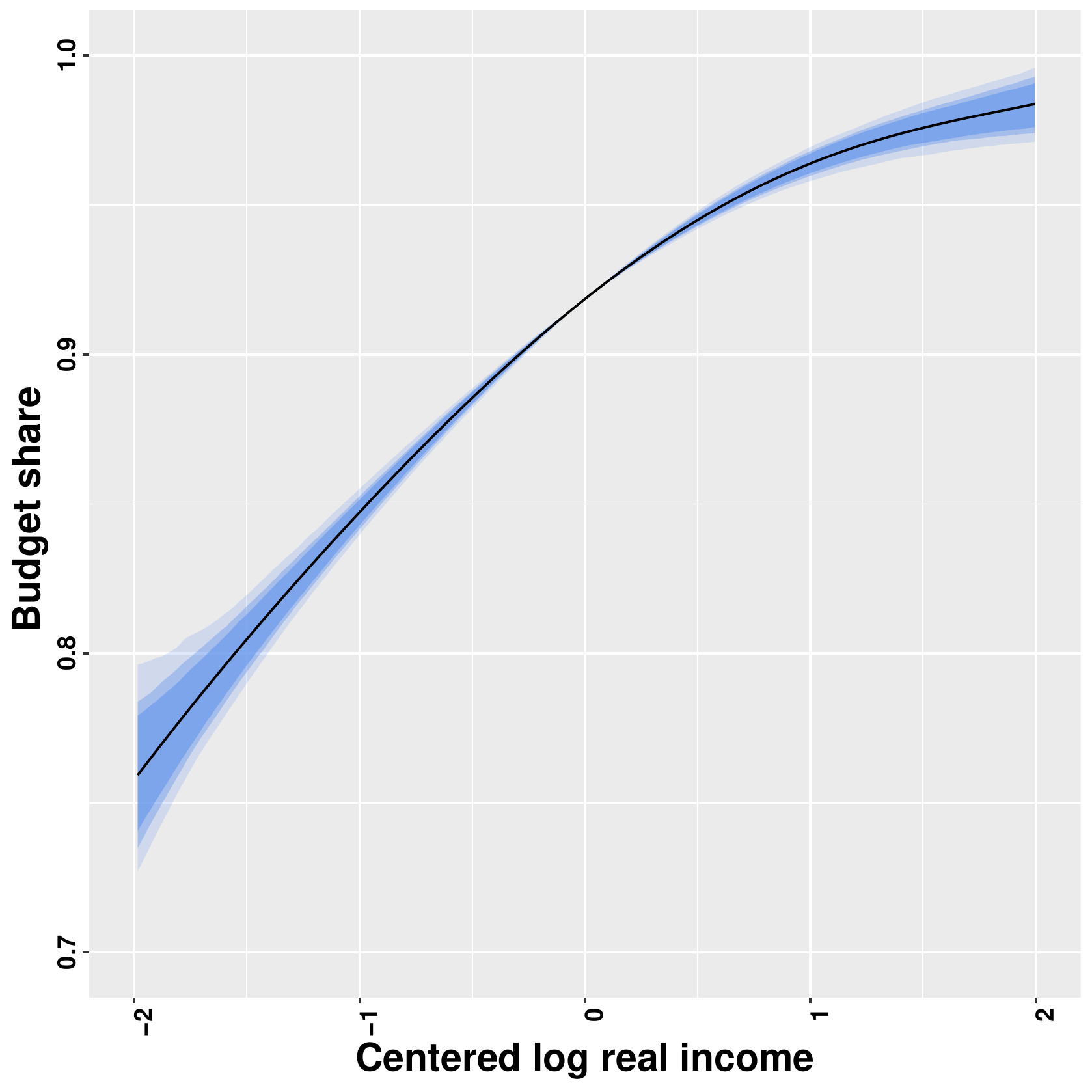}
      \caption{Numeraire}
   \label{fig:2e2}
 \end{subfigure}%
 
   \begin{subfigure}{.45\textwidth}
  \centering
  \includegraphics[width=0.8\linewidth]{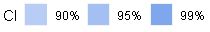}
   \label{fig:2f2}
 \end{subfigure}%
 
 {\scriptsize\caption*{\textit{Notes}: Engel curves of income shares for utilities and numeraire. Income shares for utilities are high for poor households and low for wealthy households.}}
 
\end{figure}

Figure \ref{EngelCurveRep} shows the posterior mean and the 90\%, 95\% and 99\% credible intervals of the conditional Engel curves for the representative household taking into account its unobserved preference heterogeneity cluster (see section \ref{sec:het} for cluster details). They suggest statistical significant Engel curves in most of the range of the centered log real income, except for sewerage (see Panel d). This result is intuitive as the income share for sewerage is very low, and there is a high level of uncertainty due to censoring issues. Most of the Engel curves look quadratic, except income share in sewerage. Higher order polynomials seem not to be economically significant in the middle range of income. However, if we increase the range of variability of income, we observe that polynomial order higher than 2 are meaningful, but uncertainty also increases. It seems that in our application higher polynomial orders are relevant for low income share goods, like sewerage, or in the tails of the income distribution. This may explain why previous literature found demand system ranks less or equal than three \citep{Lewbel2009}. More research should be done in this matter requiring very detailed data sets to confirm these hypothesis. 
\clearpage
We also estimate the Engel curves for representative households of each strata (see Figure \ref{EngelCurveRepS} in the Appendix \ref{subsec: descriptive}). We observe same shapes as expected, but with different scales.

In general, we observe that income share in utilities would be higher for poor households, and there is an asymptote around 1\% income share for the wealthiest households given that these households have observable characteristics equal to the representative household.

\subsection{Predictive distribution: Electricity price change}\label{sec:Pred}

We estimate the posterior predictive distribution for the representative household using the framework in section \ref{sec:Predictive}. We find that the modal predictive values of the joint distribution are 4.18\%, 1.38\%, 0.54\%, 0.18\% and 93.71\% for electricity, water, gas, sewerage and numeraire using the real income at its modal value and baseline prices. We also estimate the predictive distribution assuming a 0.8\% electricity tariff increase, which is equivalent to a new tax of 0.12 USD cents/kWh on electricity consumption taking into account an average tariff 0.16 USD/kWh for the representative household.\footnote{The Colombian government imposed a tax of 4 COP/kWh in July/2019, which is equal to 0.12 USD cents/kWh using an exchange rate equal to 3,038.26 COP/USD.} The modal predictive values assuming a 0.8\% (19\%) electricity tariff increase are 4.23\% (4.50\%), 1.39\% (1.25\%), 0.55\% (0.57\%), 0.19\% (0.26\%) and 93.63\% (93.34\%), respectively.\footnote{We also perform a predictive exercise assuming a 19\% tariff increase due to the failed fiscal tax reform proposed by the Colombian government in the first semester of 2021 suggesting this percentage.} 

\begin{figure}[h!]\caption{Predictive distribution function: Income shares for the representative household comparing no tariff change with a 0.8\% and 19\% tariff increase in electricity}
\centering
\begin{subfigure}{.5\textwidth}
  \centering
  \includegraphics[width=1\linewidth]{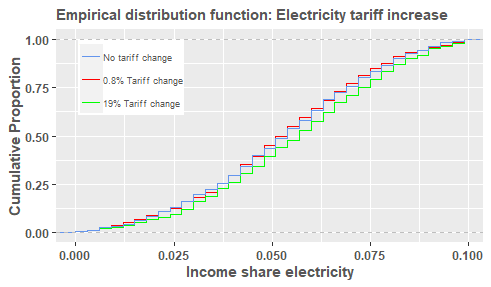}
  \caption{Electricity}
  \label{fig:Cla}
\end{subfigure}%
\begin{subfigure}{.5\textwidth}
  \centering
  \includegraphics[width=1\linewidth]{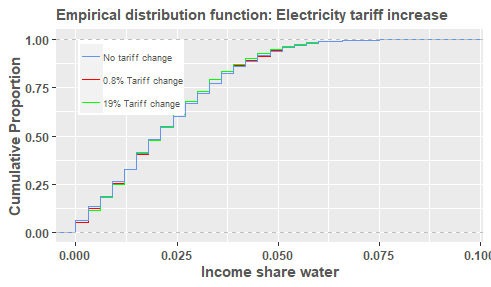}
  \caption{Water}
  \label{fig:Clb}
\end{subfigure}
\begin{subfigure}{.5\textwidth}
  \centering
  \includegraphics[width=1\linewidth]{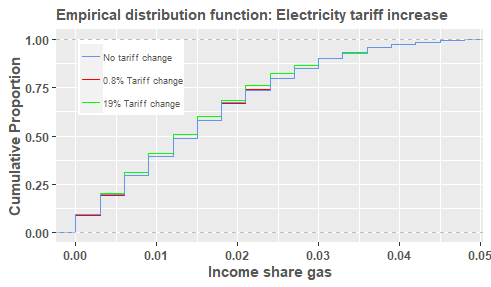}
  \caption{Gas}
  \label{fig:Clc}
\end{subfigure}%
\begin{subfigure}{.5\textwidth}
  \centering
  \includegraphics[width=1\linewidth]{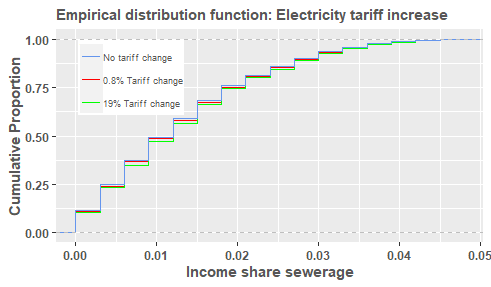}
  \caption{Sewerage}
  \label{fig:Cld}
\end{subfigure}
\begin{subfigure}{.5\textwidth}
  \centering
  \includegraphics[width=1\linewidth]{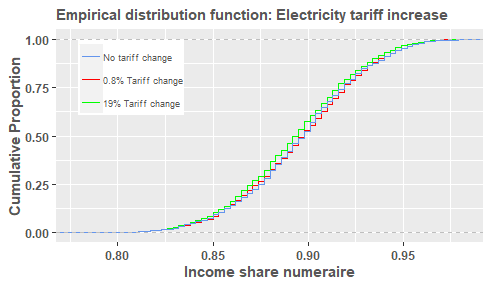}
  \caption{Numeraire}
  \label{fig:Cle}
\end{subfigure}
\label{fig:PredVarP}

{\scriptsize \caption*{\textit{Notes}: Blue line is the empirical cumulative distribution function without electricity tariff change for the representative household at baseline prices. Red and green lines are the empirical cumulative distribution functions with 0.8\% and 19\% electricity tariff increase.}}
\end{figure}

Figure \ref{fig:PredVarP} shows the empirical cumulative distribution functions associated with the marginal predictive distributions. The predictive mean values without electricity tariff increase are 5.2\%, 2.2\%, 1.5\%, 1.3\% and 89.6\% that suggest good predictive performance as observed income shares for electricity, water, gas, sewerage and numeraire for the representative household are 4\%, 2\%, 1.2\%, 1\% and 91.8\%, respectively.

In Figure \ref{fig:PredVarP} we have that higher ordinate values conditional on electricity tariff change, that is, $P(W\leq w|\Delta\bm{p}\neq\bm{0})>P(W\leq w|\Delta\bm{p}=\bm{0})$, means that there is a higher probability of spending less income on a particular good given an electricity tariff increase. Panel a in Figure \ref{fig:PredVarP} suggests that after a 4\% income share in electricity the representative household facing the 0.8\% electricity tariff variation increases the probability of spending less in electricity. We found that 63\% of time the predictive distribution for electricity with tariff increase is greater than the predictive distribution without tariff increase. This figure is equal to 35\%, 31\%, 0\% and 29\% for water, gas, sewerage and the numeraire. However, results from the predictive distribution given a small tariff change can be difficult to identify. Then, we also analyze a 19\% tariff increase in electricity demand. We have that there is potentially stochastic dominance in the marginal predictive distribution of electricity, and as a consequence, the numeraire. Panel a implies that there is a higher probability of spending more on electricity given a 19\% tariff increase over all the support compared to the baseline situation. This meant substitution effects on other non-utility goods (numeraire), and small spillover effects on gas and sewerage.

\clearpage
\subsection{Welfare analysis}

\begin{figure}[h!]
\caption{Welfare analysis: Equivalent variation distribution as percentage of income for representative households given a 0.8\% electricity tariff increase}
\centering
\begin{subfigure}{.5\textwidth}
  \centering
  \includegraphics[width=0.8\linewidth]{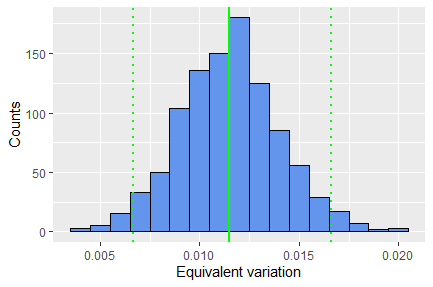}
  \caption{Sample}
  \label{fig:Cla1}
\end{subfigure}%
\begin{subfigure}{.5\textwidth}
  \centering
  \includegraphics[width=0.8\linewidth]{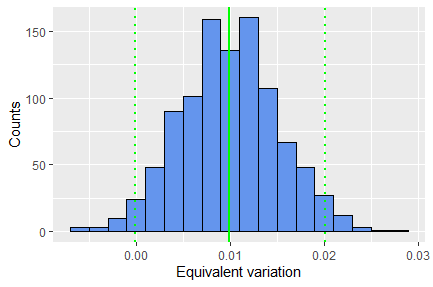}
  \caption{Stratum 4}
  \label{fig:Clb1}
\end{subfigure}
\begin{subfigure}{.5\textwidth}
  \centering
  \includegraphics[width=0.8\linewidth]{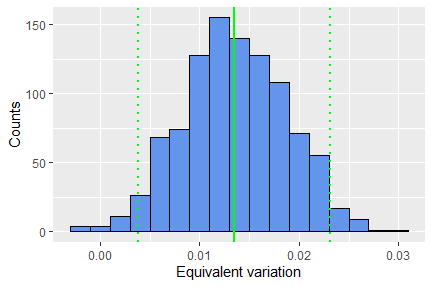}
  \caption{Stratum 5}
  \label{fig:Clc1}
\end{subfigure}%
\begin{subfigure}{.5\textwidth}
  \centering
  \includegraphics[width=0.8\linewidth]{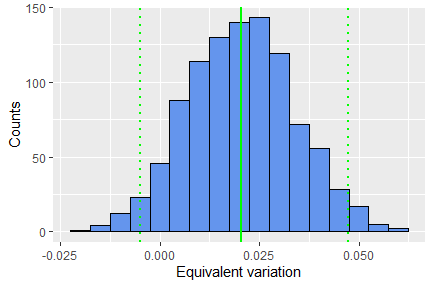}
  \caption{Stratum 6}
  \label{fig:Cld1}
\end{subfigure}
\label{fig:EqVarRepH}
{\scriptsize \caption*{\textit{Notes}: Continuous vertical green lines are posterior means, and the dotted vertical green lines are the 2.5\% and 97.5\% percentiles.}}
\end{figure}

We analyze welfare implications of the 0.12 USD cents/kWh on electricity tariff increase imposed by the Colombian government. We use the equivalent variation as percentage of the income (equation \ref{refM:eq11a}) and posterior estimates to perform inference of this change on the welfare of the representative household. Figure \ref{fig:EqVarRepH} suggests that there is a 95\% probability that the equivalent variation is between 0.60\% and 1.49\% with a mean equal to 1.02\%, that is, on average the representative household faces a utility loss equivalent to 1.02\% of its income, evaluated at the baseline prices, due to the 0.8\% increase on the electricity tariff. 

Actually, the percentage of the electricity tariff tax depends on socioeconomic strata because this is a fixed charge (0.12 US cents/Kwh) that does not discriminate strata, which in turn have different electricity tariffs depending on specific market conditions and regulation. We have that the average electricity tariff rates are 0.08\%, 0.07\% and 0.075\% for strata 4, 5 and 6, respectively.

We estimate the posterior distribution of the equivalent variation for representative households in each stratum. Observe that welfare implications vary with household as consequence of different perceptual variations in electricity tariff, income shares, prices and socioeconomic characteristics. We observe that there are higher welfare losses as socioeconomic strata increase, that is, average equivalent variations equal to 1.0\%, 1.4\% and 2.0\% for strata 4, 5 and 6. All 95\% credible intervals are positive, except for stratum 6 where there is a higher level of uncertainty. 

\begin{figure}[h!]
\caption{Welfare analysis: Sampling distribution of the equivalent variation as percentage of income given a 0.8\% electricity tariff increase}
\centering
  \includegraphics[width=8cm]{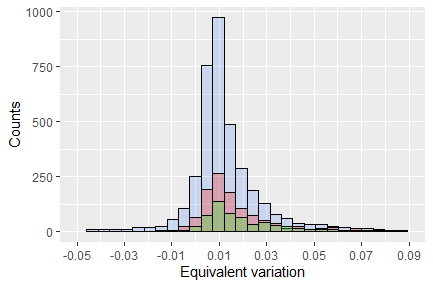}
\label{fig:EqVarSamD}

{\scriptsize \caption*{\textit{Notes}: Sampling distribution of equivalent variation by strata. Blue, red and green are histograms of strata 4, 5 and 6, respectively.}}
\end{figure}

We also estimate the sampling distribution of the equivalent variation by strata taking estimation error into account, that is, we estimate the posterior expected value of the equivalent variation for each household with electricity service, and plot the sampling histogram by stratum. Figure \ref{fig:EqVarSamD} shows that the modal value for all strata is approximately 1\%. However, the sampling trimmed average for strata 4, 5 and 6 are 1.2\%, 1.6\% and 2.1\%.\footnote{We calculated the sampling trimmed mean eliminating 2.5\% of the estimates to avoid asymmetry outliers influence.} This confirms the previous pattern that was found by representative households.  

These results highlight the potential remarkable welfare implications of taxes on inelastic goods.
\clearpage
\subsection{Unobserved preference heterogeneity clusters}\label{sec:het}

\begin{figure}[h!]
\caption{Posterior probability of households belonging to cluster 2}
\includegraphics[width=8cm]{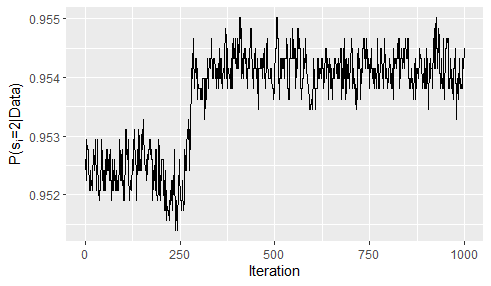}
\centering
\label{fig:PostClust2}
{\scriptsize \caption*{\textit{Notes}: Posterior probability of belonging to cluster 2 by iteration. Approximately 95\% of households belong to this cluster, and there is stability of cluster membership for each household.}}
\end{figure}

Our proposal identifies 4 clusters due to unobserved preference heterogeneity. The clusters have 233, 5517, 29 and 1 households, respectively. We can see in Figure \ref{fig:PostClust2} that approximately 95\% of the households belong to cluster 2. This suggests that there is no a lot of heterogeneity due to unobserved preferences, and potentially observable variables describe in a good way preferences in our setting. In general, there is stability regarding cluster membership for all households.

Tables \ref{sumstatCluster1} and \ref{demstatClust2} in the Appendix show descriptive statistics of shares, prices, and household characteristics by cluster. As expected, representative (modal) household in cluster 2 is very similar to the representative (modal) household in our sample. Clusters 1 and 2 are also similar regarding observable characteristics, except that on average cluster 2 has the lowest proportions of zero utility shares (excluding the singleton cluster 4), and more educated household heads with a comparative low income living in a municipality located 1,000 m.a.s.l. On the other hand, cluster 3 has the highest proportion of zero shares, except in sewerage, and the most proportion of less educated, young and women as household heads, but the highest average income. On the other hand, cluster 4 is composed by the household 1973, which has a very low income facing low utility tariffs and expending just 4\% of its income in utilities.   

\begin{figure}[h]\caption{Unobserved heterogeneity preference clusters: Income shares box plots for representative household from conditional predictive distributions}
\centering
\begin{subfigure}{.5\textwidth}
  \centering
  \includegraphics[width=1\linewidth]{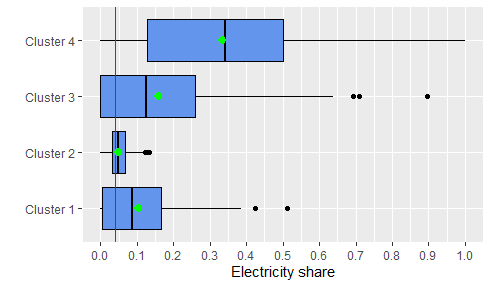}
  \caption{Electricity}
  \label{fig:Cla2}
\end{subfigure}%
\begin{subfigure}{.5\textwidth}
  \centering
  \includegraphics[width=1\linewidth]{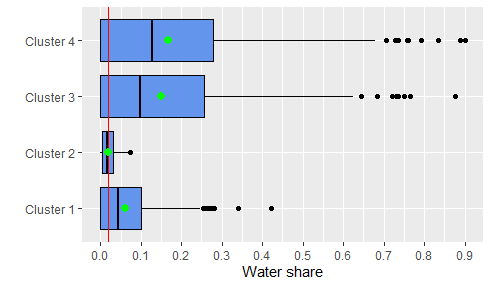}
  \caption{Water}
  \label{fig:Clb2}
\end{subfigure}
\begin{subfigure}{.5\textwidth}
  \centering
  \includegraphics[width=1\linewidth]{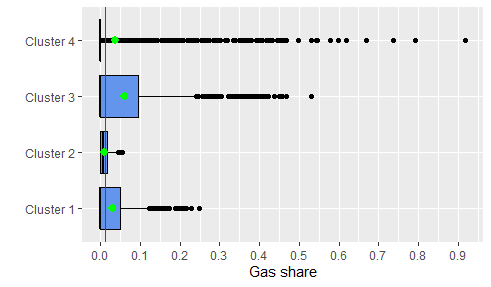}
  \caption{Gas}
  \label{fig:Clc2}
\end{subfigure}%
\begin{subfigure}{.5\textwidth}
  \centering
  \includegraphics[width=1\linewidth]{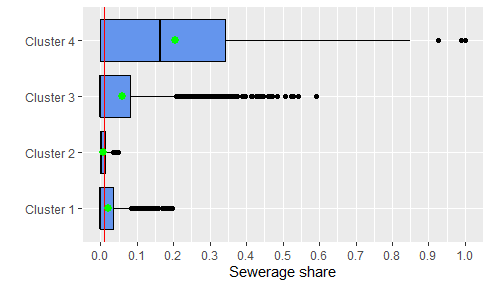}
  \caption{Sewerage}
  \label{fig:Cld2}
\end{subfigure}
\begin{subfigure}{.5\textwidth}
  \centering
  \includegraphics[width=1\linewidth]{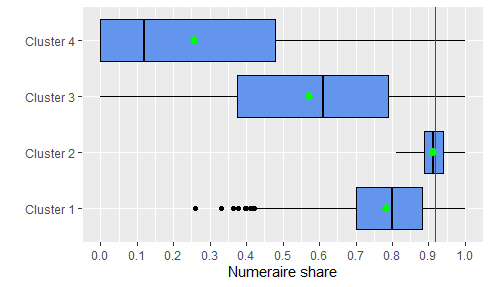}
  \caption{Numeraire}
  \label{fig:Cle2}
\end{subfigure}
\label{fig:Clust}

{\scriptsize \caption*{\textit{Notes}: Income shares by clusters based on predictive distributions conditional on unobserved heterogeneity preference membership assuming observed controls at representative household values. 

Green dot in box plots is the mean value from predictive distributions and red line is observed shares for the representative household that belongs to cluster 2.}}
\end{figure}

Figure \ref{fig:Clust} shows box plots from posterior draws of the predictive distributions conditional on cluster membership evaluated at observed values of the representative household. We can see in Figure \ref{fig:Clust} that the mean values of the representative household (red vertical lines) are very close to the mean values of the predictive distributions (green dots) of cluster 2, which is its cluster.

Observe that this is a kind of counterfactual exercise, that is, what would potentially be the posterior distribution of shares of the representative household, if this would belong to a different unobserved preference heterogeneity cluster. Using as a point predictive value the mean, we see that a hypothetical household whose observable variables would have been equal to the representative household, but with unobserved preference heterogeneity equal to cluster 1, 2, 3 and 4, would potentially have had electricity shares equal to 10.2\%, 5.0\%, 15.7\% and 33.3\%, respectively (see panel a in Figure \ref{fig:Clust}). As expected, the level of uncertainty of these predictive values increases as the number of households in each cluster decreases. Therefore, variability of cluster 2 is the lowest due to most of the households belonging to this cluster, followed by cluster 1, 3 and 4. Similar analysis can be done using the other panels in Figure \ref{fig:Clust}.

We also perform welfare analysis by cluster, the 2.5\% sampling trimmed mean of the equivalent variation is 1.8\%, 1.3\%, -11.7\% and 12.0\% for clusters 1, 2, 3 and 4, respectively. In addition, the 2.5\% and 97.5\% quantiles of the expected values of the equivalent variation are (-29.7\%, 23.0\%), (-2.2\%, 6.6\%) and (-124.8\%, 38.1\%) for clusters 1, 2 and 3. This shows that we may have unreasonable values, particularly for clusters 1, 3 and 4. These results suggest a limitation of our econometric framework; we take unobserved preference heterogeneity into account imposing the restriction of equality in location parameters. Therefore, price and income effects are heterogeneous due to observable variables, but not due to unobserved sensitivity (coefficients). We focus our attention on cluster 2 as most of households are in this group, and results seem sensible.

\clearpage
\subsection{Model assessment: Predictive p-values}

We perform model assessment conditional on membership to cluster 2 as 95\% of our sample seems to belong to it. We use predictive p-values, where our ``discrepancy functions" are the observed income shares and the normalized Slutsky matrix for the representative household that belongs to this cluster.

\begin{figure}[h!]\caption{Model assessment: Income shares histograms for the representative household from conditional predictive distributions}
\centering
\begin{subfigure}{.5\textwidth}
  \centering
  \includegraphics[width=1\linewidth]{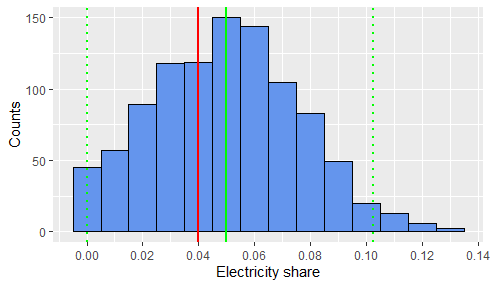}
  \caption{Electricity}
  \label{fig:pvala}
\end{subfigure}%
\begin{subfigure}{.5\textwidth}
  \centering
  \includegraphics[width=1\linewidth]{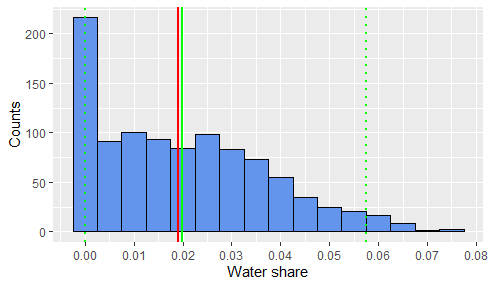}
  \caption{Water}
  \label{fig:pvalb}
\end{subfigure}
\begin{subfigure}{.5\textwidth}
  \centering
  \includegraphics[width=1\linewidth]{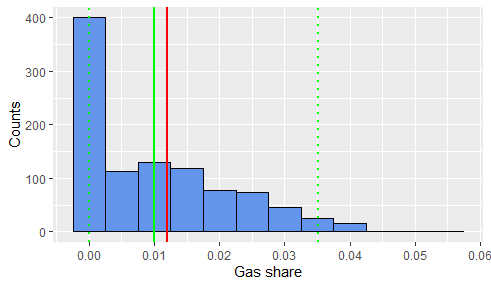}
  \caption{Gas}
  \label{fig:pvalc}
\end{subfigure}%
\begin{subfigure}{.5\textwidth}
  \centering
  \includegraphics[width=1\linewidth]{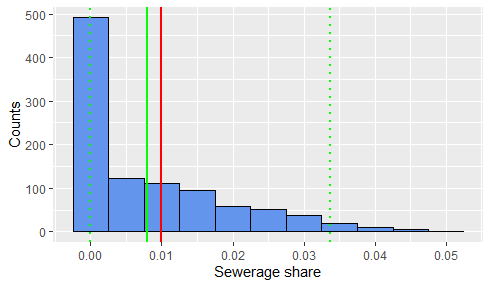}
  \caption{Sewerage}
  \label{fig:pvald}
\end{subfigure}
\begin{subfigure}{.5\textwidth}
  \centering
  \includegraphics[width=1\linewidth]{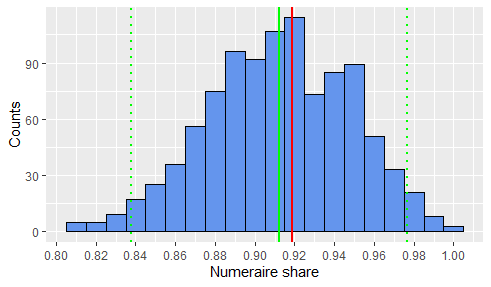}
  \caption{Numeraire}
  \label{fig:pvale}
\end{subfigure}
\label{fig:PredPval}

{\scriptsize \caption*{\textit{Notes}: Income shares for unobserved preference heterogeneity of cluster 2 based on predictive distributions conditional on observed controls at representative household values. 

Green continuous and dotted lines at mean and 95\% symmetric predictive interval from draws of the predictive distribution, and red line is observed shares for the representative household.}}
\end{figure}

Figure \ref{fig:PredPval} shows the histograms associated with the posterior predictive distributions for the representative household conditional on cluster membership due to unobserved preference heterogeneity. Predictive p-values for income shares in electricity, water, gas, sewerage and numeraire are 0.63, 0.48, 0.37, 0.29 and 0.43, respectively. These values suggest good model assessment regarding prediction for the representative household.    

\begin{figure}[h!]\caption{Scatterplot of predictive versus realized discrepancies of the main diagonal elements of the Slutsky matrix for the representative household}
\centering
\begin{subfigure}{.5\textwidth}
  \centering
  \includegraphics[width=1\linewidth]{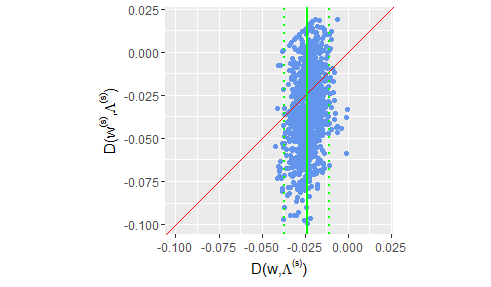}
  \caption{Electricity}
  \label{fig:pvala1}
\end{subfigure}%
\begin{subfigure}{.5\textwidth}
  \centering
  \includegraphics[width=1\linewidth]{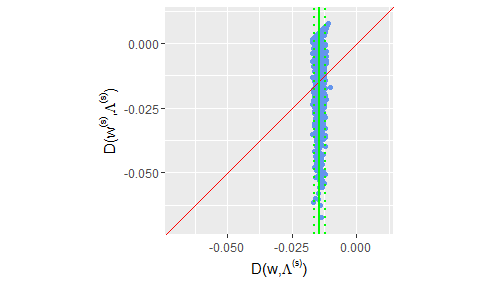}
  \caption{Water}
  \label{fig:pvalb1}
\end{subfigure}
\begin{subfigure}{.5\textwidth}
  \centering
  \includegraphics[width=1\linewidth]{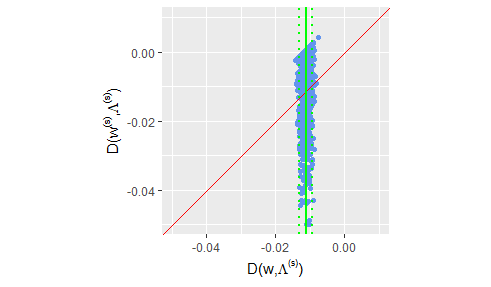}
  \caption{Gas}
  \label{fig:pvalc1}
\end{subfigure}%
\begin{subfigure}{.5\textwidth}
  \centering
  \includegraphics[width=1\linewidth]{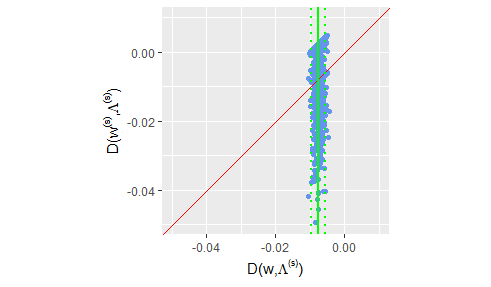}
  \caption{Sewerage}
  \label{fig:pvald1}
\end{subfigure}
\begin{subfigure}{.5\textwidth}
  \centering
  \includegraphics[width=1\linewidth]{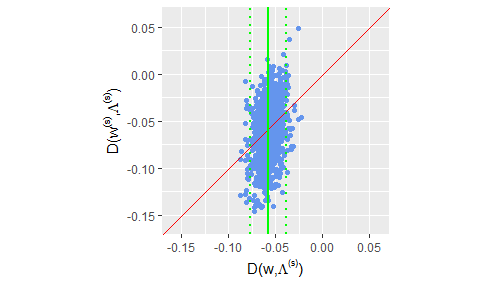}
  \caption{Numeraire}
  \label{fig:pvale1}
\end{subfigure}
\label{fig:PredPvalSl}

{\scriptsize \caption*{\textit{Notes}: The predictive p-value is estimated by the proportion of points above the 45$^{\circ}$ line. 

Red continuous line is the 45$^{\circ}$ line, green continuous and dotted lines at mean and 95\% symmetric credible interval.}}
\end{figure}

We also calculate predictive p-values for the Slutsky matrix. Figure \ref{fig:PredPvalSl} shows scatter plots of predictive versus realized discrepancies of the main diagonal elements of the Slutsky matrix for the representative household. The predictive p-value is estimated by the proportion of points above the 45$^{\circ}$ line. These are 0.32, 0.56, 0.63, 0.66 and 0.43 for electricity, water, gas, sewerage and numeraire, respectively. We have similar results using the predictive p-value based on the average discrepancy statistic. This evidence suggests good model assessment for this microeconomic concept.

We can see in Figure \ref{fig:PredPvalSl} that mean own-price Slutsky terms are all negative, and their symmetric 95\% credible intervals are also negative. This suggests that substitution effects are statistically significant.

Unobserved preference heterogeneity analysis suggests that there is not a lot of heterogeneity in this aspect, and as a consequence, observable variables seem to describe in a good way preferences in our setting. In addition, it seems that there is good model assessment. 

\clearpage
\section{Concluding remarks}\label{sec:concl}
We propose a semi-parametric inferential framework for the EASI incomplete demand system using a Dirichlet process mixture to identify clusters due to unobserved preference heterogeneity. Our application suggests that there are four clusters; although 95\% of the households belong to one of them. This suggests that observable variables describe preferences in a good way. In addition, we find that utilities seem to be inelastic normal goods in the Colombian economy, substitution effects are relevant, particularly between utilities and other goods, and Engel curves are non-linear.

Equivalent variation analysis suggests that there is potential remarkable welfare implications due taxation on inelastic services. In particular, the 0.8\% electricity tariff increase caused an average welfare loss equal to 1.02\% on the representative household. In general, predictive p-values show good model assessment regarding prediction and Slutsky matrix. However, welfare analysis based on unobserved preference heterogeneity gives non sensible results for households that have high observable discrepancies compared to the representative household. This reveals some limitations in our econometric framework.

Future research should consider a full non-parametric specification that would allow to observe price and income heterogeneous effects associated with unobserved preferences. Observe that there is variability in price and income effects just due to observable variables, and Engel curves from different unobserved preference clusters have exactly the same shape, there is just a scale change. In addition, a hypothesis testing framework under more general conditions (non-nested models) should be formulated. For instance, comparing parametric versus non-parametric models. 

\clearpage 
\bibliography{Referencias.bib} \newpage

\begin{thebibliography}{}

\bibitem[Antoniak, 1974]{Antoniak1974}
Antoniak, C. (1974).
\newblock Mixtures of {D}irichlet processes with applications to {B}ayesian
  nonparametric problems.
\newblock {\em The Annals of Statistics}, 2(6):1152--1174.

\bibitem[Banks et~al., 1997]{Banks1997}
Banks, J., Blundell, R., and Lewbel, A. (1997).
\newblock Quadratic {E}ngel curves and consumer demand.
\newblock {\em Review of Economics and Statistics}, 79(4):527--539.

\bibitem[Basu and Chib, 2003]{Basu2003}
Basu, S. and Chib, S. (2003).
\newblock Marginal likelihood and {B}ayes factors for {D}irichlet process
  mixture models.
\newblock {\em Journal of the American Statistical Association},
  98(461):224--235.

\bibitem[Baumol and Bradford, 1970]{baumol1970optimal}
Baumol, W.~J. and Bradford, D.~F. (1970).
\newblock Optimal departures from marginal cost pricing.
\newblock {\em The American Economic Review}, 60(3):265--283.

\bibitem[Blackwell and MacQueen, 1973]{Blackwell1973}
Blackwell, D. and MacQueen, J. (1973).
\newblock Ferguson distributions via {P}\'olya urn schemes.
\newblock {\em The Annals of Statistics}, 1:353--355.

\bibitem[Blundell et~al., 2007]{Blundeletal2007}
Blundell, R., Chen, X., and Kristensen, D. (2007).
\newblock {Semi-nonparametric IV estimation of shape-invariant Engel curves}.
\newblock {\em {Econometrica}}, 75(6):1613--1669.

\bibitem[Blundell and Robin, 1999]{Blundell1999}
Blundell, R. and Robin, J.~M. (1999).
\newblock {Estimation in large and disaggregated demand systems: An estimator
  for conditionally linear systems.}
\newblock {\em {Journal of Applied Econometrics}}, 14(3):209--232.

\bibitem[Burks, 1926]{burks1926inadequacy}
Burks, B.~S. (1926).
\newblock On the inadequacy of the partial and multiple correlation technique.
\newblock {\em Journal of Educational Psychology}, 17(8):532.

\bibitem[Chipman and Moore, 1980]{Chipman1980}
Chipman, J.~S. and Moore, J.~C. (1980).
\newblock {Compensating variation, consumer's surplus, and welfare.}
\newblock {\em {The American Economic Review}}, 70(5):933--949.

\bibitem[Conley et~al., 2008]{Conley2008}
Conley, T., Hansen, C., McCulloch, R., and Rossi, P. (2008).
\newblock A semi-parametric bayesian approach to the instrumental variable
  problem.
\newblock {\em Journal of Econometrics}, 144:276--305.

\bibitem[Deaton and Muellbauer, 1980]{Deaton1980}
Deaton, A. and Muellbauer, J. (1980).
\newblock An almost ideal demand system.
\newblock {\em The American Economic Review}, 70(3):312--326.

\bibitem[Di~Cosmo, 2011]{di2011consumers}
Di~Cosmo, V. (2011).
\newblock Are the consumers always ready to pay? a quasi-almost ideal demand
  system for the italian water sector.
\newblock {\em Water resources management}, 25(2):465--481.

\bibitem[D{\'\i}az and Medlock, 2021]{diaz2021price}
D{\'\i}az, A.~O. and Medlock, K.~B. (2021).
\newblock Price elasticity of demand for fuels by income level in mexican
  households.
\newblock {\em Energy Policy}, 151:112132.

\bibitem[Dickey, 1967]{Dickey1967}
Dickey, J. (1967).
\newblock Matricvariate generalizations of the {M}ultivariate t distribution
  and the {I}nverted {M}ultivariate t distribution.
\newblock {\em The Annals of Mathematical Statistics}, 38(2):511--518.

\bibitem[Dickey, 1971]{dickey1971weighted}
Dickey, J.~M. (1971).
\newblock The weighted likelihood ratio, linear hypotheses on normal location
  parameters.
\newblock {\em The Annals of Mathematical Statistics}, pages 204--223.

\bibitem[Escobar and West, 1995]{Escobar1995}
Escobar, M. and West, M. (1995).
\newblock Bayesian density estimation and inference using mixtures.
\newblock {\em Journal of the American Statistical Association},
  90(430):577--588.

\bibitem[Ferguson, 1973]{Ferguson1973}
Ferguson, T. (1973).
\newblock A {B}ayesian analysis of some nonparametric problems.
\newblock {\em The Annals of Statistics}, 1(2):209--230.

\bibitem[Galvez et~al., 2016]{Galves2016}
Galvez, P., Mariel, P., and Hoyos, D. (2016).
\newblock {Aplication of the quaids model to the residential energy demand in
  spain.}
\newblock {\em {Revista de Economía Aplicada}}, 24(72):87--108.

\bibitem[Gelman et~al., 1996]{gelman1996posterior}
Gelman, A., Meng, X.-L., and Stern, H. (1996).
\newblock Posterior predictive assessment of model fitness via realized
  discrepancies.
\newblock {\em Statistica Sinica}, pages 733--760.

\bibitem[Geweke, 1992]{Geweke1992}
Geweke, J. (1992).
\newblock {Evaluating the accuracy of sampling-based approaches to the
  calculation of posterior moments}.
\newblock In Bernardo, J.~M., Berger, J.~O., Dawid, A.~P., and Smith, A. F.~M.,
  editors, {\em Bayesian Statistics 4: Proceedings of the Fourth Valencia
  International Meeting}, pages 169--193. Oxford University Press.

\bibitem[Gundimeda and Köhlin, 2008]{Gundimeda2008}
Gundimeda, H. and Köhlin, G. (2008).
\newblock {Fuel demand elasticities for energy and environmental policies:
  Indian sample survey evidence.}
\newblock {\em {Energy Economics}}, 30(2):517--546.

\bibitem[Heidelberger and Welch, 1983]{Heidelberger1983}
Heidelberger, P. and Welch, P.~D. (1983).
\newblock Simulation run length control in the presence of an initial
  transient.
\newblock {\em Operations Research}, 31(6):1109--1144.

\bibitem[Jensen and Maheu, 2014]{Jensen2014}
Jensen, M. and Maheu, J. (2014).
\newblock Estimating a semiparametric asymmetric stochastic volatility model
  with a {D}irichlet process mixture.
\newblock {\em Journal of Econometrics}, 178:523--538.

\bibitem[Kass and Raftery, 1995]{kass1995bayes}
Kass, R.~E. and Raftery, A.~E. (1995).
\newblock Bayes factors.
\newblock {\em Journal of the American Statistical Association},
  90(430):773--795.

\bibitem[Kasteridis et~al., 2011]{Kasteridis2011}
Kasteridis, P., Yen, S.~T., and Fang, C. (2011).
\newblock Bayesian estimation of a censored linear almost ideal demand system:
  food demand in {P}akistan.
\newblock {\em American Journal of Agricultural Economics}, 93(5):1374--1390.

\bibitem[Klugkist and Hoijtink, 2007]{klugkist2007bayes}
Klugkist, I. and Hoijtink, H. (2007).
\newblock The bayes factor for inequality and about equality constrained
  models.
\newblock {\em Computational Statistics \& Data Analysis}, 51(12):6367--6379.

\bibitem[Lewbel and Pendakur, 2009]{Lewbel2009}
Lewbel, A. and Pendakur, K. (2009).
\newblock Tricks with {H}icks: {T}he {EASI} {D}emand {S}ystem.
\newblock {\em The American Economic Review}, 99(3):827--863.

\bibitem[{Mas-Colell} et~al., 1995]{Mascollel1995}
{Mas-Colell}, A., {Whinston}, M., and {Green}, J.~R. (1995).
\newblock {\em {Microeconomic Theory}}.
\newblock {Oxford university press}, {New York}.

\bibitem[Miller and Harrison, 2014]{Miller2014}
Miller, J. and Harrison, M. (2014).
\newblock Inconsistency of {P}itman-{Y}or process mixtures for the number of
  components.
\newblock {\em Journal of Machine Learning Research}, 15:3333--3370.

\bibitem[Moshiri and Santillan, 2018]{moshiri2018welfare}
Moshiri, S. and Santillan, M. A.~M. (2018).
\newblock The welfare effects of energy price changes due to energy market
  reform in mexico.
\newblock {\em Energy Policy}, 113:663--672.

\bibitem[M\"uller et~al., 2015]{Muller2015}
M\"uller, P., Quintana, F., Jara, A., and Hanson, T. (2015).
\newblock {\em {Bayesian Nonparametric Data Analysis}}.
\newblock {Springer}, {New York}.

\bibitem[Pearl, 1995]{pearl1995causal}
Pearl, J. (1995).
\newblock Causal diagrams for empirical research.
\newblock {\em Biometrika}, 82(4):669--688.

\bibitem[Pereira et~al., 2019]{Pereira2019}
Pereira, D., Squarize, A., and Gallego, J. (2019).
\newblock {Estimation of elasticities for electricity demand in Brazilian
  households and policy implications}.
\newblock {\em {Energy policy}}, 129:69--79.

\bibitem[Raftery et~al., 1992]{Raftery1992}
Raftery, A.~E., Lewis, S., et~al. (1992).
\newblock {How many iterations in the Gibbs sampler?}
\newblock {\em Bayesian Statistics}, 4(2):763--773.

\bibitem[Ramírez-Hassan, 2021]{Ramirez2021}
Ramírez-Hassan, A. (2021).
\newblock Bayesian estimation of the easi demand system: Replicating the lewbel
  and pendakur (2009) results.
\newblock {\em Journal of Applied Econometrics}, Forthcoming(.):--.

\bibitem[Ramírez-Hassan and Montoya-Blandón, 2019]{Ramirez-Hassan2019}
Ramírez-Hassan, A. and Montoya-Blandón, S. (2019).
\newblock Welfare gains of the poor: an endogenous {B}ayesian approach with
  spatial random effects.
\newblock {\em Econometric Reviews}, 38(3):301--318.

\bibitem[Renzetti, 1999]{renzetti1999municipal}
Renzetti, S. (1999).
\newblock Municipal water supply and sewage treatment: costs, prices, and
  distortions.
\newblock {\em Canadian Journal of Economics}, pages 688--704.

\bibitem[Schulte and Heindl, 2017]{Schulte2017}
Schulte, I. and Heindl, P. (2017).
\newblock {Price and income elasticities of residential energy demand in
  Germany.}
\newblock {\em {Energy Policy}}, 102:512--528.

\bibitem[Sethuraman, 1994]{Sethuraman1994}
Sethuraman, J. (1994).
\newblock A constructive definition of {D}irichlet priors.
\newblock {\em Statistica Sinica}, 4:639--650.

\bibitem[Su{\'a}rez-Varela, 2020]{suarez2020modeling}
Su{\'a}rez-Varela, M. (2020).
\newblock Modeling residential water demand: An approach based on household
  demand systems.
\newblock {\em Journal of Environmental Management}, 261:109921.

\bibitem[Tanner and Wong, 1987]{Tanner1987}
Tanner, M.~A. and Wong, W.~H. (1987).
\newblock The calculation of posterior distributions by data augmentation.
\newblock {\em Journal of the American Statistical Association},
  82(398):528--540.

\bibitem[Tovar and W\"olfing, 2018]{Tovar2018}
Tovar, M. and W\"olfing, N. (2018).
\newblock {Household energy prices and inequality: Evidence from German
  microdata based on the EASI demand system.}
\newblock {\em {Energy Economics}}, 70:84--97.

\bibitem[Verdinelli and Wasserman, 1995]{verdinelli1995computing}
Verdinelli, I. and Wasserman, L. (1995).
\newblock Computing bayes factors using a generalization of the savage-dickey
  density ratio.
\newblock {\em Journal of the American Statistical Association},
  90(430):614--618.

\bibitem[Wales and Woodland, 1983]{Wales1983}
Wales, T. and Woodland, A. (1983).
\newblock Estimation of consumer demand systems with binding non-negativity
  constraints.
\newblock {\em Journal of Econometrics}, 21:263--285.

\bibitem[Wetzels et~al., 2010]{wetzels2010encompassing}
Wetzels, R., Grasman, R.~P., and Wagenmakers, E.-J. (2010).
\newblock An encompassing prior generalization of the savage--dickey density
  ratio.
\newblock {\em Computational Statistics \& Data Analysis}, 54(9):2094--2102.

\bibitem[Wright, 1921]{Wright1921}
Wright, S. (1921).
\newblock Correlation and causation.
\newblock {\em Journal of Agricultural Research}, 20:557--585.

\bibitem[Zhen et~al., 2014]{zhen2014predicting}
Zhen, C., Finkelstein, E.~A., Nonnemaker, J.~M., Karns, S.~A., and Todd, J.~E.
  (2014).
\newblock Predicting the effects of sugar-sweetened beverage taxes on food and
  beverage demand in a large demand system.
\newblock {\em American Journal of Agricultural Economics}, 96(1):1--25.

\end{thebibliography}

\section{Appendix}\label{sec: appendix}

\subsection{Markov chain Monte Carlo sampler}\label{sec:MCMC}

\begin{itemize}
    \item Sampling $\bm\theta$
    
    Let us set $\bm u_i\equiv\begin{bmatrix} {\bm \epsilon_i}\\ \bm v_i \end{bmatrix}=\begin{bmatrix} \tilde{\bm w}_i^*\\ \bm y_i^* \end{bmatrix}-\begin{bmatrix} \bm F_i\bm\phi\\ \bm G_i\bm\psi \end{bmatrix}$, then the conditional posterior distribution of $\bm\theta_i$ is
    \begin{align*}
        \bm\theta_i|\left\{\bm\theta_{i'},\bm s_{i'}:i'\neq i\right\}, \bm u_i, \alpha & \sim \sum_{i'\neq i}\frac{N_m^{(i)}}{\alpha+N-1}f_N(\bm u_i|\bm\mu_m,\bm\Sigma_m)\\
        & +\frac{\alpha}{\alpha+N-1}\int_{\bm\mu}\int_{\bm\Sigma}f_N(\bm u_i|\bm\mu,\bm\Sigma)f_N\left(\bm\mu\Big|\bm\mu_0,\frac{1}{\tau_0}\bm\Sigma\right)f_{IW}(\bm\Sigma|r_0,\bm R_0)d\bm\Sigma d\bm\mu,
    \end{align*}
    where $N_m^{(i)}$ is the number of observations such that $s_{i'}=m$, $i'\neq i$.
    
    Let us define $g(\bm u_i)=\int_{\bm\mu}\int_{\bm\Sigma}f_N(\bm u_i|\bm\mu,\bm\Sigma)f_N\left(\bm\mu\Big|\bm\mu_0,\frac{1}{\tau_0}\bm\Sigma\right)f_{IW}(\bm\Sigma|r_0,\bm R_0)d\bm\Sigma d\bm\mu$. After collecting terms, using properties of the Inverse Wishart distribution and multivariate Gamma function, and Theorem 3.1 in \cite{Dickey1967}, we get that $g(\bm u_i)=f_{MSt}\left(\bm u_i\Big|\bm\mu_0,\left(\frac{1+\tau_0}{\tau_0}\frac{\bm R_0}{v}\right),v\right)$ where $v=r_0+1-(q+J-1)=3>2$ to have existence of the variance matrix. 
    
    Therefore, we sample $s_i$ as follows,
    \begin{equation*}
        s_i|\left\{\bm\Sigma_{i'},\bm\mu_{i'},\bm s_{i'}:i'\neq i\right\}, \bm u_i, \alpha\sim\begin{Bmatrix}P(s_i=0|\cdot)=q_0^*\\
        P(s_i=m|\cdot)=q_m^*, m=1,2,\dots,M^{(i)}\end{Bmatrix},
    \end{equation*}
    
    where $M^{(i)}$ is the number of clusters excluding $i$, which may have its own cluster (singleton cluster), $q^*_c=\frac{q_c}{q_0+\sum_m q_m}$, $q_c=\left\{q_0,q_m\right\}$, $q_m=\frac{N_m^{(i)}}{\alpha+N-1}f_N(\bm u_i|\bm\mu_m,\bm\Sigma_m)$ and $q_0=\frac{\alpha}{\alpha+N-1}f_{MSt}\left(\bm u_i\Big|\bm\mu_0,\left(\frac{1+\tau_0}{\tau_0}\frac{\bm R_0}{v}\right),v\right)$.
    
    If $s_i=0$ is sampled, then $s_i=M+1$ where $M$ is the total number of clusters in the sample, and a new $\bm\Sigma_m$ is sample from $IW\left(r_0+1,\bm R_0+\frac{\tau_0}{1+\tau_0}(\bm u_i-\bm\mu_0)(\bm u_i-\bm\mu_0)^{\top}\right)$, a new $\bm\mu_m$ is sample from $N(\tilde{\bm\mu}_i,(1+\tau_0)^{-1}\bm\Sigma_m)$ where $\tilde{\bm\mu}_i=\frac{\bm u_i+\tau_0\bm\mu_0}{1+\tau_0}$.
    
    Discarding $\bm\theta_m$'s from last step, we use $\bm s$ and $M$ to sample $\bm\Sigma_m$ from 
    
    \begin{equation*}
    IW\left(r_0+N_m,\bm R_0+\sum_{i:s_i=m}(\bm u_i-\bar{\bm u}_m)(\bm u_i-\bar{\bm u}_m)^{\top}+\frac{\tau_0N_m}{\tau_0+N_m}(\bm \mu_0-\bar{\bm u}_m)(\bm \mu_0-\bar{\bm u}_m)^{\top}\right),
    \end{equation*}
    
    where $\bar{\bm u}_m=\frac{\sum_{i:s_i=m}\bm u_i}{N_m}$. And sample $\bm\mu_m$ from
    \begin{equation*}
        N\left(\hat{\bm\mu}_m,\frac{1}{N_m+\tau_0}\bm\Sigma_m\right),
    \end{equation*}
    where $\hat{\bm\mu}_m=\frac{N_m}{N_m+\tau_0}\bar{\bm u}_m+\frac{\tau_0}{N_m+\tau_0}\bm\mu_0$, $m=1,2,\dots,M$.
    
    \item Sampling $\alpha$
    
    Following \cite{Escobar1995}, we draw $\xi|\alpha,N\sim Be(\alpha+1,N)$, and then we draw $\alpha|\xi,M,\pi_{\xi}\sim\pi_{\xi}{G}(\alpha_0+M,\beta_0-log(\xi))+(1-\pi_{\xi}){G}(\alpha_0+M-1,\beta_0-log(\xi))$, where $\frac{\pi_{\xi}}{1-\pi_{\xi}}=\frac{\alpha_0+M-1}{N(\beta_0-log(\xi))}$.
    
    \item Sampling $\bm\phi$
    
    The conditional posterior distribution for the location parameters in equation \ref{ref:eq6} is multivariate normal,
    
    \begin{align*}
      \bm\phi|(\bm\psi, \left\{\bm\Sigma_i\right\}, \left\{\bm\mu_i\right\}, \bm s, \tilde{\bm w}^*,\bm y^*)\sim MN(\bar{\bm{\phi}}, \bar{\bm{\Phi}}),  
    \end{align*}
    
    where $\bar{\bm{\Phi}}=\left(\sum_{m=1}^M\sum_{i:s_i=m} \left\{\bm F_i^{\top}(\bm\Sigma_{\epsilon\epsilon,i}-\bm\Sigma_{\epsilon v,i}\bm\Sigma_{vv,i}^{-1}\bm\Sigma_{v\epsilon,i})^{-1}\bm F_i\right\}+\bm\Phi^{-1}_0\right)^{-1}$ and 
    \begin{align*}
        \bar{\bm\phi}=&\bar{\bm\Phi}\left[\sum_{m=1}^M\sum_{i:s_i=m}\left\{\bm F_i^{\top}(\bm\Sigma_{\epsilon\epsilon,i}-\bm\Sigma_{\epsilon v,i}\bm\Sigma_{vv,i}^{-1}\bm\Sigma_{v\epsilon,i})^{-1}\right.\right.\\
        &\left.\left.(\tilde{\bm w}_i^*-\bm \mu_{\tilde{\bm w}_i^*}-\bm\Sigma_{\epsilon v,i}\bm\Sigma_{vv,i}^{-1}(\bm y_i^*-\bm\mu_{y_i^*}-\bm G_i\bm\psi))\right\}+\bm\Phi_0^{-1}\bm\phi_0\right],
    \end{align*}
    
    where $\mu_{\tilde{\bm w}_i^*}$ is the vector composed by the $J-1$ first elements of $\bm\mu_i$, and $\bm\mu_{y_i^*}$ is the vector composed by the last $q$ elements of $\bm\mu_i$.
    
    As suggested by \cite{Lewbel2009} for EASI models, and \cite{Conley2008} for semi-parametric Bayesian regression, we centered all regressors at the representative household. So, we reject any $\phi$ draw that does not satisfy the sufficient and necessary condition for concavity of the cost function, that is, $\bm A+\bm B y+\bm w\bm w^{\top}-\bm W$ should be negative semidefinite. 
    
    \item Sampling $\bm\psi$
    
    The conditional posterior distribution for the location parameters in equation \ref{ref:eq7} is multivariate normal, 
    
    \begin{equation}
      \bm\psi|(\cdot)\sim MN(\bar{\bm{\psi}}, \bar{\bm{\Psi}}),  
    \end{equation}
    
    where $\bar{\bm{\Psi}}=\left(\sum_{m=1}^M\sum_{i:s_i=m}\left\{ \bm G_i^{\top}(\bm\Sigma_{vv,i}-\bm\Sigma_{v\epsilon,i }\bm\Sigma_{\epsilon\epsilon,i}^{-1}\bm\Sigma_{\epsilon v},i)^{-1}\bm G_i\right\}+\bm\Psi^{-1}_0\right)^{-1}$ and 
    \begin{align*}
        \bar{\bm\psi}&=\bar{\bm\Psi}\left[\sum_{m=1}^M\sum_{i:s_i=m}\left\{\bm G_i^{\top}(\bm\Sigma_{vv,i}-\bm\Sigma_{v\epsilon,i}\bm\Sigma_{\epsilon\epsilon,i}^{-1}\bm\Sigma_{\epsilon v,i})^{-1}\right.\right.\\
        &\left.\left.({\bm y}_i^*-\bm\mu_{y_i^*}-\bm\Sigma_{v\epsilon,i }\bm\Sigma_{\epsilon\epsilon,i}^{-1}(\tilde{\bm w}_i^*-\mu_{\tilde{\bm w}_i^*}-\bm F_i\bm\phi))\right\}+\bm\Psi_0^{-1}\bm\psi_0\right].
    \end{align*}
    
    \item Sampling $\tilde{\bm w}^{*}$

Let us order the budget shares for unit $i$ such that the first budget shares $j\not\in J_i$, and write the covariance matrix as
\begin{equation*}
    \bm\Sigma=\begin{bmatrix} 
    \bm\Sigma_{\epsilon^-_i\epsilon^-_i} & \bm\Sigma_{\epsilon^-_i\tau_i}\\
    \bm\Sigma_{\tau_i\epsilon^-_i} & \bm\Sigma_{\tau_i\tau_i}
    \end{bmatrix},
\end{equation*}

where $\bm\Sigma_{\epsilon^-_i\epsilon^-_i}$ is the covariance matrix of $\bm\epsilon^-_{ij}=\left\{\epsilon_{ij}:j\not\in J_i\right\}$, and $\bm\tau_i=\left[\bm\epsilon_{ij}^{+\top} \ \bm v^{\top}_i\right]^{\top}$, $\bm\epsilon^+_{ij}=\left\{\epsilon_{ij}:j\in J_i\right\}$. In general, superscripts $^+$ and $^-$ indicate vector components associated with indices $j$ such that $\left\{j\in J_i\right\}$ and $\left\{j\not \in J_i\right\}$, respectively. Therefore, the conditional posterior distribution of the negative latent shares is a truncated multivariate normal,
\begin{equation*}\tilde{\bm w}_{i}^{*-}|(\cdot)\sim TMN_{(-\bm\infty,\bm 0]}\left(\widehat{\tilde{\bm w}_{i}^{*-}}, \widehat{\bm\Sigma_{\tilde{\bm w}_{i}^{*-}}}\right),
\end{equation*}

where $\widehat{\tilde{\bm w}_{i}^{*-}}=\mu_{\tilde{\bm w}_i^{*-}}+\bm F_i^-\bm\phi^- + \bm\Sigma_{\epsilon^-_i\tau_i}\bm\Sigma_{\tau_i\tau_i}^{-1}\begin{bmatrix}
\tilde{\bm w}^{*+}_{i}-\mu_{\tilde{\bm w}_i^{*+}}-\bm F_i^+\bm\phi^+\\
\bm y_i^*-\mu_{{\bm y}_i^*}-\bm G_i\bm\psi
\end{bmatrix}$, and $\widehat{\bm\Sigma_{\tilde{\bm w}_{i}^{*-}}}=\bm\Sigma_{\epsilon^-_i\epsilon^-_i}-\bm\Sigma_{\epsilon^-_i\tau_i}\bm\Sigma_{\tau_i\tau_i}^{-1}\bm\Sigma_{\tau_i\epsilon^-_i}$.

Using equation \ref{ref:eq8}, the conditional posterior distribution of the positive latent shares is a degenerate density with probability one at ${w}_{ij}^{*+}=(1-\sum_j {w}_{ij}^{*-}){ w}_{ij}$.

\item Sampling ${\bm y}^{*}$

We use the ``first stage'' system of equations just as a way to tackle endogeneity issues. Therefore, we iteratively update $\bm y$ using equation \ref{refM:eq2}, and draws of $\bm A$ and $\bm B$ from $\bm \phi$ such the conditional posterior distribution of ${\bm y}^{*}_i$ is a degenerate density with probability one at $\bm y_i^*=\left[y_i \ y_i^2 \dots y_i^R \ \bm z_{i}^{\top} y_i \ \tilde{\bm p}_i^{\top}y_i \right]^{\top}$.
\end{itemize}

\subsection{Posterior predictive p-value}\label{subsec: p-value}

We have the following algorithm to calculate the posterior predictive p--values:
\begin{itemize}
    \item Do for $s=1,2,\dots,S$:  
    \begin{enumerate}
    \item Take a draw $\bm\Lambda^{(s)}$ from the MCMC sampler.
    \item Generate a pseudo-share $\bm w^{(s)}$ using the predictive density (see Section \ref{sec:Predictive}).
    \item Calculate $D(\bm w^{(s)},\bm \Lambda^{(s)})$ and $D(\bm w,\bm \Lambda^{(s)})$. In the case of the predictive assessment $D(\bm w,\bm \Lambda^{(s)})=\bm w$, and for the Slutsky matrix using Equation \ref{refM:eq5A}  with the observed income shares.
    \end{enumerate}
    \item Estimate $p_D(\bm w)=P[D(\bm w^{(s)}, \bm\Lambda)\geq D(\bm w, \bm\Lambda)]$ using the proportion of the $S$ pairs for which $D(\bm w^{(s)}, \bm\Lambda^{(s)})\geq D(\bm w, \bm\Lambda^{(s)})$. 
    \item Use $\bm\Lambda^{(s)}$ to calculate $\mathbb{E}[D(\bm w,\bm\lambda)]\approx \frac{1}{S}\sum_{s=1}^S D(\bm w, \bm\Lambda^{(s)})$.
    \item Estimate $p_{\mathbb{E}[D]}(\bm w)=P[D(\bm w^{(s)}, \bm\Lambda)\geq D(\bm w)]$ as the proportion of $S$ draws of $D(\bm w^{(s)},\bm \Lambda^{(s)})$ that are greater than our estimate of $\mathbb{E}[D(\bm w,\bm\lambda)]$. 
\end{itemize}

\subsection{Descriptive statistics: Strata and clusters}\label{subsec: descriptive}

\begin{table}[h!]
\centering \caption{Summary statistics: Utility shares and prices by strata \label{sumstat2}}
\begin{threeparttable}
\resizebox{1\textwidth}{!}{\begin{minipage}{\textwidth}
\begin{tabular}{l c c c c c}\hline
\multicolumn{1}{c}{\textbf{Variable}} & \textbf{Mean}
 & \textbf{Std. Dev.} & \textbf{Zero Shares} & & \\
 \hline
 \textbf{Shares} &  &  &  &  &  \\
 \textbf{Stratum 4}& & & & & \\
 Electricity & 0.04 & 0.05 & 0.04 & & \\ 
 Water & 0.02 & 0.03 & 0.04 & & \\
 Sewerage & 2e-03 & 0.01  & 0.48 & & \\
 Natural Gas & 5e-03 & 0.01 & 0.20 & & \\
 Numeraire & 0.94 & 0.07 & 0.00 & & \\

  \textbf{Stratum 5}& & & & & \\
Electricity & 0.03 & 0.04 & 0.10 & & \\
Water & 0.02 & 0.02 & 0.04& & \\
Sewerage & 2e-03 & 0.01 & 0.46 & & \\ 
Natural Gas & 4e-03 & 0.01 & 0.15 & & \\
Numeraire & 0.95 & 0.06 & 0.00 & & \\

 \textbf{Stratum 6}& & & & & \\
Electricity & 0.03 & 0.05 & 0.05 & & \\
Water & 0.01 & 0.02 & 0.03 & & \\
Sewerage & 2e-03 & 0.01 & 0.50 & & \\
Natural Gas & 4e-03 & 0.01 & 0.14  & & \\
Numeraire & 0.95 & 0.07 & 0.00 & & \\

\addlinespace[.75ex]
\hline
\textbf{Prices}& & &\textbf{N} & \textbf{Min} & \textbf{Max} \\
\hline
 \textbf{Stratum 4}& & & 3,863 & & \\
 Electricity (USD/kWh) & 0.16 & 0.01 &  & 0.13 & 0.20 \\
 Water (USD/m3) & 0.55 & 0.15 & &  0.16 & 0.91 \\
 Sewerage (USD/m3) & 0.46 & 0.15 & & 0.16 & 1.05 \\
 Natural Gas (USD/m3) & 0.46 & 0.12 & & 0.09 & 0.88 \\
 \textbf{Stratum 5}& & & 1,283 & & \\
Electricity (USD/kWh) & 0.18 & 0.01 &  & 0.16 & 0.27 \\
Water (USD/m3) & 0.87 & 0.23 & & 0.23 & 1.39 \\
Sewerage (USD/m3) & 0.77 & 0.22 & & 0.24 & 1.23 \\
Natural Gas (USD/m3) & 0.57 & 0.09 & & 0.22 & 0.91 \\
 \textbf{Stratum 6}& & & 634 & & \\
 Electricity (USD/kWh) & 0.17 & 0.01 &  & 0.15 & 0.21 \\ 
 Water (USD/m3) & 1.02 & 0.21 & & 0.25 & 1.46 \\
 Sewerage (USD/m3) & 0.85 & 0.22 & & 0.38 & 1.31 \\
 Natural Gas (USD/m3) & 0.57 & 0.08 & & 0.36 & 0.83 \\
 \hline
\addlinespace[.75ex]
\end{tabular}
\begin{tablenotes}[para,flushleft]
 \footnotesize Prices are converted to dollars using the exchange rate of 30/06/2017, equivalent to COP/USD 3,038.26. \textbf{Source:} Authors' calculations based on information from ENPH, CREG, CRA, SUI and Superintendencia Financiera de Colombia.
  \end{tablenotes}
    \end{minipage}}
  \end{threeparttable}
\end{table}

\begin{table}[h]\caption{t-tests: Shares and prices \label{ttestsp}}
\begin{threeparttable}
\resizebox{0.7\textwidth}{!}{\begin{minipage}{\textwidth}
\begin{tabular}{l c c}\hline
\multicolumn{1}{c}{\textbf{Variable}} & \textbf{Difference}
 & \textbf{t statistics} \\
\hline
\textbf{Stratum 4 - 5 }   &  &   \\
\addlinespace[.75ex]

Electricity         &      0.0103$^{***}$&      (9.61)\\
Aqueduct            &    0.000147         &      (0.28)\\
Sewerage            &   -0.000161         &     (-1.17)\\
Natural Gas         &     0.00122$^{***}$&      (5.23)\\
Numeraire           &     -0.0114$^{***}$&     (-5.08)\\
Electricity(USD/kWh)&     -0.0252$^{***}$&    (-67.38)\\
Aqueduct(USD/m3 )   &      -0.320$^{***}$&    (-58.26)\\
Sewerage(USD/m3)    &      -0.312$^{***}$&    (-57.43)\\
Natural Gas(USD/m3) &      -0.103$^{***}$&    (-27.33)\\
\textbf{N}       &    \textbf{5,146} &   \\

\hline

\addlinespace[.75ex]

\textbf{Stratum 4 - 6 }   &  &     \\
Electricity         &     0.00391$^{**}$ &      (2.63)\\
Aqueduct            &     0.00157$^{*}$  &      (2.33)\\
Sewerage            &    0.000392$^{*}$  &      (2.19)\\
Natural Gas         &     0.00120$^{***}$&      (3.87)\\
Numeraire           &    -0.00787$^{*}$  &     (-2.57)\\
Electricity(USD/kWh)&     -0.0156$^{***}$&    (-31.92)\\
Aqueduct(USD/m3 )   &      -0.473$^{***}$&    (-70.66)\\
Sewerage(USD/m3)    &      -0.392$^{***}$&    (-57.59)\\
Natural Gas(USD/m3) &      -0.104$^{***}$&    (-20.37)\\
\textbf{N}       &    \textbf{4,497} &   \\

\hline

\addlinespace[.75ex]

\textbf{Stratum 5 - 6 }   &  &   \\
Electricity         &    -0.00642$^{***}$&     (-3.97)\\
Aqueduct            &     0.00142         &      (1.85)\\
Sewerage            &    0.000553$^{**}$ &      (2.65)\\
Natural Gas         &  -0.0000187         &     (-0.06)\\
Numeraire           &     0.00350         &      (1.14)\\
Electricity(USD/kWh)&     0.00950$^{***}$&     (15.82)\\
Aqueduct(USD/m3 )   &      -0.154$^{***}$&    (-14.27)\\
Sewerage(USD/m3)    &     -0.0800$^{***}$&     (-7.36)\\
Natural Gas(USD/m3) &   -0.000717         &     (-0.16)\\
\textbf{N}       &    \textbf{1,917} &   \\
\hline

\multicolumn{3}{l}{\footnotesize \textit{Ho:} mean difference is equal to 0.}\\
\multicolumn{3}{l}{\footnotesize $^{*}$ \(p<\)0.05, $^{**}$ \(p<\)0.01, $^{***}$ \(p<\)0.001}\\
\end{tabular}
    \end{minipage}}
  \end{threeparttable}
\end{table}

\begin{table}[h]
\centering \caption{Descriptive Statistics: Household characteristics by strata
\label{demstat}}
\begin{threeparttable}
\resizebox{1\textwidth}{!}{\begin{minipage}{\textwidth}
\begin{tabular}{l c c c c c}\hline
\multicolumn{1}{c}{\textbf{Variable}} & \textbf{Mean}
 & \textbf{Std. Dev.}& \textbf{Min.} &  \textbf{Max.} & \textbf{N}\\ \hline
 
\textbf{Stratum 4}& & & & & 3,863 \\
\hline
\addlinespace[.75ex]

Household head age & 54.08 & 15.75 & 18 & 97 & \\
Household head gender (female) & 0.42 &  & 0 & 1 & \\
Members & 2.91 & 1.42 & 1 & 10 & \\

\textbf{Education}& & & & & \\
Elementary school & 0.06 &  & 0 & 1 & \\
High School & 0.17 &  & 0 & 1 & \\
Vocational & 0.15 &  & 0 & 1 & \\
Undergraduate & 0.38 &  & 0 & 1 & \\
Postgraduate & 0.24 &  & 0 & 1 & \\

\textbf{Altitude}& & & & & \\
Below 1,000 m.a.s.l & 0.48 &  & 0 & 1 & \\
More than 1,000 m.a.s.l & 0.52 &  & 0 & 1 & \\

Total income & 1,653.86 & 1,810.56 & 22.65 & 43,138.32 \\

\addlinespace[.75ex]
\hline
\textbf{Stratum 5}& & & & & 1,283 \\
\hline
\addlinespace[.75ex]

Household head age & 56.66 & 16.27 & 18 & 99 \\
Household head gender (female) & 0.42 &  & 0 & 1 \\
Members & 2.69 & 1.31 & 1 & 8 \\

\textbf{Education}& & & & & \\

Elementary school & 0.04 & & 0 & 1 \\
High School & 0.11 &  & 0 & 1 \\
Vocational & 0.09 &  & 0 & 1 \\
Undergraduate & 0.43 &  & 0 & 1 \\
Postgraduate & 0.34 &  & 0 & 1 \\

\textbf{Altitude}& & & & & \\
Below 1,000 m.a.s.l & 0.30 &  & 0 & 1 \\
More than 1,000 m.a.s.l & 0.70 &  & 0 & 1 \\

Total income & 2,500.49 & 2,491.14 & 65.83 & 50,483.6 \\

\addlinespace[.75ex]
\hline
\textbf{Stratum 6}& & & & & 634 \\
\hline
\addlinespace[.75ex]

Household head age & 58.49 & 15.14 & 20 & 97 \\
Household head gender (female) & 0.36 &  & 0 & 1 \\
Members & 2.75 & 1.33 & 1 & 8 \\

\textbf{Education}& & & & & \\
Elementary school & 0.03 &  & 0 & 1 \\
High School & 0.10 &  & 0 & 1 \\
Vocational & 0.06 &  & 0 & 1 \\
Undergraduate & 0.41 &  & 0 & 1 \\
Postgraduate & 0.41 &  & 0 & 1 \\

\textbf{Altitude}& & & & & \\
Below 1,000 m.a.s.l & 0.47 &  & 0 & 1 \\
More than 1,000 m.a.s.l & 0.53 &  & 0 & 1 \\
Total income & 3,766.49 & 3,365.81 & 75.76 & 29,321.93 \\

\hline

 \hline
\end{tabular}
      \end{minipage}}
  \end{threeparttable}
\end{table}

\begin{table}[h]\caption{t-tests: Socioeconomic variables \label{ttestdem}}

\begin{threeparttable}
\resizebox{0.6\textwidth}{!}{\begin{minipage}{\textwidth}
\begin{tabular}{l c c }\hline
\multicolumn{1}{c}{\textbf{Variable}} & \textbf{Difference}
 & \textbf{t statistics} \\
\hline
\textbf{Stratum 4 - 5 }   &  &   \\ 
\addlinespace[.75ex]

Household head age             &      -2.583$^{***}$&     (-5.05)\\
Members             &       0.206$^{***}$&      (4.59)\\
Total income        &      -846.6$^{***}$&    (-13.12)\\
\textbf{N}       &    \textbf{5,146}&  \\  
\hline

\addlinespace[.75ex]

\textbf{Stratum 4 - 6 }   &  &    \\  
Household head age             &      -4.410$^{***}$&     (-6.57)\\
Members             &       0.149$^{*}$  &      (2.46)\\
Total income        &     -2,112.6$^{***}$&    (-23.47)\\
\textbf{N}       &    \textbf{4,497}&   \\

\hline
\addlinespace[.75ex]

\textbf{Stratum 5 - 6 }   &  &    \\
Household head age             &      -1.827$^{*}$  &     (-2.37)\\
Members             &     -0.058         &     (-0.91)\\
Total income        &     -1,266.0$^{***}$&     (-9.28)\\

\textbf{N}       &    \textbf{1,917}&   \\ 

\hline
\multicolumn{3}{l}{\footnotesize \textit{Ho:} mean difference is equal to 0.}\\
\multicolumn{3}{l}{\footnotesize $^{*}$ \(p<\)0.05, $^{**}$ \(p<\)0.01, $^{***}$ \(p<\)0.001}\\
\end{tabular}
    \end{minipage}}
  \end{threeparttable}
\end{table}

\begin{table}[h]\caption{$\protect{\chi}^2$ tests: Household characteristics\label{ttestdem2}}
\begin{tabular}{l c c c }\hline
\multicolumn{1}{c}{\textbf{Variable}} & $\protect{\chi}^2$
 & Pr  & df \\
 \hline
Household head gender (male)          &    8.54  &     0.014 & 2 \\
Education   &     190.05  &  0.000  & 8   \\
\textbf{N}       &    \textbf{5,780} &  &  \\

\hline
\multicolumn{3}{l}{\footnotesize \textit{Ho:} equal proportions.}\\
\multicolumn{3}{l}{\footnotesize $^{*}$ \(p<\)0.05, $^{**}$ \(p<\)0.01, $^{***}$ \(p<\)0.001}\\
\end{tabular}
\end{table}

\begin{figure}[h]
    \caption{Compensated  and  uncompensated  elasticities:  Point  estimates and 95\% credible intervals for the representative household by stratum}\label{pi_compaN}
\begin{subfigure}{.5\textwidth}
  \centering
\animategraphics[buttonsize=0.3cm,controls={play,stop, step},loop,width=\linewidth]{1}{3_Equation_Merge}{}{}
   \label{fig:2a8}
   \caption{Hicksian share semi-elasticities}
\end{subfigure}%
 \begin{subfigure}{.5\textwidth}
  \centering
\animategraphics[buttonsize=0.26cm,controls={play,stop, step},width=\linewidth]{1}{8_Equation_Merge}{}{}
   \label{fig:2b8}
   \caption{Marshallian share semi-elasticities}
\end{subfigure}%

\begin{subfigure}{.5\textwidth}
  \centering
\animategraphics[buttonsize=0.3cm,controls={play,stop, step},width=\linewidth]{1}{11_Equation_Merge}{}{}
   \label{fig:2c8}
   \caption{Marshallian quantity elasticities}
 \end{subfigure}%
\begin{subfigure}{.5\textwidth}
  \centering
\animategraphics[buttonsize=0.3cm,controls={play,stop, step},width=\linewidth]{1}{10_Equation_Merge}{}{}
   \label{fig:2d8}
   \caption{Marshallian income elasticities}
 \end{subfigure}%
{\scriptsize \caption*{\textit{Notes}: Blue, green and black are for strata 4, 5 and 6. Circles are posterior mean values, and bars are 95\% symmetric credible intervals. Notation \textit{wupj} (\textit{qupj}) indicates the effect of percent change in price of good $j$ on share (quantity) for good $u$. For instance, \textit{wepw} in panel a indicates that 1 percent price increase in water would imply 0.016 percent points less income share for electricity.}}
 \end{figure}
 

\begin{figure}[h]
    \caption{Engel cuves: Point estimates and 90\%, 95\% and 99\% credible intervals by stratum}\label{EngelCurveRepS}
\begin{subfigure}{.4\textwidth}
  \centering
\animategraphics[buttonsize=0.3cm,controls={play,stop, step},width=\linewidth]{1}{Electricity_Merge}{}{}
    \caption{Electricity}
   \label{fig:2a9}
\end{subfigure}%
 \begin{subfigure}{.4\textwidth}
  \centering
\animategraphics[buttonsize=0.3cm,controls={play,stop, step},width=\linewidth]{1}{Water_Merge}{}{}
      \caption{Water}
   \label{fig:2b9}
\end{subfigure}%

\begin{subfigure}{.4\textwidth}
  \centering
\animategraphics[buttonsize=0.3cm,controls={play,stop, step},width=\linewidth]{1}{Gas_Merge}{}{}
      \caption{Gas}
   \label{fig:2c9}
 \end{subfigure}%
\begin{subfigure}{.4\textwidth}
  \centering
\animategraphics[buttonsize=0.3cm,controls={play,stop, step},loop,width=\linewidth]{1}{Sewerage_Merge}{}{}
      \caption{Sewerage}
   \label{fig:2d9}
 \end{subfigure}%

\begin{subfigure}{.4\textwidth}
  \centering
\animategraphics[buttonsize=0.3cm,controls={play,stop, step},width=\linewidth]{1}{Numeraire_Merge}{}{}
      \caption{Numeraire}
   \label{fig:2e9}
 \end{subfigure}%
 
   \begin{subfigure}{.45\textwidth}
  \centering
  \includegraphics[width=0.8\linewidth]{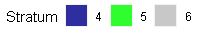}
   \label{fig:2f9}
 \end{subfigure}%
 
 {\scriptsize\caption*{\textit{Notes}: Black line is the mean value of the Engel curve. Red horizontal line is the income share at 0 centered log real income. Blue, green and gray are credible intervals for strata 4, 5 and 6.}}
 
\end{figure}

\begin{table}[h]
\centering \caption{Summary statistics: Utility shares and prices by unobserved heterogeneity clusters \label{sumstatCluster1}}
\begin{threeparttable}
\resizebox{0.90\textwidth}{!}{\begin{minipage}{\textwidth}
\begin{tabular}{l c c c c c}\hline
\multicolumn{1}{c}{\textbf{Variable}} & \textbf{Mean}
 & \textbf{Std. Dev.} & \textbf{Zero Shares} & & \\ 
 \hline
 \textbf{Shares} &  &  &  &  &  \\
 \hline
 \textbf{Cluster 1}& & & & & \\
 Electricity & 0.05 & 0.06 & 0.08 & & \\ 
 Water & 0.01 & 0.02 & 0.09 & & \\
 Sewerage & 8e-04 & 1e-03  & 0.63 & & \\
 Natural Gas & 5e-03 & 9e-03 & 0.30 & & \\
 Numeraire & 0.94 & 0.08 & 0.00 & & \\

  \textbf{Cluster 2}& & & & & \\
Electricity & 0.03 & 0.05 & 0.05 & & \\
Water & 0.02 & 0.03 & 0.05 & & \\
Sewerage & 3e-03 & 8e-03 & 0.51 & & \\ 
Natural Gas & 6e-03 & 0.01 & 0.20 & & \\
Numeraire & 0.94 & 0.07 & 0.00 & & \\

 \textbf{Cluster 3}& & & & & \\
Electricity & 0.06 & 0.06 & 0.14 & & \\
Water & 0.01 & 0.01 & 0.17 & & \\
Sewerage & 8e-04 & 3e-03 & 0.55 & & \\
Natural Gas & 6e-03 & 6e-03 & 0.38  & & \\
Numeraire & 0.92 & 0.07 & 0.00 & & \\

 \textbf{Cluster 4}& & & & & \\
Electricity & 0.03 & NA & 0.00 & & \\
Water & 2e-03 & NA & 0.00 & & \\
Sewerage & 3e-03 & NA & 0.00 & & \\
Natural Gas & 0.00 & NA & 1.00  & & \\
Numeraire & 0.96 & NA & 0.00 & & \\

\addlinespace[.75ex]
\hline
\textbf{Prices}& & &\textbf{N} & \textbf{Min} & \textbf{Max} \\ 
\hline
 \textbf{Cluster 1}& & & 233 & & \\
 Electricity (USD/kWh) & 0.16 & 0.01 & & 0.13 & 0.20 \\
 Water (USD/m3) & 0.66 & 0.29 & &  0.16 & 1.46 \\
 Sewerage (USD/m3) & 0.48 & 0.27 & & 0.16 & 1.32 \\
 Natural Gas (USD/m3) & 0.51 & 0.13 & & 0.30 & 0.91 \\
 \textbf{Cluster 2}& & & 5,517 & & \\
Electricity (USD/kWh) & 0.16 & 0.01 & & 0.13 & 0.27 \\
Water (USD/m3) & 0.67 & 0.25 & & 0.16 & 1.46 \\
Sewerage (USD/m3) & 0.57 & 0.24 & & 0.16 & 1.31 \\
Natural Gas (USD/m3) & 0.50 & 0.12 & & 0.09 & 0.91 \\
 \textbf{Cluster 3}& & & 29 & & \\
 Electricity (USD/kWh) & 0.17 & 0.01 & & 0.14 & 0.19 \\ 
 Water (USD/m3) & 0.65 & 0.26 & & 0.16 & 1.20 \\
 Sewerage (USD/m3) & 0.59 & 0.27 & & 0.24 & 1.31 \\
 Natural Gas (USD/m3) & 0.50 & 0.16 & & 0.29 & 0.85 \\
 \textbf{Cluster 4}& & & 1 & & \\
 Electricity (USD/kWh) & 0.15 & NA & & 0.15 & 0.15 \\ 
 Water (USD/m3) & 0.60 & NA & & 0.60 & 0.60 \\
 Sewerage (USD/m3) & 0.43 & NA & & 0.43 & 0.43 \\
 Natural Gas (USD/m3) & 0.33 & NA & & 0.33 & 0.33 \\
 \hline
\addlinespace[.75ex]
\end{tabular}
\begin{tablenotes}[para,flushleft]
 \footnotesize \textit{Notes}: Mean and standard deviation figures do not take zero shares into account. 
 
 \textbf{Source:} Authors' calculations based on posterior estimates.
  \end{tablenotes}
    \end{minipage}}
  \end{threeparttable}
\end{table}

\begin{table}[h]
\centering \caption{Descriptive Statistics: Household characteristics by unobserved heterogeneity clusters \label{demstatClust2}}
\begin{threeparttable}
\resizebox{0.65\textwidth}{!}{\begin{minipage}{\textwidth}
\begin{tabular}{l c c c c c}\hline
\multicolumn{1}{c}{\textbf{Variable}} & \textbf{Mean}
 & \textbf{Std. Dev.}& \textbf{Min.} &  \textbf{Max.} & \textbf{N}\\ \hline
 
\textbf{Cluster 1}& & & & & 233 \\
\hline
\addlinespace[.75ex]

Household head age & 55.13 & 17.82 & 18 & 95 & \\
Household head gender (female) & 0.48 &  & 0 & 1 & \\
Members & 2.71 & 1.60 & 1 & 9 & \\

\textbf{Education}& & & & & \\
Elementary school & 0.12 &  & 0 & 1 & \\
High School & 0.24 &  & 0 & 1 & \\
Vocational & 0.09 &  & 0 & 1 & \\
Undergraduate & 0.28 &  & 0 & 1 & \\
Postgraduate & 0.26 &  & 0 & 1 & \\

\textbf{Altitude}& & & & & \\
Below 1,000 m.a.s.l & 0.53 &  & 0 & 1 & \\
More than 1,000 m.a.s.l & 0.47 &  & 0 & 1 & \\

\textbf{Strata}& & & & & \\
Stratum 4 & 0.62 &  & 0 & 1 \\
Stratum 5 & 0.20 &  & 0 & 1 \\
Stratum 6 & 0.18 &  & 0 & 1 \\

Total income & 3,775.12 & 5,428.75 & 54.61 & 29,321.93 \\

\addlinespace[.75ex]
\hline
\textbf{Cluster 2}& & & & & 5,517 \\
\hline
\addlinespace[.75ex]

Household head age & 55.17 & 15.76 & 18 & 99 \\
Household head gender (female) & 0.41 &  & 0 & 1 \\
Members & 2.88 & 1.38 & 1 & 10 \\

\textbf{Education}& & & & & \\

Elementary school & 0.05 & & 0 & 1 \\
High School & 0.16 &  & 0 & 1 \\
Vocational & 0.12 &  & 0 & 1 \\
Undergraduate & 0.40 &  & 0 & 1 \\
Postgraduate & 0.28 &  & 0 & 1 \\

\textbf{Altitude}& & & & & \\
Below 1,000 m.a.s.l & 0.43 &  & 0 & 1 \\
More than 1,000 m.a.s.l & 0.57 &  & 0 & 1 \\

\textbf{Strata}& & & & & \\
Stratum 4 & 0.67 &  & 0 & 1 \\
Stratum 5 & 0.22 &  & 0 & 1 \\
Stratum 6 & 0.11 &  & 0 & 1 \\

Total income & 1,976.81 & 1,714.84 & 165.50 & 23,135.80 \\

\addlinespace[.75ex]
\hline
\textbf{Cluster 3}& & & & & 29 \\
\hline
\addlinespace[.75ex]

Household head age & 48.62 & 19.82 & 18 & 79 \\
Household head gender (female) & 0.52 &  & 0 & 1 \\
Members & 2.31 & 1.61 & 1 & 7 \\

\textbf{Education}& & & & & \\
Elementary school & 0.03 &  & 0 & 1 \\
High School & 0.41 &  & 0 & 1 \\
Vocational & 0.14 &  & 0 & 1 \\
Undergraduate & 0.28 &  & 0 & 1 \\
Postgraduate & 0.14 &  & 0 & 1 \\

\textbf{Altitude}& & & & & \\
Below 1,000 m.a.s.l & 0.40 &  & 0 & 1 \\
More than 1,000 m.a.s.l & 0.60 &  & 0 & 1 \\

\textbf{Strata}& & & & & \\
Stratum 4 & 0.69 &  & 0 & 1 \\
Stratum 5 & 0.24 &  & 0 & 1 \\
Stratum 6 & 0.07 &  & 0 & 1 \\

Total income & 6,869.55 & 14,714.88 & 22.65 & 50,483.60 \\

\addlinespace[.75ex]
\hline
\textbf{Cluster 4}& & & & & 1 \\
\hline
\addlinespace[.75ex]

Household head age & 52.00 & NA & 52 & 52 \\
Household head gender (female) & 1.00 &  & 1 & 1 \\
Members & 2.00 & NA & 2 & 2 \\

\textbf{Education}& & & & & \\
Elementary school & 0.00 &  & 0 & 0 \\
High School & 0.00 &  & 0 & 0 \\
Vocational & 1.00 &  & 1 & 1 \\
Undergraduate & 0.00 &  & 0 & 0 \\
Postgraduate & 0.00 &  & 0 & 0 \\

\textbf{Altitude}& & & & & \\
Below 1,000 m.a.s.l & 1.00 &  & 1 & 1 \\
More than 1,000 m.a.s.l & 0.00 &  & 0 & 0 \\

\textbf{Strata}& & & & & \\
Stratum 4 & 1.00 &  & 1 & 1 \\
Stratum 5 & 0.00 &  & 0 & 0 \\
Stratum 6 & 0.00 &  & 0 & 0 \\

Total income & 38.42 & NA & 38.42 & 38.42 \\

\hline

 \hline
\end{tabular}
\begin{tablenotes}[para,flushleft]
 \footnotesize \textbf{Source:} Authors' calculations based on posterior estimates.
  \end{tablenotes}
      \end{minipage}}
  \end{threeparttable}
\end{table}

\end{document}